\documentclass[11pt,a4paper]{article}
\pdfoutput=1
\usepackage{jheppub}

\usepackage{amsmath}
\usepackage{verbatim}
\usepackage{amssymb}
\usepackage{inputenc}
\usepackage{textcomp}
\usepackage{appendix}
\usepackage{floatrow}
\usepackage{longtable}

\newcommand{\e}{\epsilon}
\newcommand{\be}[1]{\begin{equation}\label{#1} }
\newcommand{\ee}{\end{equation}}
\newcommand{\bea}[1]{\begin{eqnarray}\label{#1} }
\newcommand{\eea}{\end{eqnarray}}
\newcommand{\p}{\partial}
\newcommand{\refb}[1]{(\ref{#1})}

\renewcommand{\L}{{\mathcal{L}}}

\newcommand{\bL}{\bar{{\mathcal{L}}}}

\newcommand{\D}{\Delta}

\renewcommand{\a}{\alpha}

\renewcommand{\b}{\beta}

\newcommand{\s}{\sigma}

\usepackage{multirow}
\newcommand{\bs}{\begin{split}}
\newcommand{\es}{\end{split}}
\newcommand{\bes}{\begin{subequations}}
\newcommand{\ees}{\end{subequations}}
\newcommand{\sss}{\scriptscriptstyle}

\newcommand{\mtD}{\mathcal{D}}

\renewcommand{\i}{\Big}
\newcommand{\non}{\nonumber}

\renewcommand{\a}{\alpha}

\newcommand{\Lim}[1]{\raisebox{0.5ex}{\scalebox{0.8}{$\displaystyle \lim_{#1}\;$}}}

\title{Field Theories with Conformal Carrollian Symmetry}

\author[a, b]{Arjun Bagchi,} \author[a,c]{Aditya Mehra,} \author[a]{and Poulami Nandi.} \author{\\}
\affiliation[a]{Indian Institute of Technology Kanpur, Kalyanpur, Kanpur 208016. INDIA\\} 

\affiliation[b]{Centre for Particle Theory, Department of Mathematical Sciences,
Durham University, \\ South Road, Durham DH1 3LE, UK \\}

\affiliation[c]{International Institute of Physics, Federal University of Rio Grande do Norte,
\\ Campus Universitario, Lagoa Nova, Natal, RN 59078-970, BRAZIL\\}

\emailAdd{abagchi@iitk.ac.in, amehra@iip.ufrn.br, poulamin@iitk.ac.in}

\abstract{Conformal Carrollian groups are known to be isomorphic to Bondi-Metzner-Sachs (BMS) groups that arise as the asymptotic symmetries at the null boundary of Minkowski spacetime. The Carrollian algebra is obtained from the Poincare algebra by taking the speed of light to zero, and the conformal version similarly follows. In this paper, we construct explicit examples of Conformal Carrollian field theories as limits of relativistic conformal theories, which include Carrollian versions of scalars, fermions, electromagnetism, Yang-Mills theory and general gauge theories coupled to matter fields. Due to the isomorphism with BMS symmetries, these field theories form prototypical examples of holographic duals to gravitational theories in asymptotically flat spacetimes. The intricacies of the limiting procedure leads to a plethora of different Carrollian sectors in the gauge theories we consider. Concentrating on the equations of motion of these theories, we show that even in dimensions $d=4$, there is an infinite enhancement of the underlying symmetry structure. Our analysis is general enough to suggest that this infinite enhancement is a generic feature of the ultra-relativistic limit that we consider. }

\preprint{}

\begin{document}
\maketitle
\vfill

\section{Introduction}

\subsection{Effective theories and singular limits}

Physics is all about length scales, and most theories that we use to describe Nature are effective theories, including the spectacularly successful Standard Model of particle physics. Effective theories are often formulated by going to a certain sub-sector of a higher theory. Hence, often there is a lot to be learnt by focussing on a sub-sector of a particular theory. One of the very obvious examples that come to mind in this regard is the non-relativistic sector of well defined relativistic theories, which become important for most of the physics around us. Galilean physics is obtained from relativistic physics by taking the speed of light to infinity. Group theoretically, the symmetries contract from the relativistic Poincare group to the Galilean group as one takes $c\to \infty$. Various counter-intuitive things happen in this singular limit. The light-cones open up, the spacetime metric degenerates and the Riemannian structure of spacetime manifold is lost. New mathematical structures emerge, viz. the spacetime manifold becomes what is called a Newton-Cartan manifold (for a brief introduction see e.g. \cite{Misner:1974qy}), and often the underlying symmetries get enhanced. Some of these ideas have come into light only recently and this is a very fertile area of current research \cite{Nishida:2007pj,Son:2013rqa,Bagchi:2014ysa,Bagchi:2015qcw, Bagchi:2017yvj, Bergshoeff:2015uaa, Bergshoeff:2014uea,Festuccia:2016caf, Hartong:2015wxa,Bleeken:2015ykr}, some of which have been motivated by understanding the notion of holography for non-relativistic systems \cite{Son:2008ye, Balasubramanian:2008dm, Kachru:2008yh, Maldacena:2008wh, Bagchi:2009my}.   

Out of mathematical curiosity, one could ask if similar new structures may emerge when one takes the speed of light to go to zero instead of infinity. This peculiar $c\to0$ limit has been dubbed the Carrollian limit in literature \cite{Leblond65}. This defines a contraction of the Poincare group different from the Galilean one, and the group so generated is called the Carrollian group. In this paper, we would be interested in constructing field theoretic examples that exhibit Carrollian symmetry. A selection of previous related work in this direction includes \cite{Duval:2014uoa, Bagchi:2016bcd, Bergshoeff:2017btm, Ciambelli:2018xat,  Basu:2018dub}. 

Purely mathematical curiosity is perhaps not very good justification for research in theoretical physics, although many very important breakthroughs, like the formulation of Yang-Mills theories that later have led to the understanding of the weak and strong nuclear forces, have come via this path. Our investigations in this paper are driven by some very strong motivations, which we elaborate on below.    

\subsection{Holographic duals to flatspace} 
It is more than 20 years since the advent of the famed AdS/CFT correspondence \cite{Maldacena:1997re}, which has given a firm footing to the idea of holography \cite{tHooft:1993dmi,Susskind:1994vu}. This has opened up an avenue of studying quantum gravity by looking at quantum field theories and vice versa. Our physical world is, however, clearly not AdS. For many applications, especially astrophysical ones, the universe can be well approximated by an asymptotically flat spacetime. It is thus of great importance to extend the notion of holography from its original setting in asymptotically AdS spacetimes to flat backgrounds. 

A natural way to construct a holographic quantum field theory for a general gravitational theory is to consider the symmetry structure at the boundary of the spacetime in which the gravitational theory lives. These symmetries at the boundary are formally given by the Asymptotic Symmetry Group (ASG) and its associated algebra, the Asymptotic Symmetry Algebra. In mathematical terms, given a set of boundary conditions, 
\be{}
\mbox{Asymptotic Symmetry Group} = \frac{\mbox{Group of all allowed diffeomorphisms}}{\mbox{Group of trivial diffeomorphisms}}\ .
\ee
One can then propose that the dual field theory lives on the asymptotic boundary of the spacetime and inherits the symmetry of the ASG. This analysis of the ASG of AdS$_3$ famously led Brown and Henneaux \cite{Brown:1986nw} to two copies of the Virasoro algebra which are of course the symmetries of a 2d conformal field theory. This could be looked upon as the principle precursor to the discovery of AdS/CFT. Recently, similar asymptotic symmetry analysis for theories of higher spin in AdS$_3$ \cite{Campoleoni:2010zq} has led to the uncovering of higher spin dualities \cite{Gaberdiel:2010pz}. Earlier, similar ideas were used to propose a holographic dual to de Sitter spacetimes \cite{Strominger:2001pn}. 

We wish to extend this analogy to the case of asymptotically flat spacetimes. For Einstein gravity in 3 and 4 dimensional Minkowski spacetime, the asymptotic symmetries at the null boundary of spacetime are given, not by the Poincare group, but by the infinite dimensional Bondi-Metzner-Sachs (BMS) groups. The associated algebras are  
\paragraph{{\em{BMS}}$_3$:} 
\bea{bms3}
&&[L_n, L_m] = (n-m)L_{m+n}, \quad [L_n, M_m] = (n-m)M_{m+n} + \frac{3}{G} \delta_{n+m,0} (n^3-n), \\
&& [M_n, M_m] = 0. \nonumber
\eea
The structure at null infinity in 3d is ${\rm I\!R} \times \mathbb{S}^1$. $L_n$ form the diffeomorphisms of the circle at $\mathcal{I}^+$, while the $M_n$'s are angle-dependent translations called supertranslations. The central charge $3/G$ to the $[L,M]$ commutator is the value for Einstein gravity \cite{Barnich:2006av}.  One could in general have a non-zero central extension to the $[L,L]$ commutator. For Einstein gravity, this is zero. But one could add a topologically massive term to generate a non-zero central extension for theories of gravity with higher derivative interactions \cite{Bagchi:2012yk}. 

\medskip

\noindent In 4d, $\mathcal{I}^+$ becomes ${\rm I\!R} \times \mathbb{S}^2$. The structure on $\mathbb{S}^2$ enhances to two copies of the Witt algebra, very much like 2d CFTs \cite{Barnich:2010eb} and the algebra becomes
\paragraph{{\em{BMS}}$_4$:} 
\bea{bms4}
&&[L_n, L_m] = (n-m)L_{m+n}, \quad [\bar{L}_n, \bar{L}_m] = (n-m)\bar{L}_{m+n}\nonumber\\  
&& [L_n, M_{r,s}] = \left(\frac{n+1}{2} -r \right)M_{n+r, s}, \quad [\bar{L}_n, M_{r,s}] = \left(\frac{n+1}{2} -s \right)M_{r, n+s} \\
&& [M_{r,s}, M_{p,q}] = 0. \nonumber
\eea
One cannot find physical boundary conditions (e.g. without excluding gravitational radiation and the Schwarzschild black holes in 4d) so as to limit these infinite dimensional groups to the finite Poincare group in dimensions 3 and 4{\footnote{It is important to mention here that another infinite dimensional extension of the BMS$_4$ exists where the infinite dimensional group is taken to be the semi-direct product of the supertranslations with $Diff(S^2)$, i.e. the group of all smooth diffeomorphisms on the conformal sphere S$^2$ \cite{Campiglia:2014yka}. We shall however not be interested in this or its possible generalisations (or the lack of it) to higher dimensions.}}. 

This is however not true in $d>4$. Here with boundary conditions strict enough one can reduce the ASG to $ISO(d-1, 1)$ \cite{Hollands:2003ie,Tanabe:2009va,Tanabe:2011es}. Interestingly though, one can find looser boundary conditions so that there are infinite enhancements in $d>4$ \cite{Kapec:2015vwa}. This begs the question as to which set of boundary conditions is the more physically relevant. 

Of late, there has been a resurgence in the study of physics in flat-spacetimes, initiated by Strominger and his collaborators. An incomplete list of interesting work in this direction includes \cite{Strominger:2013jfa, Strominger:2013lka, He:2014laa, Cachazo:2014fwa, Kapec:2014opa, Strominger:2014pwa,Pasterski:2015tva,He:2014cra,Hawking:2016msc,Laddha:2017ygw,Campiglia:2015kxa,Campiglia:2015yka,Banerjee:2018gce}. A very intriguing picture has emerged linking asymptotic symmetries with soft theorems in QFT in asymptotically flat spacetimes and to memory effects. The interested reader is pointed to \cite{Strominger:2017zoo} for a review of the current developments in this field. The asymptotic BMS symmetries arise at null infinity and it is thus very natural that these should play an important role in scattering theory in flat space. Weinberg's soft graviton theorem \cite{Weinberg:1965nx} relates the S-matrix element of a theory of quantum gravity to another which differs from the original one by the addition of a soft graviton. Recently it has been shown that this soft-graviton theorem actually arises out of a supertranslation Ward identity \cite{He:2014laa}. Weinberg's theorem does not depend on the dimension of spacetime where the quantum theory of gravity lives. So, if we were to assume that the relation between asymptotic symmetries and soft theorems exist in all dimensions, then these infinitely extended supertranslation symmetries should also exist in all dimensions. It is thus perhaps of greater physical significance to consider looser boundary conditions in $d>4$ which generate an infinite dimensional asymptotic symmetry algebra. 

If we are to draw inspiration from AdS/CFT, holography in asymptotically flat spacetimes should involve these infinite dimensional symmetry algebras and the putative dual theories should be non-gravitational quantum field theories living on the null boundary and invariant under the infinite extended BMS algebras. A natural avenue to explore flat holography is to investigate the singular limit where the bulk theory goes from AdS to flat space, i.e. taking the radius of AdS to infinity. It can be shown that this leads to an ultra-relativistic contraction of the boundary CFT \cite{Bagchi:2012cy}. So these conformal versions of Carrollian theories that we discussed in the beginning are putative duals of flat space. It has also been shown that the conformal Carrollian symmetries are isomorphic to BMS symmetries \cite{Bagchi:2010eg,Duval:2014uva}.  

A relativistic CFT in $d$ dimensions has a finite symmetry algebra, viz. $so(d,2)$. The ultra-relativistic contraction of this also obviously leads to a finite symmetry algebra $iso(d,1)$. This, clearly, is also the process of taking radius of AdS$_{d+1}$ to infinity to get to flat space, which has $(d+1)$ dimensional Poincare symmetry. Even before speaking of infinite extensions, it is clear that these $d$-dimensional field theories are exotic theories. For example, for asymptotically flat spacetimes in 4 dimensions, we are seeking a field theory in $d=3$ that is invariant under $iso(3,1)$. These are clearly not usual relativistic field theories and as we have stated just above, one of the natural ways to generate them is to consider Carrollian limits of relativistic CFTs. 
 
The case of $d=2$ is special. Here, as is very well known, the relativistic conformal algebra enhances to two copies of the infinite dimensional Virasoro algebra, which under the Carrollian limit, generate the ultra-relativistic algebra \refb{bms3}: 
\be{}
L_n = \L_n - \bL_{-n}, \quad M_n = \e\left( \L_n + \bL_{-n}\right), \quad \e \to 0.
\ee
Here $\e$ is the speed of light in the 2d theory. In holographic terms, the identification is as follows: $\e = G/\ell$, with $\ell$ being the AdS radius and $G$ is the Newton's constant. This is thus also the flatspace limit. The ultra-relativistic contraction and its link to the large radius limit of AdS is explored in detail in  \cite{Bagchi:2012cy}. The symmetries in both the parent AdS theory and its contracted flat space version remain infinite dimensional and there is a lot to be learnt by exploiting this singular limit. This has been used to understand features of flat holography in 3 bulk and 2 boundary dimensions. An incomplete list of references include \cite{Bagchi:2010eg,Bagchi:2012cy,Bagchi:2012yk,Bagchi:2012xr,Barnich:2012xq,Cornalba:2003kd,Bagchi:2013qva,Barnich:2012rz,Bagchi:2013lma,Bagchi:2014iea,Bagchi:2015wna,Barnich:2014kra,Krishnan:2013wta,Barnich:2015uva,Hartong:2015usd,Campoleoni:2016vsh,Duval:2014uva}. The interested reader is pointed to \cite{Bagchi:2016bcd,Riegler:2016hah} for a summary of work in this direction. Some of the more recent interesting developments include \cite{Bagchi:2016geg,Jiang:2017ecm,Hijano:2017eii,Hijano:2018nhq}.

In boundary dimensions $d>3$, the story is very different. The limit from AdS yields just finite dimensional symmetries. But, in flat space, we need more,- we need to generate infinite dimensional supertranslation symmetries in order to relate to Weinberg's theorem. Starting out with finite symmetries on the boundary via the process of contraction that we have just outlined above, we thus need a mechanism to generation these infinite supertranslations. Following earlier work \cite{Bagchi:2016bcd}, we describe an algebraic process which achieves this. We will focus specifically in $d=4$ and find that generic field theories generated by this process of contraction naturally have these infinite dimensional symmetries that emerge in this singular limit.   

\subsection{Other potential applications} 

Taking the speed of light to zero focuses us on a subsector of a parent theory that is very highly energetic. One could think of doing this on the worldsheet of the relativistic string. This singular limit, interestingly, is also the tensionless limit of string theory \cite{Schild:1976vq} where a greater symmetry structure has been conjectured to exist and this has been in focus for its relation to recent higher spin holographic theories \cite{Gaberdiel:2012uj}. The residual symmetry on the tensionless string worldsheet after fixing the equivalent of the conformal gauge turns out to be \refb{bms3} \cite{Isberg:1993av, Bagchi:2013bga,Duval:2014lpa}. The construction of tensionless strings has been recently revisited in connection to these symmetry structures \cite{Bagchi:2015nca, Bagchi:2016yyf, Bagchi:2017cte}. It is very likely that when one is interested in taking the tensionless limit of membranes, higher BMS or equivalently conformal Carrollian field theories would have a decisive role to play \cite{Duval:2014lpa}. 

\subsection{Outline of this paper}
In this paper, we systematically construct Carrollian field theories which are further invariant under the Conformal Carrollian group or equivalently the BMS group (in one higher dimension). We start from well-known relativistic conformal field theories and show that in the ultra-relativistic limit, the contraction of these field theories become invariant under the infinitely-extended conformal Carrollian group. In Sec.~\ref{Conformal Carrollian Symmetry}, we begin with a review of the algebraic aspects of conformal Carrollian symmetry. We discuss the process of contraction that gets one from the relativistic conformal algebra to first the finite conformal Carrollian  Algebra (CCA) and then proceed to give the CCA an infinite dimensional lift. We then provide a detailed survey of the representation theory of this algebra. 

In Sec.~\ref{sec3}, we begin our construction of field theories invariant under the CCA. We start with scalar fields and outline the process of finding the symmetries of the limiting field theory by focusing on equations of motion (EOM). We are thus discussing on-shell symmetries throughout this work. We then move on to fermions and then to the Carrollian version of Yukawa theory. We focus on dimensions $d=4$. In each case, the study of EOM reveals an infinite symmetry enhancement of the underlying theory to the (infinitely) extended conformal Carrollian  group. 

In Sec.~\ref{sec4}, we revisit the construction of Carrollian electrodynamics and the emergence of the Electric and Magnetic limits here \cite{Duval:2014uoa,Bagchi:2016bcd}. We then add matter to these theories. We showcase the example of Carrollian scalar electrodynamics with the most general scalings of the (massless) matter fields and show how consistency of the theory reduces the allowed parameter space of scalings to a particular patch in each of the Electric and Magnetic limits. All these allowed values of scalings give rise to EOM which are invariant under the infinite CCA. We show this for a theory with an arbitrarily chosen set of parameters. We then add fermions to the $U(1)$ theory and perform a similar analysis and find infinite symmetries of the EOM. 

In Sec.~\ref{cfdf}, we move on to non-Abelian theories with Carrollian invariance. First we review pure Carrollian Yang-Mills theory following earlier work \cite{Bagchi:2016bcd}. We then go on to add fermionic matter to these theories and investigate the various limits arising from the parent theory. The (gory!) details of the $SU(2)$ theory are worked out in Appendix \ref{Carrollian $SU(2)$ Yang-Mills with Fermions}. In the main body of the paper, we present the general analysis of SU$(N)$ Carrollian Yang-Mills theory coupled to fermionic matter. We end up with numerous possible sectors in the UR limit which we systematically analyse. A number of these sectors can be discarded for the absence of kinetic terms in some equations of motion. We again obtain that in all of these sectors, the symmetries of the EOM get enhanced infinitely and the CCA turns out to be the relevant symmetry algebra. This enhancement of symmetries in the ultra-relativistic limit of conformal field theories thus seems to be generic and we propose that this would be true at the classical level for all dimensions and all conformal field theories.  

We conclude with a summary of our results and a number of possible directions of future work. Appendix \ref{Galilean $SU(N)$ Yang-Mills Theory with matter} contains details of the cousin of the Carrollian Yang-Mills theory, viz. the Galilean SU$(N)$ Yang-Mills theory and associated additional fermionic matter. We present the similarities and differences between the two theories in Appendix \ref{Comparison between Galilean and conformal Carrollian  field theories}. 

\newpage

\section{Conformal Carrollian Symmetry} 
\label{Conformal Carrollian Symmetry}

We begin with some algebraic preliminaries. We will first consider how to obtain conformal Carrollian algebras from relativistic conformal algebras and then go on to a detailed review of the representation theory which we would use throughout the rest of the paper. 

\subsection{Contraction and algebra}

The ultra-relativistic limit of a relativistic conformal field theory is obtained by performing an In{\"o}n{\"u}-Wigner contraction on the generators of conformal group. Let us consider a $d$ dimensional CFT. The limit that one needs to take in terms of the spacetime coordinates is: 
\be{}\label{stc}
x_i\to x_i,~ t \to \e t,~ \e \to 0,
\ee
where $i= 1, \ldots, (d-1)$. This means we are sending the speed of light, $c \to 0$. Under this limit, the relativistic conformal algebra contracts and the generators of the contracted algebra, which is the finite conformal Carrollian  algebra (CCA), are written as:
\bea{}
&&\non B_i=x_i \p_t, ~ J_{ij}=(x_i \p_j-x_j \p_i), ~ H=\p_t, ~ P_i=\p_i,\\
 &&D= (t \p_t+x_i \p_i),~ K= x_i x_i \p_t,~ K_j=2x_j(t\p_t+x_i\p_i)-(x_i x_i)\p_j.
\eea
The non vanishing brackets of the finite CCA generators are given by:
\bea{}\label{algebra}
&&\non [J_{ij}, B_k ]=\delta_{k[j}B_{i]}, ~ [J_{ij}, P_k ]=\delta_{k[j}P_{i]}, ~ [B_i,P_j]=-\delta_{ij}H, ~ [ J_{ij},J_{kl}]=J_{i[l}\delta_{k]j}+J_{j[k}\delta_{l]i},\\
&&\non [D,K]=K,~[K,P_i]=-2B_i, ~ [K_i,P_j]=-2D\delta_{ij}-2J_{ij},~ [H,K_i]=2B_i,\\
&&[D,H]=-H, ~[D,P_i]=-P_i,~[D,K_i]=K_i.
\eea
where $J_{ij},H,P_i,B_i$ are the rotation generators in $(d-1)$ spatial dimensions, time translation, spatial translations, and Carrollian boosts respectively. These form the Carrollian group, which is the ultra-relativistic analogue of the Galilean group. The conformal generators are dilatation $D$, and the Carrollian versions of temporal and spatial conformal generators $K, K_i$. The set of the generators $\{J_{ij},P_i,D,K_i\}$ close to form a $so(d-1)$ subalgebra of the finite conformal Carroll algebra, which unsurprising is $iso(1,d-1)$. This $so(d-1)$ subalgebra can be looked upon as the conformal generators of the $\mathbb{S}^{d-1}$ at the null boundary of flat spacetime, which as stated earlier, has a structure ${\rm I\!R} \times \mathbb{S}^{d-1}$. 

In \cite{Bagchi:2016bcd}, it was shown that it is possible to give the CCA infinite dimensional extensions. These are different for different dimensions. The infinite extensions in the $d=2$ case yields \refb{bms3} and for $d=3$ gives \refb{bms4}. Here we are interested in the $d=4$ construction. We have three spatial directions and one temporal one. Of course, for $d\geq4$, there is no infinite enhancement of the (relativistic) conformal symmetries on the sphere at infinity like in the 2d and 3d examples. The proposed infinite extension for the generators is thus only in the supertranslation part: 
\be{}
M^{m_1,m_2,m_3} =x^{m_1}y^{m_2}z^{m_3}\p_t.
\ee
The finite part of the algebra described above fit into this scheme as follows: 
\bea{}
&& H=M^{0,0,0},~
B_x=M^{1,0,0}, \ B_y=M^{0,1,0}, \ B_z=M^{0,0,1}, \nonumber\\
&& K=M^{2,0,0}+M^{0,2,0}+M^{0,0,2}. 
\eea 
The infinite part of CCA algebra are given by the lie brackets \cite{Bagchi:2016bcd,Basu:2018dub}:
\bea{}
&&\non [P_x, M^{m_1,m_2,m_3}]=m_1 M^{m_1-1,m_2,m_3},\\
&&\non [D,M^{m_1,m_2,m_3}]=(m_1+m_2+m_3-1) M^{m_1,m_2,m_3},\\
&&\non [K_x,M^{m_1,m_2,m_3}]=(m_1+2m_2+2m_3-2) M^{m_1+1,m_2,m_3}\\&&\non\hspace{3.2cm} -m_1(M^{m_1-1,m_2+2,m_3}+M^{m_1-1,m_2,m_3+2}),\\
&&[J_{xy},M^{m_1,m_2,m_3}]=m_2 M^{m_1+1,m_2-1,m_3}-m_1 M^{m_1-1,m_2+1,m_3}.
\eea
The other commutators (e.g. for the other components of $P_i$ or $K_i$) follow similarly. The supertranslations form an abelian sub-algebra among themselves:
\bea{}
[M^{m_1,m_2,m_3}, M^{n_1,n_2,n_3}]=0.
\eea

\medskip

\paragraph{Generalisation to higher dimensions:}
The above construction has a natural generalisation to dimensions $d>4$. Again, we would look to give an infinite lift to only the supertranslation generators:
\begin{equation}\label{higherD}
M^{m_1,m_2,\hdots ,m_n}=(x_1^{m_1}x_2^{m_2} \hdots x_n^{m_n})\partial_t.
\end{equation}
The conformal structure on the $\mathbb{S}^d$ remains finite dimensional and generated by the $so(d)$ algebra mentioned above. The algebra of generators follows immediately and is a direct generalisation of the $d=4$ case: 
\begin{eqnarray}
&&\non [P_i, M^{m_1,m_2,\hdots,m_n}] =m_i  M^{m_1,\hdots,m_i-1,\hdots,m_n},\\
&&\non [D,M^{m_1,m_2,\hdots, m_n }] =- M^{m_1,m_2,\hdots ,m_n}+\displaystyle\sum_{m_i=m_1}^{m_n} m_i M^{m_1,\hdots,m_i,\hdots,m_n},\\
&&\non [K_j,M^{m_1,m_2,\hdots,m_n}]=-2 M^{m_1,\hdots,m_j+1,\hdots,m_n}+m_j M^{m_1,\hdots,m_j+1,\hdots,m_n},\\
&&\non \hspace{3.7cm}+\displaystyle\sum_{m_i \neq m_j}( 2m_i M^{m_1,\hdots,m_i,m_j+1,\hdots,m_n}-m_jM^{m_1,\hdots,m_i+2,m_j-1,\hdots m_n} ),\\
&&[J_{ij},M^{m_1,m_2,\hdots,m_n}]=m_j M^{m_1,\hdots,m_i+1,m_j-1,\hdots m_n}-m_i M^{m_1,\hdots,m_i-1,m_j+1,\hdots m_n}.
\end{eqnarray}
Again, the supertranslations form an abelian sub-algebra by themselves:
\be{}
[M^{m_1,m_2,\hdots,m_n}, M^{p_1,p_2,\hdots,p_n}] = 0.
\ee
 The infinite supertranslations can also be written as:
\be{}
M_f=f(x_i)\p_t, ~~\text{where}~~f(x_i)=(x_1^{m_1}x_2^{m_2} \hdots x_n^{m_n})
\ee
and the above algebra can be written in a compact form for $d\geq4$ \cite{Basu:2018dub}:
\bea{}
&&\non [P_i, M_f] =M_{\p_i f},\quad  [D,M_f] =M_h,~\text{where}~h=x_i \p_i f-f,\\
&&\non [K_i,M_f]= M_{\tilde{h}},~\text{where}~\tilde{h}=2x_i h-x_k x_k\p_i f,\\
&&[J_{ij},M_f]= M_g,~\text{where}~{g}=x_{[i}\p_{j]}f.
\eea

\medskip

\subsection{Representation Theory} \label{Representation Theory}
We now discuss the building blocks of infinite conformal Carrollian  algebra, the representation theory based on highest weights. Unless otherwise stated, our construction would be for $d=4$. As with the algebra described above, it is expected that the representation theoretic aspects would also have a natural generalisation to higher dimensions. This subsection is primarily intended to be a detailed review of earlier work \cite{Bagchi:2016bcd} along with an extension in case of different integer and half-integer spins. We aim to study the conformal Carrollian limit of different free and interacting field theories in $d=4$. Each of these field theories can be classified based on their spins. Hence, a natural choice of constructing the representation theory is to label the states by the eigenvalues of dilatation and rotation operators. Thus, the states will be labelled with definite scaling dimension $\D$ and spin $j$. This particular choice of representation is known as the scale-spin representation. Below we describe in detail the scale-spin representation of spin 0, spin $\frac{1}{2}$ and spin 1 fields.

\paragraph{Scale-spin representation:} As just stated above, we would label the states by their scaling dimension $\D$ and spin $j$.
\be{}
D |\Phi \rangle= \D |\Phi\rangle, ~~ J^2 |\Phi\rangle=j (j+1)|\Phi\rangle.
\ee
where, $J^2$ is the sum of square of the three rotation generators in three spatial dimensions. We propose that similar to 2d CFT, in these conformal Carrollian theories, there exists a state-operator correspondence:
\be{stateop}
\Lim{(x_i,t) \to (0,0)} \Phi(x_i,t) |0 \rangle = |\Phi \rangle.
\ee
This is not strictly necessary, but helps us in translating from operator language to states. It follows that
\be{stateops} 
[D ,\Phi(0,0)]= \D \Phi(0,0), ~ [J^2 \Phi(0,0)]=j (j+1)\Phi(0,0).
\ee
In \eqref{stateops} the generators have been promoted to operators and the brackets imply commutators henceforth. 

Next, we introduce the notion of primaries in a way similar to CFTs. We demand that the spectrum of these field theories (i.e. the value of $\D$) is bounded from below. Following \cite{Bagchi:2016bcd}, the conformal Carrollian  primary is defined as an operator annihilated by all lowering operators. This means for the finite algebra:
\be{}
[K_i, \Phi(0,0) ]=0,~~[K, \Phi(0,0) ]=0. 
\ee
This translates to 
\be{}
[M^{m_1,m_2,m_3},\Phi(0,0) ]=0\: \text{for at least one of } m_i  > 1,
\ee
for the infinite dimensional algebra. 
Under rotation, 
\be{} 
[J_{ij}, \Phi(0,0) ]=\Sigma_{ij}  \Phi(0,0).
\ee 
where $\Sigma_{ij}$  is the spin operator. The spatial and time translation of a generic field at any spacetime point is generated by the Hamiltonian $H$ and momentum  operators $P_i$ as
\be{}
[H, \Phi(t,x) ]=\p_t \Phi(t,x),~~ [P_i, \Phi(t,x) ]=\p_i \Phi(t,x). 
\ee
From \eqref{algebra}, it is clear that the primaries are not eigenstates of Carrollian boosts $B_i$. Hence, following previous work \cite{Bagchi:2016bcd} we use the Jacobi identity to examine the action of boosts on the primaries:
\be{}
[J_{ij},[B_k,\Phi(0,0)]]=[B_k,\Sigma_{ij}\Phi(0,0)]+\delta_{k[i}[B_{j]},\Phi(0,0)].
\ee
The most general transformation is
\be{}
[B_k,\Phi(0,0)]=r\varphi_k+\, f \sigma_k\phi + f^{\prime} \sigma_k\chi\, + a A_t \delta_{ik}+b A_k+ \hdots
\ee
Here, $\varphi, \,\{\phi,\chi \},\,  \{A_t , A_k\}$ are primaries of different spins and $r,\{f,f^\prime\},\{a,b\}$ are some constants yet to be determined. For the purposes of this paper, we will truncate the above expansions at spin 1. In principle, we could have mixing with higher spins as well. 

\medskip

Now we review the action of the finite and infinite CCA generators on the primaries at any general spacetime point. A CCA primary $\Phi(t,x)$ is related to its own version at the origin by  
\be{}
\Phi(t,x)=U \Phi(0,0) U^{-1},~ \text{where, } ~ U=e^{-tH-x_i P_i}.
\ee
Hence, for any finite CCA generator $\mathcal{O}$
\be{}
[ \mathcal{O}, \Phi(t,x)]= U [ U^{-1}\mathcal{O}U, \Phi(0,0)] U^{-1}.
\ee
We employ Baker-Campbell-Hausdorff formula along with the algebra in \eqref{algebra} to recast the above expression for finite and infinite generators in a simplified form as 
\bea{repgen}
&&\non\hspace{-.25cm}  [J_{ij}, \Phi(t,x)]
= (x_i \p_j-x_j \p_i ) \Phi(t,x)+\Sigma_{ij}\Phi(t,x),\\
&&\hspace{-.25cm} \non [B_j, \Phi(t,x)]
=x_j\p_t \Phi(t,x)+U[ B_j, \Phi(0,0)]U^{-1},\\ 
&&\non\hspace{-.25cm}  [D, \Phi(t,x) ]
= (\D+t\p_t+x_i \p_i) \Phi(t,x),\\ 
&&\non\hspace{-.25cm}  [K_j, \Phi(t,x) ]
=  (2\Delta x_j+2x_jt\partial_t+2x_i x_j \partial_i-2x_i \Sigma_{ij}- x_i x_i \partial_j )\,\Phi(t,x)+2t \, U [B_j, \Phi(0,0)]U^{-1},\\ 
&&\hspace{-.25cm}  [M^{m_1,m_2,m_3}, \Phi(t,x) ]= (x^{m_1}y^{m_2}z^{m_3})\p_t \Phi(t,x)+\p_i (x^{m_1}y^{m_2}z^{m_3})\:U [ B_i, \Phi(0,0)]U^{-1}.
\eea
The equations \eqref{repgen} are the action of finite and infinite symmetry generators on a CCA primary. Below we explicitly write down the transformations for scalars, fermions and spin-1 vector bosons. 

\medskip

\paragraph{Scalars $(\varphi)$:} 
\be{scalar} 
[B_j,\varphi(0,0)]=  r\varphi_j(0,0),~~ [\Sigma_{ij},\varphi(t,x)]=0.\\
\ee
The representation of CCA for scalars is given by  $\{\D,j,r\}$.  These would be fixed by input from dynamics, e.g. when we will demand that the Carrollian scalar theory emerges as a limit of the relativistic scalar field theory, as we do in the next section, we would be able to fix the value of $r$ to be zero. \\

\paragraph{Fermions $(\Psi)$:} We would be working with two component Dirac spinors $\phi$ and $\chi$,
\begin{equation}\label{ddc}
\Psi=
\begin{pmatrix}
\phi \\
\chi
\end{pmatrix}.
\end{equation}
From \eqref{repgen} the representation of CCA for Dirac fermions is given by the quadruplet  $\{\D,j,f,f^\prime\}$.
\bea{}
&&\non[B_i,\phi(0,0)]= f\sigma_i \chi (0,0),~~[B_i,\chi(0,0)]=  f'\sigma_i \phi(0,0),\\
&&[\Sigma_{ij},\phi(t,x)]= \frac{1}{4} [\sigma_{i},\sigma_{j}]\phi(t,x),~~[\Sigma_{ij},\chi(t,x)]= \frac{1}{4} [\sigma_{i},\sigma_{j}] \chi(t,x).
\eea
Again, as stated before, merely invariance under the underlying symmetries is not enough to fix these values. We would need input from dynamics. 

\medskip

\paragraph{Vectors:} The representation is given by the quadruple $\{\D,j,a,b\}$
\bea{}
&&\non[B_i,A_t(0,0)]= a A_i(0,0),~ [B_i,A_j(0,0)]=  b\delta_{ij}A_t(0,0),\\
&&[\Sigma_{ij},A_t(t,x)]= 0,~~[\Sigma_{ij},A_k(t,x)]= \delta_{ik}A_j(t,x)-\delta_{jk}A_i(t,x).
\eea
Having detailed the features of the scale-spin representation, we are now in a position to deal with conformal Carrollian field theories. Our construction of these theories in the next few sections will crucially depend on the algebraic details outlined in this section. We will start with relativistic field theories and through a process of contraction, generate Carrollian field theories. We will then use the representation theory developed in this section to examine the symmetries of the equations of motion of these Carrollian theories. 

We should emphasise that all Carrollian theories need not to be constructed in this way and there should be an independent way of arriving at these peculiar field theories without recourse to any limiting procedure. The advantage of constructing theories in this particular way from the ultra-relativistic limit of relativistic conformal field theories is that the values of the constants undetermined by symmetries that we discussed in this section would be fixed by the limit. 

\newpage


\section{Carrollian field theories: spin $0$ and spin $\frac{1}{2}$}
\label{sec3}
In this section we will begin our analysis of constructing conformal Carrollian field theories for spin-0 and spin-$\frac{1}{2}$ fields. We will be looking at the conformal Carrollian versions of free scalars, free fermions and the interacting Yukawa theory.
\subsection{Scalar Fields}
The simplest relativistic field theory is of course the theory of a massless spin-0  real scalar field.  The relativistic Lagrangian of massless scalars and the corresponding equation of motion are invariant under conformal transformations in all dimensions. While dealing with Carrollian field theories, we would also be initiating our discussion with the spin-0 field. We wish to see whether the finite conformal Carrollian  symmetry remains in the ultra-relativistic limit of real scalar. Afterwards, we would extend our analysis in search of the infinite enhancement of conformal Carrollian symmetry.\\

\noindent The equation of motion of a relativistic massless real scalar field is
\begin{equation}\label{relsca}
\partial_\mu \partial^\mu \varphi=0.
\end{equation}
We take appropriate spacetime scaling to obtain the ultra-relativistic limit of massless scalar field.
\begin{equation}\label{stsc}
x^i \rightarrow  x^i, \, \, t \rightarrow \epsilon t , \, \, \epsilon \rightarrow 0.
\end{equation}
 Plugging eq\eqref{stsc} in \eqref{relsca}, the Carrollian scalar equation of motion can be expressed as:
\begin{equation}\label{ureq}
\partial_t \partial_t \varphi=0.
\end{equation}
We will now find the symmetries of Carrollian scalar fields at the level of its equation of motion. 

The philosophy that we would be employing to understand the symmetries of the equations of motion of a particular theory is the following. 
Suppose we have a generic equation: 
\be{}
\mathcal{D} \circ \Phi(t,x)= \mathcal{J}
\ee
where $\mathcal{D}$ is some differential operator acting on a field $\Phi(t,x)$ of arbitrary spin, and $\mathcal{J}$ is a source term. We wish to examine the symmetries of this equation under a symmetry group $\mathcal{G}$, whose associated algebra has generators $\mathcal{Q}_\alpha$. We will say that the equation is invariant under this symmetry when: 
\be{}
\delta_{\varepsilon} \left(\mathcal{D} \circ  \Phi(t,x)\right) = \mathcal{D} \circ \delta_{\varepsilon} \Phi(t,x)= \mathcal{D} \circ [\varepsilon^\a \mathcal{Q}_\alpha,\Phi(t,x)]=0,
\ee
where $\varepsilon^\a$ is the infinitesimal parameter of the particular transformation. 

\newpage

\noindent {\em{Fixing the values of the representation theory from the limit}} 

\medskip

\noindent We now need to fix the values of the constants $\{\D, j, r\}$ which remained unfixed in our symmetry analysis by input from the dynamics. We will follow the limit from the relativistic massless scalar theory in order to do this. 
The relativistic field $\Phi$ had a scaling dimension 
\be{deltafix}
[D^{\mbox{\tiny{rel}}}, \Phi(0,0)] = \D^{\mbox{\tiny{rel}}} \ \Phi(0,0), \quad \mbox{where} \quad \D^{\mbox{\tiny{rel}}}= \frac{d-2}{2}
\ee
The relativistic generator does not change in form in the ultra-relativistic limit. So its action on the field is also unchanged. We thus will work with the input 
\be{deltafix2}
\D = \D^{\mbox{\tiny{rel}}}= \frac{d-2}{2} = 1 \quad \mbox{for} \ d=4. 
\ee
We are dealing with a scalar field. So, of course, we have $j=0$. 

\medskip
\noindent For fixing the boost labels, we start with the relativistic boost generator
\be{}
J_{i0}=x_i \p_t +t \p_i.
\ee
The action of the relativistic boost operator on a generic relativistic field $\Phi$ is 
\be{genboost}
J_{i0}\Phi(t,x)=(x_i \p_t+t\p_i)\Phi(t,x)+\Sigma_{i0}\Phi(t,x)
\ee
where $\Sigma_{i0}$ describes analogue of the spin for the spatial rotation. As, we are dealing with a scalar field, $\Sigma_{i0}=0$. 
When we take the ultra-relativistic limit, as stated in eq\eqref{stsc}, we get
\be{}
B_i = \lim_{\e\to0} \e J_{i0} = x_i \partial_t
\ee
Hence, the action of Carrollian boost generator on a Carrollian scalar field is obtained as,
\be{rlabel}
B_i \varphi(t,x)=x_i \partial_t \varphi(t,x)
\ee
From the representation theory, the action of Carrollian boost on Carrollian scalar fields using equation \eqref{scalar} can be written as
\be{}
\Big[B_i, \varphi(t,x)\Big]=x_i \p_t \varphi(t,x)+r \varphi_i(t,x)
\ee
Comparing the above equation with \eqref{rlabel}, we fix the value of boost label `$r$' for the Carrollian scalar field $\varphi$ to be zero.\\

\medskip

\noindent {\em{Symmetries of the EOM}} 

\medskip

\noindent Under finite conformal Carrollian generators, the scalar field equation transforms as:
\bes \label{scalardynamics}
\begin{eqnarray}&&
\Big[D, \partial_t \partial_t \varphi\Big]= (t \partial_t +2 + x_i \partial_i +\Delta)(\partial_t \partial_t \varphi)=0,\\&&
\Big[K_{i},\partial_t \partial_t \varphi\Big]=(2\Delta x_i+2x_i t\p_t+4 x_i+2 x_i x_j \partial_j- x_j x_j \partial_i)(\partial_t \partial_t \varphi)=0,\\&&
\Big[K,\p_t\p_t \varphi \Big]=(x_i x_i \p_t)\p_t \p_t \varphi=0 
\end{eqnarray}
\ees
Hence,  real  scalar field  respects the finite  conformal Carrollian  symmetry. Furthermore, it also has the infinite conformal Carrollian symmetry in $d=4$. 
\be{}
[M^{m_1,m_2,m_3},\partial_t \partial_t \varphi]=x^{m_1} y^{m_2} z^{m_3}\p_t(\partial_t \partial_t \varphi)=0.
\ee
It can also be checked that the Carrollian scalar obeys conformal Carrollian invariance in all dimensions.

\subsection{Fermionic fields}\label{fft}
In this section, we will be looking at the ultra-relativistic limit of spin-$\frac{1}{2}$ field or  free massless Dirac fermions in 4 dimensions. The Dirac fermion $\Psi$ can be written as two component Dirac spinors $\phi$ and $\chi$:
\begin{equation}\label{ddc}
\Psi=
\begin{pmatrix}
\phi \\
\chi
\end{pmatrix}.
\end{equation}
We will be working in the Pauli-Dirac representation in which the gamma matrices are given by
$$
\gamma^0=
\begin{pmatrix}
1 & 0\\
0 &-1
\end{pmatrix},
\; \;
\gamma^i=
\begin{pmatrix}
0 & \sigma^i\\
-\sigma^i & 0\\
\end{pmatrix}.
$$
The $\sigma^i$'s are the Pauli matrices. The relativistic free Dirac equation is 
\bea{}&&
i \gamma^\mu \partial_\mu \Psi=0.
\eea
In terms of the spinors $(\phi, \chi)$ this becomes
\bea{}&& i\p_t \phi +i\sigma^{i}\p_{i}\chi =0,~~  i\p_t \chi +i\sigma^{i}\p_{i}\phi =0.\eea

\medskip

\noindent {\em{Carrollian scalings for Fermions}}

\medskip

\noindent For the ultra-relativistic massless fermionic theory, we scale the spinor fields as
\bea{}\label{sdf}&&
\phi \to \phi,\:  \chi \to \epsilon \chi.
\eea
Taking the scaling \eqref{sdf} on the relativistic equations of motion, the ultra-relativistic Dirac equation can be recast as,
\begin{eqnarray} \label{eomdf}&&
i\partial_t \phi=0,~~
i\partial_t \chi + i\sigma_{i}\partial_i \phi=0.
\end{eqnarray}
We could have considered a general scaling for the spinors:
\be{}
\phi \to \e^r \phi,~\chi \to \e^s \chi.
\ee
In this scaling the ultra-relativistic Dirac equation takes the form
\begin{eqnarray} 
&& i\partial_t \phi +i \e^{s+1-r}\sigma_{i}\partial_i \chi=0,\quad i\partial_t \chi + i\e^{r+1-s}\sigma_{i}\partial_i \phi=0.
\end{eqnarray}
We wish to see an interaction term apart from the usual kinetic term in the EOM. This can be achieved if 
\be{genscafermions}
s+1-r=0,~~ \text{or,}~~r+1-s=0.
\ee
We have used the second condition, without the loss of generality 
\be{}
s=1+r.
\ee
This condition justifies the scaling in \eqref{sdf}, if we put $r=0.$ Considering, any higher value of $r$ will also result in the same Carrollian Dirac equations  \eqref{eomdf}.
It is worth mentioning here that choosing the first condition in \eqref{genscafermions}, will give the Carrollian Dirac equations where the role of $\phi$ and $\chi$ will be exchanged among each other in \eqref{eomdf}. 

\medskip

\noindent {\em{Fixing coefficients}}

\medskip

\noindent Now, we will fix the values of the undetermined constants $\{\Delta,f,f^\prime \}$ by taking the ultra-relativistic limit on the parent relativistic dilatation and boost transformation. We start with \eqref{genboost} and write down  the action of $\Sigma_{i0}$ on the relativistic fermion $\Psi$.
\begin{equation}\label{ffp}
\Sigma^{i0}\Psi=-\frac{1}{4}[\gamma^i,\gamma^0] \Psi \implies \Sigma_{i0}\Psi=
\begin{bmatrix}
0 & -\frac{\sigma_i}{2}\\
-\frac{\sigma_i}{2} & 0
\end{bmatrix}
\begin{bmatrix}
\phi \\
\chi
\end{bmatrix}.
\end{equation}
where we have lowered the indices in the final result. Also, we mention here the action of relativistic spin operator $\Sigma_{ij}$ on $\Psi$,
\be{}
\Sigma^{ij} \Psi=-\frac{1}{4}[\gamma^i,\gamma^j] \Psi
\implies \Sigma_{ij} \Psi= \frac{1}{4}
\begin{bmatrix}
[\sigma_i,\sigma_j] & 0\\
0 & [\sigma_i,\sigma_j]
\end{bmatrix}
\begin{bmatrix}
\phi \\
\chi
\end{bmatrix}.
\ee
Here, $\sigma_i$'s are the Pauli Matrices and $[\sigma_i,\sigma_j]=2i\epsilon_{ijk}\sigma_k$. For a generic spacetime point, we obtain the transformation of  the relativistic spinors under relativistic boost as in \eqref{genboost}
\bea{}
&&J_{i0} \ \phi(t,x)=(x_i \p_t+t \p_i)\phi(t,x)-\frac{\sigma_i}{2}\chi(t,x),\\
&&J_{i0} \ \chi(t,x)=(x_i \p_t+t \p_i)\chi(t,x)-\frac{\sigma_i}{2}\phi(t,x).
\eea
In the ultra-relativistic limit $(t\to \epsilon t,x^i\to x^i,\phi\to \phi,\chi\to \epsilon\chi,\epsilon \to 0)$, the transformation of the spinors under Carrollian boost becomes:
\be{labelfer}
[B_i, \phi(t,x)]=x_i \p_t \phi(t,x),\quad [B_i, \chi(t,x)]=x_i \p_t\chi(t,x)-\frac{\sigma_i}{2}  \phi(t,x).
\ee
Taking input from the dynamics \eqref{labelfer}, we are able to fix the values of the boost labels in the representation theory to $\{f=0,f'=-\frac{1}{2}\}$. Similarly, following the same argument as in \eqref{deltafix} and \eqref{deltafix2}, we fix the  scaling dimension $\D$ for the Carrollian spinors to be
\be{}
\D=\D^{\text{\tiny{rel}}}=\frac{d-1}{2}
\ee
For 4 dimensions, $\Delta$ takes the value of $3\over2$.

\medskip

\noindent {\em{Symmetries of EOM}}
\medskip

\noindent Next, we are interested in checking the invariance under conformal Carrollian algebra. Under scale transformation, 
\bea{}&& [D,i\partial_t \phi]=0,~[D,i\partial_t \chi + i\sigma_{i}\partial_i \phi]=0.\eea 
Similarly, under the spatial part of the Carrollian special conformal transformation (SCT)
\bes
\bea{}&&[K_i ,i\partial_t \phi]=2if\sigma_i \chi +2itf\sigma_i \p_t \chi, \\&&
[K_l,i\partial_t \chi + i\sigma_{i}\partial_i \phi]= 2i\sigma_l (f' +\D -1)\phi +2itf\sigma_{i}\sigma_l \p_i \chi .\eea
\ees
Here, we rewrite the representation of the massless Carrollian fermion theory for convenience  \be{} \label{repf}
\i\{\Delta=\frac{3}{2},j=\frac{1}{2},f=0,f^\prime=-\frac{1}{2}\i\}.\ee
It is thus clear that the equations are invariant under spatial SCT, for these particular values. It can be easily checked that the equations are invariant under the temporal SCT. 
Under $M^{m_1,m_2,m_3}$, the equations of motion are trivially invariant.
\bea{}&&[M^{m_1,m_2,m_3}, i\partial_t \phi]=0,~[M^{m_1,m_2,m_3},i\partial_t \chi + i\sigma_{i}\partial_i \phi]=0.  \eea
So, we have seen that the EOM for Carrollian fermions are also invariant under the infinite CCA. It should be emphasised here that we have focussed on $d=4$. The analysis is expected to hold for all higher dimensions as well. 

\medskip

\subsection{Yukawa Theory}
In the previous sections, we described in detail, the theory of Carrollian version of the free spin-0 and spin-$\frac{1}{2}$ fields. We are now equipped to analyse ultra-relativistic interacting theories. In this section, we will construct the Carrollian version of Yukawa theory.

 We will consider relativistic massless Yukawa theory and then consider its ultra-relativistic limit. The parent  theory consists of a scalar field $\varphi$ and a fermion $\Psi$. Massless Yukawa theory is classically conformally invariant in $d=4$. The action for relativistic massless Yukawa theory is
\be{aY} S =\int d^{4}x~ \i[-\frac{1}{2}\p^{\mu}\varphi\p_{\mu}\varphi +i \bar{\Psi}\gamma^{\mu}\p_{\mu}\Psi -g\bar{\Psi}\varphi\Psi \i].  \ee   
 The equations of motion are
\be{eomyt} 
\p^{\mu}\p_{\mu}\varphi -g\bar{\Psi}\Psi =0,\qquad
i\gamma^{\mu}\p_{\mu}\Psi -g\varphi\Psi=0.
\ee
We will now decompose the Dirac spinor into $(\phi,\chi)$ and use Pauli-Dirac representation of gamma matrices. The relativistic EOM  can be written down  in terms of $(\phi,\chi)$ as:
\be{eomyt1} \p^{\mu}\p_{\mu}\varphi -g(\phi^{\dagger}\phi -\chi^{\dagger}\chi)=0,~
i\p_{t}\phi +i\sigma^{i}\p_{i}\chi -g\varphi\phi=0,~
i\p_{t}\chi +i\sigma^{i}\p_{i}\phi +g\varphi\chi=0.\ee
Now we want to see the ultra-relativistic limit of massless Yukawa theory and check for its symmetries.

\medskip

\newpage

\noindent {\em{Carrollian scalings for Yukawa theory}}

\medskip
\noindent Along with the usual coordinates scaling, the scalar and fermion are scaled as
\bea{} \label{yusc}
&&\varphi \to \varphi,\,\, \phi \to \e^\a \phi,\,\, \chi \to \e^\b \chi.
\eea
We consider this general scaling in order to make sure that we are not missing out on interesting interaction terms by restricting our scalings. 
Plugging the scaling \eqref{yusc} in the equations of motion \eqref{eomyt1} we obtain the scaled equations:
\bes \label{cayu}
\begin{eqnarray}
&&-\p_t\p_t \varphi+\e^2\p_i \p_i \varphi-g(\e^{2\a+2}\phi^{\dagger}\phi -\e^{2\b+2}\chi^{\dagger}\chi)
=0,\\&&
i\p_t \phi+\e^{\b-\a+1}i\s_{i}\p_i\chi -\e g\varphi \phi=0,\\&&
i\p_t \chi+\e^{\a-\b+1}i\s_{i}\p_i\phi +\e g\varphi \chi=0.
\end{eqnarray}
\ees 
For consistency of the theory, when the fermions are turned off, the EOM \eqref{cayu} need to reduce to free Carrollian scalar EOM and similarly, when the scalar is turned off the EOM should reduce to the free fermionic EOM. These put restrictions on the allowed values of $(\a, \b)$ in our scalings above. The constraints on $\a,\b$ are summarised in Table[\ref{yukawacon}]. Combining  all the constraints, we finally arrive at the following relations:
\begin{equation}
\alpha \geq -1,\,\,\,\b=\a+1.
\end{equation}
Hence the possible values of $\alpha,\beta$ are

\medskip

\begin{center}
\begin{tabular}{ | c |  c c c c c c|}
\hline
$\alpha$ & $-1$ & $-\frac{1}{2}$ &0 &$\frac{1}{2}$ & 1& $\hdots$ \\ [0.5ex]
\hline 
$\beta$  &  $0$ & $\frac{1}{2}$ & 1 & $\frac{3}{2}$ & 2 & $\hdots$ \\
\hline
\end{tabular}
\end{center}

\begin{table}[t]
\centering 
\begin{tabular}{| p{5cm} | | p{5cm} |}
\hline
\multicolumn{2}{|c|}{Carrollian massless Yukawa theory}\\
\hline
Free equations & Constraints\\ [0.5ex]
\hline 
$\p_t\p_t \varphi=0$ & $2\alpha+2\geq 0, \, 2\beta+2 \geq 0$ \\
$i\p_t \phi=0$ & $\beta-\alpha+1>0$ \\
$i\p_t \chi+i\sigma_i \partial_i\phi =0$ & $\alpha-\beta+1=0$ \\
\hline
\end{tabular}
\caption{Constraints on $\a,\b$ for massless Carrollian Yukawa theory}
\label{yukawacon}
\end{table}

\bigskip

\noindent Interestingly, the parameter space of different values of $\a, \b$ splits into two distinct cases: 
\begin{itemize}
\item{Case 1: ($\alpha=-1,\beta=0$):} 
\bea{}\label{uryuk}&&\p_t \p_t \varphi +g\phi^{\dagger}\phi=0,~~
i\p_t \phi=0,~~ i\p_t \chi +i\s_{i}\p_i \phi =0. 
\eea
This yields an interesting interacting theory. We will be primarily interested in this and will call this Carrollian Yukawa theory. 

\item{Case 2: (Reproduces free theory) ($\alpha\neq-1,\beta=\a+1$):} 
\begin{eqnarray} 
\p_t\p_t\varphi=0,~
i \partial_t \phi
=0,~
i\partial_t \chi+i\sigma_i \partial_i \phi
=0.
\end{eqnarray}
This just gives back the free equations of the scalar and fermions. Since there are no interaction terms, we will not be interested in these class of scalings. 
\end{itemize}

\medskip
\noindent {\em{Symmetries of EOM}}

\medskip
\noindent As just mentioned, we will choose $\a=-1,\b=0$. Now we investigate the symmetries for the equations \eqref{uryuk}. For this, we use the values of the constants of the individual free  scalar and fermionic theories {\footnote{Since we are concerned with the classical invariance of the EOM in this work, interaction terms would not affect the values of these constants and hence using the constants obtained from the free theories is a justified choice.}}. They are given as
\bea{} \label{repsed}&&
\i\{\underbrace{\Delta_1=1,r=0}_{\textbf{Scalar Field}},\underbrace{\Delta_2=\frac{3}{2},f=0,f'=-\frac{1}{2}}_
{\textbf{Fermionic Field}}\i\}.\eea
The transformations of the EOM under scale transformation are given by
\bes
\bea{}&& [D, \p_t \p_t \varphi +g\phi^{\dagger}\phi]=(\D_1 -1)\p_t\p_t \varphi +(2\D_2 -3)g\phi^{\dagger}\phi ,\\&&
[D,i\p_t \phi]=0,~~[D,i\p_t \chi +i\s_{i}\p_i \phi] =0.\eea \ees
Similarly, under special conformal transformation ($K_l$), we have
\bes
\bea{}&&[K_l ,i\partial_t \phi]=0,
[K_l,i\partial_t \chi + i\sigma_{i}\partial_i \phi]= 2i\sigma_l \i(\D_2 -\frac{3}{2}\i)\phi,\\&&
[K_l, \p_t \p_t \varphi +g\phi^{\dagger}\phi]=(2\D_1 -2)x_l (\p_t\p_t \varphi) +(4\D_2-6)x_l (g\phi^{\dagger}\phi).
\eea
\ees
So we see that with the values of parameters as in \refb{repsed}, the EOM are invariant under the finite CCA{\footnote{Invariance under $K$ is trivial.}}. The equations of motion are trivially invariant under $M^{m_1,m_2,m_3}$. 
\bea{}
&&[M^{m_1,m_2,m_3}, i\partial_t \phi]=0,~[M^{m_1,m_2,m_3},i\partial_t \chi + i\sigma_{i}\partial_i \phi]=0,\\
 && [M^{m_1,m_2,m_3},  \p_t \p_t \varphi +g\phi^{\dagger}\phi]=0 \eea
So we have constructed our first example of an interacting conformal Carrollian theory in this paper and have checked that its EOM have infinitely enhanced symmetries for $d=4$. 

\newpage

\section{Carrollian Electrodynamics and massless matter}
\label{sec4}

Having investigated Carrollian scalars and fermions, the natural next step is to consider spin-1 theories. We begin with a brief review of Carrollian electrodynamics \cite{Duval:2014uoa, Bagchi:2016bcd} and then we go on to add massless scalars and fermions to the $U(1)$ theory and investigate its different sectors and the associated symmetry structure of the equations of motion in each sector, like in the previous section.

\subsection{Carrollian Electrodynamics: A quick look back}
This section is intended to be a quick review of the earlier work \cite{Bagchi:2016bcd} before stepping onto the next part of this paper. The Carrollian limits of Abelian ($U(1)$ Electrodynamics) and Non-abelian ($SU(N)$ Yang-Mills) gauge theories were discussed in detail in \cite{Bagchi:2016bcd}. We will summarise the details of Electrodynamics in this current section.

The relativistic Electrodynamics and Yang-Mills theory are classically conformally invariant in $d=4$. The starting point of studying gauge fields in ultra-relativistic limit concerns breaking the Lorentz invariance by treating the spatial and temporal part of a four vector differently, in addition to the scaling of the underlying spacetime. The corresponding scaling results in two distinct sectors:
\bes \label{lims}
\bea{}
\mbox{Electric sector:} \quad A_t \to A_t, \, A_i \to \e A_i \label{elimcont},\\
\mbox{Magnetic sector:} \quad A_t \to \e A_t, \, A_i \to A_i \label{mlimcont}.
\eea
\ees 
The existence of two different Carrollian sectors of electrodynamics can be traced to the fact that there is no non-degenerate metric for the entire Carrollian spacetime. For this precise reason, the covariant and contravariant vectors form two different inequivalent characterisation of the spin-1 theory. On taking the UR limit on the spacetime, contravariant vectors transform as
\bea{contravary}
A^{\prime \:i}&=&\Big(\frac{\partial x^{\prime\: i}}{\partial x^j} \Big)A^j=\delta^i_j A^j,\quad A^{\prime \: 0}=\Big(\frac{\partial t^{\prime}}{\partial t} \Big)A^0=\epsilon A^0.
\eea
We associate this with the magnetic sector \eqref{mlimcont}:
\be{}
A_t\to  \epsilon A_t,~A_i \to  A_i.
\ee
 On the other hand, covariant vectors transform as
\bea{}
A_{ i}^\prime=\Big(\frac{\partial x^{ j}}{\partial x^{\prime \: i}} \Big)A_j=\delta^j_i A_j,\quad A^{\prime}_0=\Big(\frac{\partial t}{\partial t^{\prime}} \Big)A_0={1 \over \epsilon} A_0,
\eea
We can write the transformation of the covariant vectors (after an overall multiplication by a factor of $\e$ which does not change anything) as
\be{covary}
A_t\to  A_t,~A_i \to \epsilon A_i.
\ee
This is the electric sector. 
The dynamics of the two sectors are governed by different equations of motion. These EOM for the electric and magnetic sector are respectively given as:
\bes \label{edeom}
\bea{} 
\label{urseleom} && \mbox{Electric Sector:} \quad \p_{i}\p_{i}A_{t}-\p_{i}\p_{t}A_{i}=0,~~ \p_{t}\p_{i}A_{t}-\p_{t}\p_{t}A_{i}=0, \\
\label{ursmageom} &&  \mbox{Magnetic Sector:} \quad \p_{i}\p_{t}A_{i}=0,~~ \p_{t}\p_{t}A_{i}=0.
\eea
\ees
As we have done previously, we can check for the symmetries of the EOM in these sectors, based on the representation theory discussed in Sec[\ref{Representation Theory}]. We will also need input from dynamics, which as before, can be obtained by looking carefully at the limit from the relativistic parent theory. Both sets of EOM turn out to have finite and infinite conformal Carrollian  symmetry. For the detailed analysis of the symmetries of the EOM, the reader is pointed to \cite{Bagchi:2016bcd}. We will use this knowledge of the pure gauge theory to go beyond and study Carrollian gauge fields coupled to matter in the next section.

\subsection{Adding scalars}

 Scalar electrodynamics is the theory of $U(1)$ gauge field coupled to a complex charged scalar field. The Lagrangian density of the massless scalar electrodynamics theory is given by
\bea{}&&
\mathcal{L}=-\frac{1}{4}F^{\mu \nu} F_{\mu \nu}-(D_\mu \varphi)^{\dagger} (D^\mu \varphi),
\eea
where $F_{\mu \nu}$ is the field strength defined as 
$
F_{\mu \nu}=\partial_\mu A_\nu-\partial_\nu A_\mu
$ and
 $D_{\mu}$ is the gauge covariant derivative
$
D_{\mu}=\partial_\mu + ie A_\mu
$, with
$e$ being the coupling parameter and $A_\mu$ being the $U(1)$ gauge field. $\varphi$ is the massless complex scalar field charged under the $U(1)$ gauge group. The gauge transformation are given as,
\bea{}&&
A_\mu(x)\to A_\mu(x)+\p_\mu \alpha(x),~~ \varphi(x) \to e^{-ie\alpha(x)}\varphi(x).
\eea
Here, $\alpha$ is a parameter of gauge transformation. The equations of motion of the massless complex scalar field $\varphi$ and the gauge field $A_{\mu}$ is given as,
 \bea{} \label{SED}&&
\partial_\mu F^{\mu \nu}+ie[\varphi^\dagger( D^\nu \varphi)- \varphi (D^\nu \varphi)^\dagger]=0,~~D_\mu D^\mu \varphi=0.
\eea
Massless scalar electrodynamics preserves relativistic conformal symmetry in $d=4$. The complex scalar field $\varphi$ can be expressed in terms of two real scalar fields $\varphi_1$ and $\varphi_2$. 
\bea{}
\varphi=\varphi_1 +i \varphi_2
\eea
The equations of motion \eqref{SED} in terms of the components are: 
\bes
\bea{}\label{fsed}&&
\p_{\mu} F^{\mu\nu}+2e[  \varphi_2 \p^{\nu} \varphi_1-\varphi_1\p^{\nu} \varphi_2- e A^{\nu} (\varphi_1^2+\varphi_2^2)]=0,\\&&
\p_\mu \p^\mu \varphi_1 -e(\p_\mu A^\mu)\varphi_2-2eA_\mu (\p^\mu \varphi_2)-e^2 A_\mu A^\mu \varphi_1=0,\\&&
\p_\mu \p^\mu \varphi_2 +e(\p_\mu A^\mu)\varphi_1+2eA_\mu (\p^\mu \varphi_1)-e^2 A_\mu A^\mu \varphi_2=0.
\eea\ees

\medskip

\subsection*{Carrollian scalar electrodynamics} 
As we have just seen, in Carrollian limit, we get two different sectors of electrodynamics  depending on the scaling of  gauge field. Similarly, the scalar electrodynamics will also exhibit these two different sectors, viz. the Electric and Magnetic sectors. 
\paragraph{Electric sector:} In this sector, we scale the gauge field in Carrollian scalar electrodynamics in the fashion similar to electric sector of electrodynamics. Along with usual spacetime scaling, the scalings of the gauge field and the scalar field are chosen as (we are considering the most generic situation by scaling the two real scalar fields individually):
\begin{equation}
A_t \to A_t, \; A_i \to \epsilon A_i,\; \varphi_1 \to \epsilon^{p_1} \varphi, \; \varphi_2 \to \epsilon^{p_2} \varphi_2.
\end{equation}
where $p_1,p_2$ are two arbitrary constants which will be determined later.
The scalar and gauge field equations in the electric sector are then given by: 
\bes  \label{esed}
\begin{eqnarray}&&\hspace{-1 cm}
\partial_i \partial_i A_t - \p_i \partial_t A_i+2e[\e^{p_1+p_2-1} (\varphi_2 \p_t \varphi_1 - \varphi_1 \p_t \varphi_2)
- e A_t (\e^{2p_1} \varphi_1^2+\e^{2p_2} \varphi_2^2)]=0,\\&&\non\hspace{-1 cm}
 \partial_t \partial_t A_j -\p_t \partial_j A_t -\e^2 (\p_i \p_i A_j -\p_i \p_j A_i)- 2e[\e^{p_1+p_2+1} (\varphi_2 \p_j \varphi_1 - \varphi_1 \p_j \varphi_2) \\&&\hspace{6.5cm}
- e A_j (\e^{2p_1+2} \varphi_1^2+\e^{2p_2+2} \varphi_2^2)] =0,\\&&\non\hspace{-1 cm}
 \partial_t \partial_t \varphi_1 -\e^2 \p_{i}\p_{i}\varphi_1- \e^{p_2-p_1+1}[ 2e A_t (\p_t \varphi_2) +e (\p_t A_t)\varphi_2]
 +\e^{p_2-p_1+3}[2e A_i (\p_{i}\varphi_2)\\ &&\hspace{4.2cm}+e(\p_{i}A_{i})\varphi_2]-\e^2 e^2 A_t A_t \varphi_1 +\e^{4}e^2 A_{i}A_i \varphi_1 =0,\\&&\non\hspace{-1 cm}
 \partial_t \partial_t \varphi_2 -\e^2 \p_{i}\p_{i}\varphi_2 +\e^{p_1-p_2+1}[ 2e A_t (\p_t \varphi_1) +e (\p_t A_t)\varphi_1]
 -\e^{p_1-p_2+3}[2e A_i (\p_{i}\varphi_1)\\ &&\hspace{4.2cm}+e(\p_{i}A_{i})\varphi_1]-\e^2 e^2 A_t A_t \varphi_2 +\e^{4}e^2 A_{i}A_i \varphi_2 =0.
\end{eqnarray}
\label{sedeom}
\ees
This of course are just the equations with the scalings put in. The actual EOM will be determined when the appropriate scalings have been performed and the terms with higher powers of $\e$ as compared to the leading term are set to zero. 

The above equations must reduce to free scalar and electrodynamics equations when the gauge and the scalar field are respectively turned off. This leads to constraints on $p_1,p_2$. The constraints that we obtain for electric sector is:
\bea{}
-1 \leq p_1-p_2 \leq 1,~~
p_1\geq 0, ~~p_2 \geq 0,~~
p_1+p_2 \geq 1.
\eea
\begin{figure}[t]
\centering
\includegraphics[width=8cm]{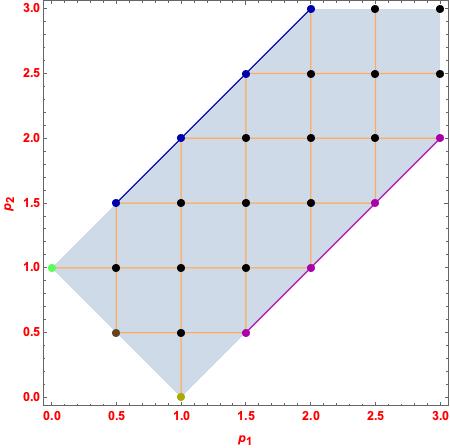}
\floatfoot{\small {{\bf{Key:}} Points in black: free Carrollian theories, Coloured points: non-trivial interacting sectors.}}
\caption{Allowed region for electric  sector of Carrollian scalar electrodynamics.}

\label{ScEDelecplot}
\end{figure}
The allowed values of $p_1, \: p_2$ are shown in  Fig[\ref{ScEDelecplot}]. The marked points represent different sectors within electric limit of Carrollian scalar electrodynamics. 

\medskip

\noindent{\em{Different interacting sub-sectors}}

\medskip

\noindent Restricting ourselves to integral and half integral values of $(p_1, p_2)$, we find that there are five distinct interacting sub-sectors within the Electric sector of Carrollian scalar electrodynamics. These different sectors for different $(p_1,p_2)$ are:
\begin{itemize}
\item Case 1 : {$p_1=1, p_2=0:$}
 \bes 
\bea{}&&
\p_i \p_i  A _t- \p_{i}\p_t A_i+2e[(\varphi_2 \p_t \varphi_1-\varphi_1  \p_t \varphi_2)-eA_t \varphi^{2}_{2}]=0, \p_t  \p_t \varphi_2=0 ,\\&&
\p_t\p_t A_j- \p_t\p_j  A_t=0, \p_t  \p_t \varphi_1-e[2A_t  (\p_t \varphi_2)+\varphi_2  (\p_t A_t)]=0.
\eea
\ees
\item Case 2: {$p_1=\frac{1}{2}, p_2=\frac{1}{2}:$}
\bes
\bea{}&&
\p_i  \p_i  A_t- \p_{i}\p_t  A_i +2e (\varphi_2 \p_t \varphi_1-\varphi_1  \p_t \varphi_2)=0,\\&& \p_t \p_t  A_j-\p_t \p_j  A_t=0, \p_t  \p_t \varphi_{1,2}=0.
\eea
\ees
\item Case 3: {$p_1=\frac{3}{2}, p_2=\frac{1}{2}:$}
\bes
\bea{}&&
 \p_i \p_i  A_t- \p_i\p_t  A_i=0, \p_t \p_t A_j- \p_t \p_j  A_t=0,\\&&
\p_t  \p_t \varphi_1-e[2A_t  (\p_t \varphi_2)+\varphi_2  (\p_tA_t)]=0, \p_t  \p_t \varphi_2=0.
\eea
\ees
\item Case 4: {$p_1=0, p_2=1:$} 
\bes
\bea{}&&
\p_i \p_i  A _t- \p_{i}\p_t A_i+2e[(\varphi_2 \p_t \varphi_1-\varphi_1  \p_t \varphi_2)-eA_t \varphi^{2}_{1}]=0, \p_t  \p_t \varphi_1=0 ,\\&&
\p_t\p_t A_j- \p_t\p_j  A_t=0, \p_t  \p_t \varphi_2+e[2A_t  (\p_t \varphi_1)+\varphi_1  (\p_t A_t)]=0.
\eea
\ees
\item Case 5: {$p_1=\frac{1}{2}, p_2=\frac{3}{2}$}: 
\bes
\bea{}&&
 \p_i \p_i  A_t- \p_i\p_t  A_i=0, \p_t \p_t A_j- \p_t \p_j  A_t=0,\\&&
\p_t  \p_t \varphi_2+e[2A_t  (\p_t \varphi_1)+\varphi_1  (\p_tA_t)]=0, \p_t  \p_t \varphi_1=0.
\eea
\ees
\end{itemize}

 We have demarcated the distinct sectors with coloured dots in Fig[\ref{ScEDelecplot}]. The points shown in black reduce to free Carrollian scalar and electric sector of Carrollian electrodynamics. The values of $p_1,p_2$ which result in same sectors are connected with each other in the same figure. It is of interest to note here that the non-trivial interacting sub-sectors within the Electric sector of Carrollian scalar electrodynamics all lie along the corners of the allowed region. This is a feature that will be true for all cases we consider after this, where we look at the parameter space of these allowed field theories. 

\medskip

\noindent{\em{Symmetries of EOM}}

\medskip

\noindent Now we want to check the symmetries the above equations posses. 
We will examine an arbitrary sector, say ($p_1=0,p_2=1$) to discuss the symmetry. It will require the following inputs from the representation theory,
\bea{} \label{repsed}
\i\{\underbrace{\Delta=1,j=0,r=0}_{\textbf{Scalar Field}},\underbrace{\Delta^\prime=1,j^\prime=1,a=0,b=1}_
{\textbf{Gauge Field}}\i\}.\eea
 In this sector the equations of motion are:
\bes
\bea{}&&
\label{ursedeomel}\p_i \p_i  A _t- \p_{i}\p_t A_i+2e[(\varphi_2 \p_t \varphi_1-\varphi_1  \p_t \varphi_2)-eA_t \varphi^{2}_{1}]=0,\\&&
\p_t\p_t A_j- \p_t\p_j  A_t=0, \p_t  \p_t \varphi_2+e[2A_t  (\p_t \varphi_1)+\varphi_1  (\p_t A_t)]=0,\p_t  \p_t \varphi_1=0.
\eea
\ees
Let us now check for the invariance under scale transformation:
\bes
\begin{eqnarray}&&\hspace{-1 cm}
[D,(\ref{ursedeomel})]=(\Delta' -1)(\p_i \p_i  A _t- \p_{i}\p_t A_i -2e^2 A_t \varphi^{2}_{1})\non\\&&\hspace{2.3cm}+(2\Delta -2)2e[(\varphi_2 \p_t \varphi_1-\varphi_1  \p_t \varphi_2)-eA_t \varphi^{2}_{1}],\\&&\hspace{-1 cm}
[D,\p_t  \p_t \varphi_2+e\lbrace2A_t  (\p_t \varphi_1)+\varphi_1  (\p_t A_t)\rbrace]=(\Delta'-1)e[2A_t  (\p_t \varphi_1)+\varphi_1  (\p_t A_t)],\\&&\hspace{-1 cm}
[D,\p_t\p_t A_j- \p_t\p_j  A_t]=0, [D,\p_t  \p_t \varphi_1]=0.
\end{eqnarray}
\ees
Similarly, under $K_i$, we have
\bes
\begin{eqnarray}&&\hspace{-1 cm}\non
[K_l,(\ref{ursedeomel})]= (4\Delta' +2-2\delta_{ii})\p_l A_t -(2\Delta' +4-2\delta_{ii})\p_t A_l \\&&\non\hspace{1.4cm}+2x_l (\Delta' -1)(\p_i\p_i A_t-\p_t\p_i A_i -2e^2 A_t \varphi^{2}_{1})\\&&\hspace{1.4cm}+4x_l(\Delta -1)[2e\lbrace(\varphi_2 \p_t \varphi_1-\varphi_1  \p_t \varphi_2)-eA_t \varphi^{2}_{1}\rbrace], \\&&\hspace{-1 cm}
[K_i,\p_t  \p_t \varphi_2+e\lbrace2A_t  (\p_t \varphi_1)+\varphi_1  (\p_t A_t)\rbrace]=2x_i (\Delta'-1)e[2A_t  (\p_t \varphi_1)+\varphi_1  (\p_t A_t)],\\&&\hspace{-1 cm}
[K_i,\p_t\p_t A_j- \p_t\p_j  A_t]=-(2\Delta' -2)\delta_{ij}\p_tA_t,~[K_i,\p_t  \p_t \varphi_1]=0.
\end{eqnarray} 
\ees
Hence the scale invariance and special conformal invariance is preserved in $d=4$. There is invariance under $M^{m_1,m_2,m_3}$ for all the equations. \bes
\bea{}
&&\hspace{-.5cm}[M^{m_1,m_2,m_3},(\ref{ursedeomel})]=0,~[M^{m_1,m_2,m_3},\p_t  \p_t \varphi_2+e\lbrace 2A_t  (\p_t \varphi_1)+\varphi_1  (\p_t A_t)\rbrace]=0,~~~~~~~\\
 &&\hspace{-.5cm} [M^{m_1,m_2,m_3},\p_t\p_t A_j- \p_t\p_j  A_t]=0, ~[M^{m_1,m_2,m_3},\p_t  \p_t \varphi_1]=0. \eea\ees
It can be shown in an identical fashion that in all the other sectors mentioned in our previous analysis, the EOM would also have this infinite dimensional conformal Carrollian invariance.

\bigskip

\begin{figure}[t]
\centering
\includegraphics[width=8cm]{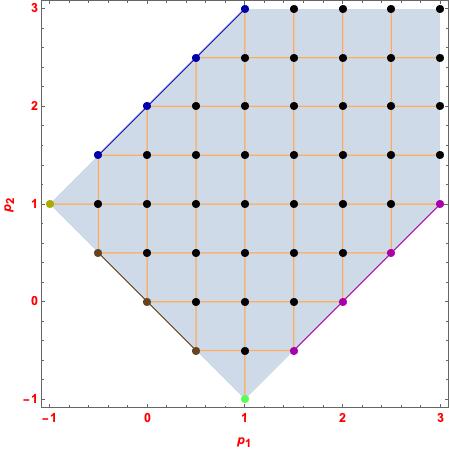}
\caption{Allowed region for magnetic  sector.}
\floatfoot{\small {{\bf{Key:}} Points in black: free Carrollian theories, Coloured points: non-trivial interacting sectors.}}
\label{ScEDmagplot}
\end{figure}

\paragraph{Magnetic sector:} Carrollian scalar electrodynamics also has a magnetic sector, named after the particular sector in Carrollian electrodynamics with magnetic field dominating over the electric field. In the magnetic sector, the scaling are given by
\bea{}
A_t\to \e A_t,\, A_i\to A_i,
\varphi_1\to \e^{p_1} \varphi_1,\,\, \varphi_2\to \e^{p_2} \varphi_2.
\eea
The equations of motion in this limit are:
\bes \label{msed}
\begin{eqnarray}&&\hspace{-1 cm}
-\e^{2}\p_{i}\p_i A_t +\partial_i \p_t A_i - 2e[\e^{p_1+p_2} (\varphi_2  \p_t  \varphi_1-\varphi_1  \p_t  \varphi_2)\non\\&&\hspace{6cm}
- e A_t (\e^{2p_1+2} \varphi_1^2+\e^{2p_2+2} \varphi_2^2)]  =0,\\&&\hspace{-1 cm}
\p_t \p_t A_j- \e^2 (\p_t \p_j A_t +\p_i \p_i A_j-\p_i \p_j A_i)- 2e[\e^{p_1+p_2+2} (\varphi_2 \p_j \varphi_1-\varphi_1 \p_j \varphi_2)\non\\&&\hspace{6cm}
- e A_j (\e^{2p_1 +2} \varphi_1^2+\e^{2p_2+2} \varphi_2^2)]=0,\\&&\hspace{-1 cm}
 \p_t \p_t \varphi_1 -\e^2 \p_i \p_i \varphi_1 - \e^{p_2-p_1+2}[ e(\p_t A_t) \varphi_2 -e (\p_i A_i) \varphi_2
 +2e (\p_t\varphi_2) A_t \non\\&& \hspace{3.5cm} - 2e (\p_i\varphi_2) A_i]-\e^4 e^2 A_tA_t\varphi_1 +\e^2 e^2 A_iA_{i}\varphi_1=0,\\&&\hspace{-1 cm}
 \p_t \p_t \varphi_2 -\e^2 \p_i \p_i \varphi_2 + \e^{p_1-p_2+2}[ e(\p_t A_t) \varphi_1 -e (\p_i A_i) \varphi_1
 +2e (\p_t\varphi_1) A_t \non\\&& \hspace{3.5cm} - 2e (\p_i\varphi_1) A_i] -\e^4 e^2 A_tA_t\varphi_2 +\e^2 e^2 A_iA_{i}\varphi_2=0.
\end{eqnarray}
\ees
Studying the equations we find the constraints on $p_1,p_2$ as,
\be{}
-2 \leq p_1-p_2 \leq 2,\quad p_1+p_2\geq 0, \quad p_1\geq-1, \quad p_2\geq-1.
\ee
The allowed values of $p_1,\: p_2$ are easily realised in Fig[\ref{ScEDmagplot}].
The marked points represent different sectors within magnetic limit of Carrollian scalar electrodynamics. 

\newpage

\medskip

\noindent{\em{Different interacting sub-sectors}}

\medskip

\noindent We again find 5 distinct sub-sectors within the Magnetic sector. These different possible sub-sectors are:
\begin{itemize}
\item Case 1: {$p_1=1, p_2=-1$}:
\bes
\bea{}&&\hspace{-1.2cm}
\p_i \p_t A_i -2e[(\varphi_2 \p_t \varphi_1-\varphi_1 \p_t \varphi_2)-e A_t \varphi_2^2]=0,
\p_t \p_t A_j+2e^2 A_j \varphi_2^2=0,\\&&\hspace{-1.5cm}
\p_t \p_t \varphi_1-e[(\p_t A_t)\varphi_2 -(\p_iA_i) \varphi_2+2A_t(\p_t\varphi_2)-2A_i(\p_i\varphi_2)]=0,
\p_t \p_t \varphi_2 =0.
\eea
\ees
\item Case 2: {$p_1=\frac{1}{2},p_2=-\frac{1}{2}$}:
\bea{}
\p_i \p_t A_i -2e(\varphi_2\p_t\varphi_1-\varphi_1\p_t \varphi_2)=0,
\p_t\p_t A_j=0,
\p_t \p_t \varphi_{1,2} =0.
\eea
\item Case 3: {$p_1=\frac{3}{2},p_2=-\frac{1}{2}$}:
\bes
\bea{}&&
\p_i \p_t A_i =0,
\p_t\p_t A_j=0,\p_t \p_t \varphi_2 =0,\\&&
\p_t \p_t \varphi_1 -e[(\p_tA_t) \varphi_2-(\p_iA_i) \varphi_2+2A_t (\p_t\varphi_2)-2A_i (\p_i\varphi_2)]=0.
\eea
\ees
\item Case 4: {$p_1= -1,p_2= 1$}:
\bes
\bea{}&&\hspace{-1.2cm}
\p_i \p_t A_i -2e[(\varphi_2 \p_t \varphi_1-\varphi_1 \p_t \varphi_2)-e A_t \varphi_1^2]=0,
\p_t \p_t A_j+2e^2 A_j \varphi_1^2=0,\\&&\hspace{-1.5cm}
\p_t \p_t \varphi_2+e[(\p_t A_t)\varphi_1 -(\p_iA_i) \varphi_1+2A_t(\p_t\varphi_1)-2A_i(\p_i\varphi_1)]=0,
\p_t \p_t \varphi_1 =0.
\eea
\ees
\item Case 5: {$p_1=-\frac{1}{2},p_2=\frac{3}{2}$}:
\bes
\bea{}&&
\p_i \p_t A_i =0,
\p_t\p_t A_j=0,\p_t \p_t \varphi_1 =0,\\&&
\p_t \p_t \varphi_2 +e[(\p_tA_t) \varphi_1-(\p_iA_i) \varphi_1+2A_t (\p_t\varphi_1)-2A_i (\p_i\varphi_1)]=0.
\eea
\ees
\end{itemize}
  The black points in Fig[\ref{ScEDmagplot}]   reduce to free Carrollian scalar and magnetic limit of Carrollian electrodynamics. The non trivial sectors  having interaction terms are marked in different colours in  Fig[\ref{ScEDmagplot}]. The values of $p_1,p_2$ which result in same sectors are connected with coloured lines in the same figure.
\medskip

\noindent{\em{Symmetries of EOM}}

\medskip

\noindent 
Before discussing the symmetries, we require the values of the constants $\lbrace\Delta, \Delta', r, a, b\rbrace$ from the representation theory. \bea{} \label{repsed2}
\i\{\underbrace{\Delta=1,j=0,r=0}_{\textbf{Scalar Field}},\underbrace{\Delta^\prime=1,j^\prime=1,a=1,b=0}_
{\textbf{Gauge Field}}\i\}.
\eea
We are choosing a  sector randomly, $p_1=\frac{1}{2},p_2=-\frac{1}{2}$. The equations in this sector are:
\bea{}
\p_i \p_t A_i -2e(\varphi_2\p_t\varphi_1-\varphi_1\p_t \varphi_2)=0,
\p_t\p_t A_j=0,
\p_t \p_t \varphi_{1,2} =0.
\eea
We will now check for invariance of the above equations. Under scale transformations, we have
\bea{} &&[D,\p_i \p_t A_i -2e(\varphi_2\p_t\varphi_1-\varphi_1\p_t \varphi_2)]=(\Delta'-1)\p_i \p_t A_i +(2\Delta -2)[2e(\varphi_1 \p_t \varphi_2 -\varphi_2 \p_t \varphi_1)],\non\\&&
[D,\p_t\p_t A_j]=0, [D,\p_t \p_t \varphi_{1,2}]=0.
\eea
Similarly, under  $K_i$, the transformation is given as
\bes
\bea{}&&
[K_l ,\p_i \p_t A_i -2e(\varphi_2\p_t\varphi_1-\varphi_1\p_t \varphi_2)]= (2\Delta' +4-2\delta_{ii})\p_t A_l +2x_l (\Delta' -1)(\p_i \p_tA_i)\non\\&& \hspace{6.4cm}+(4\Delta -4)x_l [2e(\varphi_1 \p_t \varphi_2 -\varphi_2 \p_t \varphi_1)],\\&&
[K_i,\p_t\p_t A_j]=0, [K_i, \p_t \p_t \varphi_{1,2}] =0.
\eea
\ees
The right hand side of all of the above expression vanishes as $\{\Delta=1,\D^\prime=1\}$.
The equations of motion also have infinite $M^{m_1,m_2,m_3}$ symmetry.
\bea{} &&[M^{m_1,m_2,m_3},\p^i \p_t A_i -2e(\varphi_2\p_t\varphi_1-\varphi_1\p_t \varphi_2)]=\p_i(x^{m_1}y^{m_2}z^{m_3})\p_t\p_t A_i=0,\non\\&&
[M^{m_1,m_2,m_3},\p_t\p_t A_j]=0, [M^{m_1,m_2,m_3},\p_t \p_t \varphi_{1,2}]=0.
\eea
We can perform the same symmetry analysis for all the possible subsectors and we find that the emergent infinite dimensional symmetry exists in all of these sectors. 

\bigskip

\subsection{Adding Fermions}
 
The relativistic theory of $U(1)$ gauge field $A_\mu$ coupled to fermionic fields $\Psi$ exhibit conformal symmetry in $d=4$. The Lagrangian density of this theory is given as
\bea{} \mathcal{L}=-\frac{1}{4}F_{\mu\nu}F^{\mu\nu} + i\bar{\Psi}\gamma^{\mu}D_{\mu}\Psi, \eea
where $D_{\mu}$ is the covariant derivative and $F_{\mu\nu}$ is the electromagnetic field strength:
\bea{} D_{\mu}=\p_{\mu}-ieA_{\mu}, F_{\mu\nu}= \p_{\mu}A_{\nu}-\p_{\nu}A_{\mu}.\eea  
The relativistic equations of motion are:
\bea{spedeom} \p_{\mu}F^{\mu\nu}+e\bar{\Psi}\gamma^{\nu}\Psi=0,~
i\gamma^{\nu}D_{\nu}\Psi=0.\eea
We will decompose the Dirac fermion in two component spinors as in Sec[\ref{fft}]. The relativistic equations of motion get recasted as,
\bes \label{eomdsed} 
\bea{}&&\hspace{-.5cm} \p_{i}F_{it}-e(\phi^{\dagger}\phi+\chi^{\dagger}\chi)=0,~
-\p_{t}F_{tj}+\p_{i}F_{ij}+e(\phi^{\dagger}\sigma_{j}\chi
+\chi^{\dagger}\sigma_{j}\phi)=0,\\&&\hspace{-.5cm}
 (i\p_{t}+eA_{t})\phi +(i\sigma_{i}\p_{i}+e\sigma_{i}A_{i})\chi=0 ,~
(i\sigma_{i}\p_{i}+e\sigma_{i}A_{i})\phi+ (i\p_{t}+eA_{t})\chi=0.
\eea
\ees
Next, we are interested in studying the Carrollian limit of the same theory.

\newpage

\subsection*{Carrollian electrodynamics with fermions}
We have seen that the gauge fields can be scaled in two ways,  electrically and magnetically. Here, we will also scale the spinors accordingly. For easier understanding, we will be using the conventional names of the different limits in a  way similar to Carrollian electrodynamics.
\paragraph{Electric sector:}
In this sector, the scaling of gauge fields and fermions are given as:
\begin{equation}
\phi \to \epsilon^\alpha \phi,  \, \chi \to \epsilon^\beta \chi,  \,A_t \to A_t,  \, A_i\to \epsilon A_i .
\end{equation}
Here $\alpha$ and $\beta$ are two arbitrary constants which will be determined later. In this sector, the equations of motion become:
\bes{}\label{efed}
\bea{}&&
 \p_i\partial_i A_t-\p_i\p_t A_i- e(\epsilon^{2\alpha}\phi^{\dagger}\phi+\epsilon^{2\beta}\chi^{\dagger}\chi)=0,\\&&
-\p_t \p_t A_j+\p_t \partial_j A_t+\e^2 \p_i(\p_i A_j-\partial_j A_i)+ \epsilon^{\alpha+\beta+1}e(\phi^{\dagger}\sigma_j\chi
+\chi^{\dagger}\sigma_j\phi)=0,\\&&
i\p_t \phi+\epsilon^{\beta-\alpha+1}i \sigma_i \partial_i \chi + \epsilon^{\beta-\alpha+2} e\sigma_i A_i \chi+\epsilon eA_t \phi
=0,\\&&
i\p_t \chi+\epsilon^{\alpha-\beta+1}i\sigma_i \partial_i \phi + \epsilon^{\alpha-\beta+2} e\sigma_i A_i \phi+\epsilon eA_t \chi
=0.
\eea \ees
The above equations must reduce to the  free Carrollian equations of the electric sector of electrodynamics and fermions separately. These constraints put some restrictions on the arbitrary constants $\alpha,\beta$. This can be seen in Table[\ref{electrc}].

\begin{table}[t]
\centering 
\begin{tabular}{|c| |c|}
\hline
\multicolumn{2}{| c |}{Electric sector of $U(1)$ + Fermion}\\
\hline
Free equation & Constraints\\ [0.5ex]
\hline 
$\partial_i\partial_i A_t-\p_i\p_t A_i=0$ & $2\alpha\geq 0, \, 2\beta \geq 0$ \\
$\p_t\p_t A_i-\p_t \partial_i A_t=0$ & $\alpha+\beta+1 \geq 0$ \\
$i\p_t \phi=0$ & $\beta-\alpha+1>0$ \\
$i\p_t \chi+i\sigma_i\partial_i\phi=0$ & $\alpha-\beta+1=0, \, \alpha-\beta+2\geq 0$ \\
\hline
\end{tabular}
\caption{Constraints on $\alpha,\beta$ for Electric sector of electrodynamics with fermions.}
\label{electrc}
\end{table}

\medskip

\noindent Combining all the constraints we finally arrive at the following relations:
\begin{equation}
\alpha=\beta-1, \, \alpha \geq 0, \, \beta\, \geq 0.
\end{equation}
Hence the possible values of $\alpha,\beta$ are:
\begin{center}
\begin{tabular}{ | c |   c c c c c|}
\hline
$\alpha$ & $0$ & $\frac{1}{2}$ &1 &$\frac{3}{2}$ &  $\hdots$ \\ [0.5ex]
\hline 
$\beta$  &  $1$ & $\frac{3}{2}$ & 2 & $\frac{5}{2}$ &  $\hdots$ \\
\hline
\end{tabular}
\end{center} 

\bigskip

\noindent{\em{Different sub-sectors in the theory}}

\medskip

\noindent We will now write down different sectors for the Electric sector of Carrollian electrodynamics coupled to fermions:
\begin{itemize}
\item Case 1: {$\alpha=0,\beta=1$}:
\bes \label{c1efed}\begin{eqnarray}\label{efed1}&&
 \p_i (\partial_i A_t-\p_t A_i)- e(\phi^{\dagger}\phi)=0,~~
\p_t (\p_t A_j-\partial_j A_t)=0,\\&&
i\p_t \phi=0,~~
i\p_t \chi+ i\sigma_{i}\p_i \phi=0.
\end{eqnarray}\ees
\item Case 2: {$\alpha> 0,\beta=\alpha+1$}: 
\bes \label{c2efed}
\begin{eqnarray}&&
\partial_i(\partial_i A_t-\p_t A_i)=0,~~
\partial_t(\partial_t A_j-\partial_j A_t)=0,\\&&
i \p_t \phi=0,~~
i\p_t \chi+i\sigma_{i}\partial_i \phi=0.
\end{eqnarray}
\ees
\end{itemize}
We see that in Case 2, the equations \eqref{c2efed} reduces to Carrollian equations of  electrodynamics and free Dirac fermions. 

\medskip

\noindent{\em{Symmetries of EOM}}

\medskip

\noindent The next task is to find the invariance of these equations under conformal Carrollian symmetries. For that, we will only consider Case 1, since these equations contain interaction terms. For this case, the values of the constants  are mentioned here:
\bea{} \label{repsced}
\i\{\underbrace{\Delta=\frac{3}{2},f=0,f'=-\frac{1}{2}}_{\textbf{Fermionic Field}},\underbrace{\Delta^\prime=1,a=0,b=1}_
{\textbf{Gauge Field}}\i\}.\eea
The invariance under scale transformation is given by
\bes
\bea{}&&\hspace{-1cm}[D, \p_i (\partial_i A_t-\p_t A_i)- e(\phi^{\dagger}\phi)]=(\Delta' -1)[\p_i \partial_i A_t-\p_{i}\p_t A_i]-(2\Delta -3)e\phi^{\dagger}\phi,\\&&\hspace{-1cm}
[D,\p_t (\p_t A_j-\partial_j A_t)]=0,~~[D,i\p_t \phi]=0,~~
[D,i\p_t \chi+ i\sigma_{i}\p_i \phi]=0.
\eea
\ees
Under $K_{l}$, we have
\bes
\bea{}&&\hspace{-1.2cm} [K_{l}, \p_i (\partial_i A_t-\p_t A_i)- e(\phi^{\dagger}\phi)]=(4\Delta' -2\delta_{ii}+2)\p_l A_t +(2\delta_{ii}-2\Delta' -4)\p_t A_l \non\\&&\hspace{4cm}+(2\Delta'-2)x_l (\p_{i}\p_{i}A_t -\p_{t}\p_{i}A_i)-(4\Delta -6)x_l (e\phi^{\dagger}\phi),~~~\\&&\hspace{-1.2cm}
[K_l,\p_t (\p_t A_j-\partial_j A_t)]=-2(\Delta' -1)\delta_{lj}\p_t A_t,~~[K_l,i\p_t \phi]=0,\\&&\hspace{-1.2cm}
[K_l,i\p_t \chi+ i\sigma_{i}\p_i \phi]= 2i\sigma_l \i(\Delta -\frac{3}{2}\i)\phi.
 \eea
 \ees
 Similarly, the invariance under $M^{m_1,m_2,m_3}$ can be worked out in a similar way. 
 \bes
\bea{}&&\hspace{-1cm} [M^{m_1,m_2,m_3}, \p_i (\partial_i A_t-\p_t A_i)- e(\phi^{\dagger}\phi)]=-\p_t (\p_t A_j-\partial_j A_t)=0,\\
&&\hspace{-1cm}[M^{m_1,m_2,m_3},i\p_t \phi]=0,
[M^{m_1,m_2,m_3},\p_t (\p_t A_j-\partial_j A_t)]=0,\\&&\hspace{-1cm}
[M^{m_1,m_2,m_3},i\p_t \chi+ i\sigma_{i}\p_i \phi]=0.
\eea
\ees
We see that the Electric sector comes out to be invariant under finite and infinite conformal Carrollian symmetries in $d=4$.
\paragraph{Magnetic sector:}
The scaling of the fields in this sector are given by
\begin{equation}\label{mefed}
\phi \to \epsilon^\alpha \phi,  \, \chi \to \epsilon^\beta \chi,  \,A_t \to \epsilon A_t,  \, A_i\to  A_i.
\end{equation}
 In this limit, the equations of motion becomes
 \bes{}\label{mfed}
\bea{}&&
 \p_i (\e^2 \partial_i A_t-\p_t A_i)- e(\epsilon^{2\alpha+1}\phi^{\dagger}\phi+\epsilon^{2\beta+1}\chi^{\dagger}\chi)=0,\\&&
-\p_t(\p_t A_j-\e^2\partial_j A_t)+\e^2 \p_i(\p_i A_j-\partial_j A_i)+ \epsilon^{\alpha+\beta+2}e(\phi^{\dagger}\sigma_j\chi
+\chi^{\dagger}\sigma_j\phi)=0,\hspace{1cm}\\&&
i\p_t \phi+\e^2 eA_t \phi+ \e^{\b-\a+1}(i\sigma_{i}\p_i \chi +e\sigma^{i}A_i \chi)=0,\\&&
i\p_t \chi+\e^{\a-\b+1} i\sigma_{i}\p_i \phi+\e^2 eA_t \chi+ \e^{\a-\b+1}e\sigma^{i}A_i \phi=0.
\eea \ees
Comparing the above equations with the free equations of  magnetic sector of Carrollian electrodynamics and free fermions we obtain restrictions on the arbitrary constants $\alpha,\beta$  mentioned in Table \ref{magSED}.
\begin{table}[t]
\centering 
\begin{tabular}{|c| |c|}
\hline
\multicolumn{2}{| c |}{ Magnetic sector of Fermion+ED}\\
\hline
Free equation & Constraints\\ [0.5ex]
\hline 
$\p_t \p_t A_i=0$ & $\alpha+\beta+2\geq 0$ \\
$\partial_i\p_t A_i=0$ & $2\alpha+1 \geq 0, \,2\beta+1 \geq 0 $ \\
$i\p_t \phi=0$ & $\beta-\alpha+1>0$ \\
$i\p_t \chi+i\sigma_i\partial_i\phi=0$ & $\alpha-\beta+1=0$ \\
\hline
\end{tabular}
\caption{Constraints on $\alpha,\beta$ for Magnetic sector of electrodynamics with fermions.}
\label{magSED}
\end{table} 
\noindent Analysing all the constraints we get the following relations:
\begin{equation}
\alpha=\beta-1, \, \alpha \geq -\frac{1}{2}, \, \beta\, \geq -\frac{1}{2}.
\end{equation}

\noindent Hence the possible values of $\alpha,\beta$ are:

\bigskip

\begin{center}
\begin{tabular}{ | c |   c c c c c|}
\hline
$\alpha$ &  $-\frac{1}{2}$ &0 &$\frac{1}{2}$ & 1& $\hdots$ \\ [0.5ex]
\hline 
$\beta$  &    $\frac{1}{2}$ & 1 & $\frac{3}{2}$ & 2 & $\hdots$ \\
\hline
\end{tabular}
\end{center} 
%
%

\bigskip

\medskip

\noindent{\em{Different subsectors}}

\medskip

\noindent The different subsectors within the Magnetic sector are the following:
 \begin{itemize}
 \item Case 1: {$\alpha=-\frac{1}{2},\beta=\frac{1}{2}$}:
 \bes{} \label{c1mfed}
\begin{eqnarray}&&
\partial_i\p_t A_i+e\phi^{\dagger}\phi=0,~~\p_t\p_t A_i=0,\\&&
i \p_t \phi=0,~~
i\p_t \chi+i\sigma^{i}\partial_i \phi+e\sigma_{i}A_i \phi=0.
\end{eqnarray} 
\ees
 \item Case 2: {$\alpha>-\frac{1}{2},\beta=\a+1$}:
\bes{} \label{c2mfed}
 \begin{eqnarray}&&
\partial_i\p_t A_i=0,~~\p_t\p_t A_i=0,\\&&
i \p_t \phi=0,~~
i\p_t \chi+i\sigma_{i}\partial_i \phi+e\sigma_{i}A_i \phi=0.
\end{eqnarray} 
\ees
 \end{itemize}
Unlike the electric sector, Case 2 does not reduce to just the free equations of Carrollian gauge fields and fermions. There are interaction terms as well. 
 
\medskip

\noindent{\em{Symmetries of EOM}}

\medskip

\noindent The values of the constants  under Carrollian symmetries are given as
\bea{} \label{repscmd}
\i\{\underbrace{\Delta=\frac{3}{2},f=0,f'=-\frac{1}{2}}_{\textbf{Fermionic Field}},\underbrace{\Delta^\prime=1,a=1,b=0}_
{\textbf{Gauge Field}}\i\}.\eea
We will consider Case 1 as the representative sector to examine the symmetries. The same analysis can be repeated for Case 2, with similar results. The transformations of the equations of motion under scale transformations are given by
\bes
\bea{}&&\hspace{-1cm}[D,\partial_i\p_t A_i+e\phi^{\dagger}\phi]=(\Delta' -1) \p_i \partial_t A_i+(2\Delta -3)e\phi^{\dagger}\phi,\\&&\hspace{-1cm}
[D,\p_t \p_t A_i]=0,~~\Big[D,i\p_t \phi]=0,~
[D,i\p_t \chi+ i\sigma_{i}\p_i \phi +e\sigma_{i} A_i \phi]=(\Delta' -1)e\sigma_{i} A_i \phi.~~~
\eea
\ees
Putting in the values of $ \{\Delta, \Delta^\prime\}$, we find that the equations are invariant. 
Similarly, under $K_{l}$, we have
\bes
\bea{}&& [K_{l},\partial_i\p_t A_i+e\phi^{\dagger}\phi]=(4+2\Delta' -2\delta_{ii}) \p_t A_l+(2\Delta'-2)x_l (\p_{i}\p_t A_i)\non\\&&\hspace{3.6cm}+(4\Delta -6)x_l (e\phi^{\dagger}\phi),\\&&
[K_l,i\p_t \chi+ i\sigma_i\p_i \phi +e\sigma_i A_i \phi]=2(\Delta' -1)x_l (e\sigma_i A_i \phi) +(2\Delta -3) i\sigma_l \phi,\\&&
[K_l,\p_t \p_t A_i]=0,~~[K_l,i\p_t \phi]=0.
\eea
\ees
Again, putting in the values of the constants, the EOM turn out to be invariant. Under the infinite dimensional $M^{m_1,m_2,m_3}$, the invariance of the EOM are straight-forward. 
\bes
\bea{}&& [M^{m_1,m_2,m_3},\partial_i\p_t A_i+e\phi^{\dagger}\phi]=0,~~[M^{m_1,m_2,m_3},\p_t \p_t A_i]=0\\&&
[M^{m_1,m_2,m_3},i\p_t \chi+ i\sigma_i\p_i \phi +e\sigma_i A_i \phi]=-i \Big({\sigma_i\over 2}\Big)\p_i(x^{m_1}y^{m_2}z^{m_3})\p_t \phi=0,\\&&
[M^{m_1,m_2,m_3},i\p_t \phi]=0.
\eea
\ees 
So, we can conclude that the Magnetic sector of  Carrollian $U(1)$ gauge field coupled to fermion respects the conformal Carrollian  symmetry in $d=4$.

\newpage

\subsubsection*{Lessons from this section}
For the reader who does not wish to be buried in the details of the rather tedious analysis we have done so far in this section, we will provide a brief recapitulation of the main results. In this section, we have carried out a detailed analysis of  Carrollian Electrodynamics theory with scalars and fermions. In case of Carrollian Electrodynamics, there were only two sectors, viz. the electric and magnetic sectors. However, the inclusion of matter into the theory led to many diverse subsectors inside each of the two above mentioned sectors. Massless scalars led to the splitting of each sector into 5 different interacting sub-sectors. The inclusion of fermions made the splitting a bit more restrictive and there were two subsectors in each of the Electric and Magnetic sectors. 

\medskip

Next, we checked for the symmetries of the theories under conformal Carrollian generators. We found that the field theories possess finite as well as infinite conformal Carrollian symmetries in $d=4$. The infinite extension of symmetries makes Carrollian Electrodynamics with matter prominently distinct from its relativistic counterpart.

\newpage

\section{Carrollian Yang-Mills theory and massless fermions}\label{cfdf}
So far we have constructed Carrollian field theories with Abelian gauge symmetry. We have found that these theories respect finite and infinite conformal Carrollian  symmetry in $d=4$. At this point, it is natural to ask if we can extend our analysis to field theories with non-Abelian gauge symmetry as well. The ultra-relativistic limit of $SU(N)$ Yang-Mills has been described in detail in \cite{Bagchi:2016bcd}. In this section, we would construct Carrollian $SU(N)$ Yang-Mills theory coupled to massless fermionic matter  in $d=4$.

\subsection{Carrollian Yang-Mills Theory: A brief review} 
This section is going to be a quick review of pure Carrollian Yang-Mills (YM) theory presented in \cite{Bagchi:2016bcd}. The field content of Carrollian YM theory is the gauge potential
\be{}
A_{\mu}= A^a_{\mu} T_a.
\ee 
Here $T_a \, (a=1, \dots , \mathfrak{D})$ are the Lie algebra generators and the dimension of the gauge group is  $\mathfrak{D}$. In \cite{Bagchi:2016bcd}, it was shown that there are a large number of new sectors arising in the Carrollian limit, as opposed to the two sectors in Carrollian Electrodynamics. The reason lies in the projection of individual gauge fields into either individual electric or magnetic sectors. 

\medskip

\noindent{\em{Different Sectors}}

\medskip
\noindent Following \cite{Bagchi:2016bcd}, we divide the theory into $ \mathfrak{D} +1$ sectors, each characterised by a vector:
\bea{}\label{thetap}
\Xi_{(p)} = (\underbrace{0,0,\dots , 0}_{\mathfrak{D}-p}, \underbrace{1,1,\dots , 1}_{p}) \qquad ~~ p=0, \dots, \mathfrak{D} 
\eea
 $\Xi_{(p)}^a$ denotes its $a$'th component. We choose a particular sector (say, $p_0$-th) denoted by  the vector $\Xi_{(p_0)}$. In $p_0$-th sector $\Xi_{(p_0)}$, the gauge fields scale as
\bea{}
A^a_t \to \frac{\epsilon}{1+\epsilon-\Xi^a_{(p_0)}} A^a_t,\:\: A^a_i \to \frac{\epsilon}{\epsilon+\Xi^a_{(p_0)}} A^a_i,
\eea
If $\Xi^a_{(p_0)}=1$, the gauge fields transform electrically and we denote them by the index $\a,\b$:
\bea{}
A^\a_t \to A^\a_t,\:\: A^\a_i\to \epsilon A^\a_i .
\eea
On the other hand, for $\Xi^a_{(p_0)}=0$, the gauge fields transform magnetically and we denote them by the index $I,J$:
\bea{}
A^I_t \to \epsilon A^I_t,\:\: A^I_i\to  A^I_i .
\eea
It is seen from \eqref{thetap} that $I,J$ run from $1,2\hdots,\mathfrak{D}-p_0$ and $\a,\b$ are in the range $\mathfrak{D}-p_0+1,\hdots,\mathfrak{D}$. The Carrollian Yang-Mills equations are:
\subsection*{Case 1: $1 \leq a \leq \mathfrak{D}-p_0$}
The Carrollian scalar equation is:
\bea{remnant1}
\p_i \p_t A_i^I + g f^{I}{}_{JK} A^{J}_{i} \p_{t}A_i^{K}=0.
\eea
The vector one is:
\bea{remnant2}
\p _t \p_t A_j^I =0.
\eea
\subsection*{Case 2: $\mathfrak{D}-p_0+1 \leq a \leq \mathfrak{D}$}
Scalar equation:
\bea{special}
f^{\alpha}{}_{IJ}A^{I}_{i}\p_{t} A_{i}^{J}=0.
\eea
Vector equation:
\bea{special2}
\p_t( \p_t A_j^{\alpha} - \p_j A_t^{\alpha}+g f^{ \alpha}{}_{\beta I} A_t^{\beta} A_j^{I}) +  g f ^{\alpha}{}_{ \beta I} A_{t}^{\beta}\p_{t} A^I_j=0.
\eea
It is worth noting here that the equation \eqref{special} does not have a kinetic term. Hence, these sectors would not be considered when we are looking at sensible Carrollian theories. This argument turns out to be a little too hasty. Notice also that the interaction piece vanishes when there is only one Magnetic leg ($I=J=1$).  Hence, in \cite{Bagchi:2016bcd}, the next higher order term in the series of $\epsilon$ was considered and this led to the following equation:
\bea{subleadin_eqn}
&&\p_i (\p_i A^{\a}_t - \p_t A^{\a}_i+g f^{\a}_{~I\b} A^{I}_i A^{\b}_t )- g f^{\a}_{~\b I} A^\b_i \p_t A^{I}_i+gf^{\a}_{~IJ}A^{I}_{i}(gf^{J}_{~K\rho}A^{K}_iA^{\rho}_{t})\non\\
&&~~~~~~~~~~~~~~~~~~~~~~~~~~~~~~~~~~+gf^{\a}_{~I\b} A_i^I (\p_i A^{\b}_t-\p_t A^{\b}_i+gf^{\b}_{~K\rho}A^{K}_iA^{\rho}_t)=0. 
\eea
So, this is a mixed leg that we would consider as a valid sector. 

\medskip

\noindent To summarise, the sectors of pure Carrollian Yang-Mills theory that we interested in are:
\begin{itemize}
\item{\em{Purely Magnetic sector:}} The equations of relevance here are $\{\eqref{remnant1}, \eqref{remnant2} \}$;  
\item{\em{Purely Electric sector:}} The purely electric sector survives. This is the one where we consider $\{\eqref{special2}$ (with $I=0$), $\eqref{subleadin_eqn}\}$. There are no magnetic legs, and hence we have to put $A^I=0$ in these equations. This leads to:  
\be{}
\p_t( \p_t A_j^{\alpha} - \p_j A_t^{\alpha}) =0,\quad \p_i( \p_i A_t^{\alpha} - \p_t A_i^{\alpha}) =0.
\ee
\item{\em{Mixed sector (with $J=1$):}} Only one mixed sector survives. This corresponds to $p_0=\mathfrak{D}-1$ and equations $\{\eqref{special2}$  (with $I=1$),  $\eqref{subleadin_eqn}$ (with $I,J,K=1$)\}. 
\bea{}
&&\p_t( \p_t A_j^{\alpha} - \p_j A_t^{\alpha}+g f^{ \alpha}{}_{\beta 1} A_t^{\beta} A_j^{1}) +  g f ^{\alpha}{}_{ \beta 1} A_{t}^{\beta}\p_{t} A^1_j=0,\\
 &&\p_{i}(\p_{i}A^{\alpha}_{t}-\p_{t}A^{\alpha}_{i}+gf^{\alpha}_{\hspace{.2cm}1\beta}A^{1}_{i}A^{\beta}_{t})-gf^{\alpha}_{\hspace{.2cm}\beta 1}A^{\beta}_i \p_{t}A^{1}_{i}  \non\\
&&~~~~~~~~~~~~~~ +gf^{\alpha}_{\hspace{.2cm}1\beta}A^{1}_i(\p_{i}A^{\beta}_{t}-\p_{t}A^{\beta}_{i}+gf^{\beta}_{\hspace{.2cm}1\rho} A^{1}_{i}A^{\rho}_{t})=0
\eea
\end{itemize}
It was also shown in \cite{Bagchi:2016bcd} that these the sectors in Carrollian Yang-Mills theory respect finite and infinite conformal Carrollian symmetries in $d=4$.

\subsection*{Yang-Mills with Fermions}
 The Lagrangian density of  the relativistic $SU(N)$ Yang-Mills with fermions is given by
\bea{}\label{spnyml} \mathcal{L}=-\frac{1}{4}F_{\mu\nu}^{a}F^{\mu\nu a} + i\bar{\psi}_m \gamma^{\mu}(D_{\mu}\psi)_m ,\eea
where $ D_\mu \equiv \partial_\mu-igT^{a}A_{\mu}^{a}$ is the non abelian gauge covariant derivative  and $F_{\mu\nu}^{a}= \partial_{\mu} A_{\nu}^{a} - \partial_\nu A_{\mu}^{a}+gf^{abc}A^{b}_{\mu}A^{c}_{\nu}$ is the non abelian field strength tensor. The label $a$ is the colour index, $m$ is an internal symmetry index and $f^{abc}$ are the structure constants of the underlying gauge group with generators following the algebra $[T^{a},T^{b}]=if^{abc}T^{c}$.
The equations of motion for this theory are 
\be{}\label{eomsymm}  \p_{\mu}F^{\mu \nu a}+gf^{abc}A^{b}_{\mu}F^{\mu\nu c}+g\bar{\psi}_{m} \gamma^{\nu}T^{a}_{mn}\psi_{n}=0,~~
i\gamma^{\nu}(D_{\nu}\psi)_n =0.  \ee
The relativistic equations of motion  under the decomposition of Dirac fermion (\ref{ddc}) are  
\bes{}\label{gsymeomc}
\bea{} 
&& {\hspace{-0.7cm}} \p_{t}F_{tj}^{a}-\p_{i}F_{ij}^{a} + gf^{abc}(A^{b}_{t}F_{tj}^{c} - A^{b}_{i}F_{ij}^{c}) - g(\phi^{\dagger}_{m} \sigma_{j}T^{a}_{mn}\chi_{n} +\chi^{\dagger}_{m} \sigma_{j}T^{a}_{mn}\phi_{n})=0,~~\\
&&{\hspace{-0.7cm}} \p_{i}F_{it}^{a}+gf^{abc}A^{b}_{i}F_{it}^{c} -g(\phi^{\dagger}_m T^{a}_{mn}\phi_{n} +\chi^{\dagger}_{m} T^{a}_{mn}\chi_{n})=0,~~\\
&&{\hspace{-0.7cm}} i\p_{t}\phi_m +gT^{a}_{mn}A_{t}^{a}\phi_{n} +i\sigma_{i}\p_{i}\chi_{m} +g\sigma_{i}T^{a}_{mn}A_{i}^a \chi_{n} =0,\\
 &&{\hspace{-0.7cm}} i\p_{t}\chi_{m}+gT^{a}_{mn}A_{t}^{a}\chi_{n} +i\sigma_{i}\p_{i}\phi_{m} +g\sigma_{i}T^{a}_{mn}A_{i}^a \phi_{n} =0.
\eea\ees 
The relativistic Yang-Mills coupled to fermions have conformal symmetry in $d=4$.

\subsection{Carrollian $SU(N)$ Yang-Mills with Fermions}\label{eer}
\label{Carrollian $SU(N)$ Yang-Mills with Fermions}
The simplest example of a non-Abelian group is, of course, $SU(2)$. We have discussed the details of the Carrollian $SU(2)$ Yang-Mills with fermionic matter extensively in Appendix \ref{Carrollian $SU(2)$ Yang-Mills with Fermions}, to which the interested reader is pointed. Here we proceed to present the general analysis for Carrollian $SU(N)$ Yang-Mills with matter. The discussion can also be generalised immediately to any other compact gauge group. 

Let us now move to the scaling of  fields in $SU(N)$ case. The scaling of  gauge fields is very similar to the scaling  in \cite{Bagchi:2016bcd} and is briefly explained in the previous section. 
Next step will be to take scaling on the fermions ($\phi_m,\chi_m$) where $m=1,\hdots,N$. The scaling is taken as,
\be{}
\phi_m\to \epsilon^{a_m}\phi_m,\:\: \chi_m\to \epsilon^{b_m}\chi_m.
\ee
We will look at the equations of motion of different sectors of $SU(N)$ Carrollian Yang-Mills coupled to fermions:
\paragraph{$\bullet~~ 1\leq a\leq \mathfrak{D}-p_0 $:} The equations of motion in this case are given as
\bes{}
\bea{} \label{mtA}
&&\hspace{-1cm}\mathcal{A}=\p_i \p_t A_i^I+gf^{I}_{~JK} A_i^J  \p_t  A^{K}_i+g[ \e^{a_m+a_n+1}\phi^{\dagger}_m T^{I}_{mn}\phi_n+\e^{b_m+b_n+1}\chi^{\dagger}_m T^{I}_{mn}\chi_n]=0,\\
&&\hspace{-1cm}\mathcal{B}= \p_t \p_t A^{I}_j-g[\e^{a_m+b_n+2}\phi^{\dagger}_m\sigma_j T^{I}_{mn}\chi_n+ \e^{b_m+a_n+2}\chi^{\dagger}_m\sigma_j T^{I}_{mn}\phi_n ]=0.
\eea\ees
For simplicity, we will denote the LHS of first equation by $\mathcal{A}$ and second by $\mathcal{B}$.
\paragraph{$\bullet~~ \mathfrak{D}-p_0+1\leq a\leq \mathfrak{D} $:} The equations of motion in this case are given as
\bes{}
\bea{}
&&\hspace{-1cm}\mathcal{C}=\p_t(\p_t A^{\a}_j-\p_j A^\a_t+gf^{\a}_{~\b I}A^{\b}_t A^{I}_j) +gf^{\a}_{~\b I}A^{\b}_t \p_t A^{I}_j-g[ \e^{a_m+b_n+1}\phi^{\dagger}_m \sigma_j T^{\a}_{mn}\chi_n   \non\\&&\hspace{7.5cm}+
\e^{b_m+a_n+1}\chi^{\dagger}_m\sigma_jT^{\a}_{mn}\phi_n]=0,\\ \label{zeroeq} 
&&\hspace{-1cm} \mathcal{E}=g f^{\a}_{~IJ}A^I_i \p_t A^{J}_i+ g[\e^{a_m+a_n+1}\phi^{\dagger}_mT^{\a}_{mn}\phi_n+ \e^{b_m+b_n+1}\chi^{\dagger}_mT^{\a}_{mn}\chi_n]=0.\label{ffd}
\eea\ees
We will denote the LHS of the third and fourth equation by $\mathcal{C}$ and $\mathcal{E}$. When there is only one magnetic leg $I=J=1$, the first term in \eqref{zeroeq} goes to zero by the antisymmetric property of the structure constant $f^{\a}_{~IJ},\:\: (f^{\a}_{~11}=0)$. For that reason we consider the next order term in $\epsilon$ (that is $\e^{0}$), and the equation becomes
\bea{}\label{fff}&&\mathcal{F}= \p_i (\p_i A^{\a}_t - \p_t A^{\a}_i+g f^{\a}_{~I\b} A^{I}_i A^{\b}_t )- g f^{\a}_{~\b I} A^\b_i \p_t A^{I}_i+gf^{\a}_{~IJ}A^{I}_{i}(gf^{J}_{~K\rho}A^{K}_iA^{\rho}_{t})\non\\
&&+gf^{\a}_{~I\b} A_i^I (\p_i A^{\b}_t-\p_t A^{\b}_i+gf^{\b}_{~K\rho}A^{K}_iA^{\rho}_t)- g[\e^{a_m+a_n}\phi^{\dagger}_m T^{\a}_{mn}\phi_n+ \e^{b_m+b_n}\chi^{\dagger}_m T^{\a}_{mn}\chi_n ]=0.\non\\
\eea
The Dirac equations are same in both the cases and they are given as
\bes
\bea{}
&&\mathcal{G}=i\p_t \phi_m+\e^{b_m -a_m +1}i\sigma_{i}\p_i \chi_m+\e^{a_n -a_m+1} gT^{\a}_{mn}A^{\a}_{t}\phi_n +\e^{a_n -a_m +2} gT^{I}_{mn}A^{I}_{t}\phi_n \non \\
&&\hspace{3.5cm} + \e^{b_n -a_m +2}g\sigma_{i}T^{\a}_{mn} A^{\a}_{i}\chi_n+\e^{b_n -a_m+1}g\sigma_{i}T^{I}_{mn} A^{I}_{i}\chi_n=0,\\ \label{mtH}
&&
\mathcal {H}=i\p_t \chi_m+\e^{a_m -b_m +1}i\sigma_{i}\p_i \phi_m+\e^{b_n -b_m+1} gT^{\a}_{mn}A^{\a}_{t}\chi_n +\e^{b_n -b_m +2} gT^{I}_{mn}A^{I}_{t}\chi_n \non \\
&&\hspace{3.5cm} + \e^{a_n -b_m +2}g\sigma_{i}T^{\a}_{mn} A^{\a}_{i}\phi_n+\e^{a_n -b_m+1}g\sigma_{i}T^{I}_{mn} A^{I}_{i}\phi_n=0.~~~
\eea
\ees
Hence, in this section we are left with the equations $\mathcal{A,B,C,E,F,G,H}$ depending on different sectors of Carrollian $SU(N)$ Yang-Mills coupled to fermions. Please note that we are yet to specify $a_m$'s and $b_m$'s, which we will be doing in the next section.

\subsubsection*{Detailed discussion of different sectors} 

At this juncture, it is important to remind ourselves of the discussion in the pure Carrollian YM case. There we had discarded sectors which had no kinetic terms. This obviously would also continue to hold when we put in matter fields. Clearly, a sector where the equation $\mathcal{E}$ \refb{zeroeq} holds, is not a ``nice" sector. In the analysis that follows, we will continue to investigate all sectors, keeping in mind, that if we were to focus on ``nice" sectors of Carrollian YM coupled to matter, all the sectors which have a non-trivial realisation of \refb{zeroeq} would be discarded. 

In case of $SU(2)$ Carrollian Yang-Mills with fermions there are three gauge fields and 4 different sectors: EEE, EEM, EMM, MMM. Similarly for $SU(N)$ case, we can have $\mathfrak{D}+1$ distinct sectors. We would organise $\mathfrak{D}+1$ sectors depending on the numbers of magnetic legs.   

\paragraph{Number of Magnetic legs $J\geq 1$:} We start with the case in which one and more gauge field transform magnetically. For $J=1$, the equation $\mathcal{E}$ does not hold. Hence instead of $\mathcal{E}$, we get 
\bea{}&&\mathcal{F}= \p_i (\p_i A^{\a}_t - \p_t A^{\a}_i+g f^{\a}_{~1\b} A^{1}_i A^{\b}_t )- g f^{\a}_{~\b 1} A^\b_i \p_t A^{ 1}_i
+gf^{\a}_{~1\b} A_i^1 (\p_i A^{\b}_t-\p_t A^{\b}_i\non\\
&&\hspace{2cm}+gf^{\b}_{~1\rho}A^{1}_iA^{\rho}_t) - g[\e^{a_m+a_n}\phi^{\dagger}_m T^{\a}_{mn}\phi_n+ \e^{b_m+b_n}\chi^{\dagger}_m T^{\a}_{mn}\chi_n ]=0.
\eea
The rest of the equations $\mathcal{A,B,C,G,H}$ remains same with $J=1$. Now we want to see the constraints on $a_m~\text{and}~ b_m$. 
In Table[\ref{SU(N)1table}], we find the constraints by comparing with the free equations.
\begin{table}[t]
\centering
    \begin{tabular}{| l |  p{5cm} |}
   
    \hline
\multicolumn{2}{ |c| }{Magnetic leg $J\geq 1$} \\
\hline
  From equation & Constraints\\ [0.2cm]
\hline
\multirow{2}{*}{Equation $\mathcal{A}$ of Gauge field  $A^I$}  &(i) $a_m+a_n+1\geq0$\\[0.2cm]
 & (ii) $b_m+b_n+1>0$\\
   \hline
   Equation $\mathcal{B}$ of Gauge field  $A^{I}$ &(iii) $a_m+b_n+2>0$ [no equality to reduce it to U(1)]\\[0.2cm]
    \hline
 Equation $\mathcal{C}$ of Gauge field $A^{\a}$  &  (iv) $a_m+b_n+1>0$\\[0.2cm]
\hline
\multirow{2}{*}{Equation $\mathcal{E}$ (for $J>1$) of Gauge field $A^\a$}  &(v) $a_m+a_n+1\geq0$\\[0.2cm]
 & (vi) $b_m+b_n+1>0$\\
 \hline
\multirow{2}{*}{Equation $\mathcal{F}$ (for $J=1$) of Gauge field $A^\a$}  &(vii) $a_m+a_n\geq0$\\[0.2cm]
 & (viii) $b_m+b_n>0$\\
\hline
\multirow{3}{*}{Equation $\mathcal{G}$ for $\phi_{\sss_m}$} & (ix) $b_m-a_m+1\geq0$,\\[0.2cm]
    & (x) $a_n-a_m+1\geq0$ \\
    & (xi) $b_n-a_m+1\geq0$\\
   \hline
   \multirow{3}{*}{Equation $\mathcal{H}$ for $\chi_{\sss_m}$}&  (xii) $a_m-b_m+1=0$\\
   & (xiii) $b_n-b_m+1\geq0$\\
   & (xiv) $a_n-b_m+1\geq0$\\
 \hline
\end{tabular}
\caption{Constraints on $a_m,b_m$ in $J\geq 1$ sector.}
\label{SU(N)1table}
\end{table}
Next, we analyze the range of $a_m$ and $b_m$. From equation  (xii) and (xiv) we can see,
\bea{sun1}&&\non
(xii) : a_m-b_m+1=0 \implies b_m=a_m+1,\\&&
(xiv)  : a_n-b_m+1 \geq 0 \implies a_n-a_m\geq0.
\eea
Eq \eqref{sun1} is valid for all $m,n=(1,2,\hdots N)$. Hence, the second equation of \eqref{sun1} says,
\be{sun2}
a_n=a_m \quad  \forall ~m,n.
\ee
We are yet to specify the range in which $a_m$ belong. They are given by
\bea{}\label{sun3}
 &&\text{For }J>1 ~~(v):  a_m+a_n+1\geq 0 \implies a_m \geq -\frac{1}{2}\\ \label{sun4}
 &&\text{For }J=1 ~~(vii):  a_m+a_n\geq 0 \implies a_m \geq 0 
 \eea
Combining equations \eqref{sun1}-\eqref{sun4}, we get the Table[\ref{ddd}].
\begin{table}[t]
\centering
    \begin{tabular}{| l | c|}
    \hline
\multirow{2}{*}{For $J=1$} & $a= 0,\frac{1}{2},1,\frac{3}{2}\hdots$\\
   & $b=1, \frac{3}{2},2,\frac{5}{2},\hdots$\\
\hline
\multirow{2}{*}{For $J>1$} & $a= -\frac{1}{2},0,\frac{1}{2},1,\frac{3}{2}\hdots$\\
   & $b=\frac{1}{2},1, \frac{3}{2},2,\frac{5}{2},\hdots$\\
\hline
\end{tabular}
\caption{Values of $a_m,b_m$ in $J\geq 1$ sector.}
\label{ddd}
\end{table}
Writing the equations of motion in each sector separately, we get
\paragraph{For the case $J> 1$,} the equations of motion are written as,
\begin{itemize}
\item Case 1: ($a_m=-\frac{1}{2},b_m=\frac{1}{2}$):
\bes{} \label{jg1}
\bea{}\label{e1}
&&\hspace{-1cm}\p_i \p_t A_i^I+gf^{I}_{~JK} A_i^J  \p_t  A^{K}_i+g\phi^{\dagger}_m T^{I}_{mn}\phi_n=0,~ \p_t \p_t A^{I}_j=0,\\ \label{j21}
&&\hspace{-1cm}\p_t(\p_t A^{\a}_j-\p_j A^\a_t+gf^{\a}_{~\b I}A^{\b}_t A^{I}_j) +gf^{\a}_{~\b I}A^{\b}_t \p_t A^{I}_j=0,\\\label{j11}
&&\hspace{-1cm} g f^{\a}_{~IJ}A^I_i \p_t A^{J}_i+ g\phi^{\dagger}_mT^{\a}_{mn}\phi_n=0,
\\&&\hspace{-1cm} 
i\p_t \phi_m=0,~
i\p_t \chi_m +i\sigma_{i}\p_i \phi_m +g\sigma_{i}T^{I}_{mn} A^{I}_{i}\phi_n=0.
\eea
\ees
\item Case 2: ($a_m > -\frac{1}{2},b_m=a_m+1$):
\bes{} \label{jg2}
\bea{}\label{e2}
&&\hspace{-1cm}\p_i \p_t A_i^I+gf^{I}_{~JK} A_i^J  \p_t  A^{K}_i=0,~ \p_t \p_t A^{I}_j=0,\\\label{j22}
&&\hspace{-1cm}\p_t(\p_t A^{\a}_j-\p_j A^\a_t+gf^{\a}_{~\b I}A^{\b}_t A^{I}_j) +gf^{\a}_{~\b I}A^{\b}_t \p_t A^{I}_j=0,\\\label{j12}
&&\hspace{-1cm} g f^{\a}_{~IJ}A^I_i \p_t A^{J}_i=0,
\\&&\hspace{-1cm} 
i\p_t \phi_m=0,~
i\p_t \chi_m +i\sigma_{i}\p_i \phi_m +g\sigma_{i}T^{I}_{mn} A^{I}_{i}\phi_n=0.
\eea
\ees
\end{itemize} 
Both of the above mentioned sectors contain no kinetic term in \eqref{j11},\eqref{j12}. But \eqref{j21},\eqref{j11},\eqref{j22} and \eqref{j12} drop off completely when $\a=0$, or in other words when there is no electric leg. It corresponds to the purely magnetic limit of $SU(N)$ Yang-Mills. Hence, for the purely magnetic limit we get two individual sectors which differ by a fermionic interaction term $g\phi^{\dagger}_m T^{I}_{mn}\phi_n$ within themselves (see \eqref{e1} and \eqref{e2}).

\newpage

\paragraph{For the case $J= 1$,} the equations of motion are written as,
\begin{itemize}
\item Case 1: ($a_m =0, b_m =1$) :
\bes{} \label{je1}
\bea{}
&&\hspace{-1.5cm}\p_i \p_t A_i^1=0,~ \p_t \p_t A^{1}_j=0.\\
&&\hspace{-1.5cm}\p_t(\p_t A^{\a}_j-\p_j A^\a_t+gf^{\a}_{~\b 1}A^{\b}_t A^{1}_j) +gf^{\a}_{~\b 1}A^{\b}_t \p_t A^{1}_j=0,\\
&&\hspace{-1.5cm} \p_i (\p_i A^{\a}_t - \p_t A^{\a}_i+g f^{\a}_{~1\b} A^{1}_i A^{\b}_t )- g f^{\a}_{~\b 1} A^\b_i \p_t A^{1}_i
\non\\&&+gf^{\a}_{~1\b} A_i^1 (\p_i A^{\b}_t-\p_t A^{\b}_i+gf^{\b}_{~1\rho}A^{1}_iA^{\rho}_t)- g\phi^{\dagger}_m T^{\a}_{mn}\phi_n=0,\\&&\hspace{-1.5cm}
i\p_t \phi_m=0,~ i\p_t \chi_m +i\sigma_{i}\p_i \phi_m +g\sigma_{i}T^{1}_{mn} A^{1}_{i}\phi_n=0.
\eea
\ees
\item Case 2: ($a_m >0, b_m =a_m+1$):
\bes{} \label{je2}
\bea{}
&&\hspace{-1cm}\p_i \p_t A_i^1=0,~ \p_t \p_t A^{1}_j=0.\\
&&\hspace{-1cm}\p_t(\p_t A^{\a}_j-\p_j A^\a_t+gf^{\a}_{~\b 1}A^{\b}_t A^{1}_j) +gf^{\a}_{~\b 1}A^{\b}_t \p_t A^{1}_j=0,\\
&&\hspace{-1cm} \p_i (\p_i A^{\a}_t - \p_t A^{\a}_i+g f^{\a}_{~1\b} A^{1}_i A^{\b}_t )- g f^{\a}_{~\b 1} A^\b_i \p_t A^{1}_i
\non\\&&+gf^{\a}_{~1\b} A_i^1 (\p_i A^{\b}_t-\p_t A^{\b}_i+gf^{\b}_{~1\rho}A^{1}_iA^{\rho}_t)=0,\\&&\hspace{-1cm}
i\p_t \phi_m=0,~ i\p_t \chi_m +i\sigma_{i}\p_i \phi_m +g\sigma_{i}T^{1}_{mn} A^{1}_{i}\phi_n=0.
\eea
\ees
\end{itemize}
From the above analysis we see that there are (2+2) different sectors in the purely magnetic ($\a=0$) and  only one magnetic leg ($J=1$) limit respectively.

\medskip

\paragraph{Number of Magnetic legs $J=0$:} This sector defines the case in which all the gauge fields transform electrically.  The equations of motion are given as
\bes{}
\bea{}
&&\hspace{-1cm}\p_t(\p_t A^{\a}_j-\p_j A^\a_t) -g[ \e^{a_m+b_n+1}\phi^{\dagger}_m \sigma_j T^{\a}_{mn}\chi_n  +
\e^{b_m+a_n+1}\chi^{\dagger}_m\sigma_jT^{\a}_{mn}\phi_n]=0,\label{edf1}
\\&&\hspace{-1cm} \p_i (\p_i A^{\a}_t - \p_t A^{\a}_i)- g[\e^{a_m+a_n}\phi^{\dagger}_m T^{\a}_{mn}\phi_n+ \e^{b_m+b_n}\chi^{\dagger}_m T^{\a}_{mn}\chi_n ]=0,\label{edf2}
\\&&\hspace{-1cm}
i\p_t \phi_m+\e^{b_m -a_m +1}i\sigma_{i}\p_i \chi_m+\e^{a_n -a_m+1} gT^{\a}_{mn}A^{\a}_{t}\phi_n +\e^{b_n -a_m +2}g\sigma_{i}T^{\a}_{mn} A^{\a}_{i}\chi_n =0,\\&&\hspace{-1cm}
i\p_t \chi_m+\e^{a_m -b_m +1}i\sigma_{i}\p_i \phi_m+\e^{b_n -b_m+1} gT^{\a}_{mn}A^{\a}_{t}\chi_n+ \e^{a_n -b_m +2}g\sigma_{i}T^{\a}_{mn} A^{\a}_{i}\phi_n =0.
\eea
\ees
Similar to the previous case, we will be writing down Table[\ref{SU(N)2table}] to have a clear understanding of the constraints on $a_m~\text{and}~b_m$.
\begin{table}[t]
\centering
    \begin{tabular}{| l |  p{5cm} |}
    \hline
\multicolumn{2}{ |c| }{Purely Electric sector ($J=0$)} \\
\hline
  From equation & Constraints\\ [0.2cm]
\hline
 Equation \eqref{edf1} of Gauge field $A^{\a}$  &  (i) $a_m+b_n+1>0$\\[0.2cm]
\hline
\multirow{2}{*}{Equation \eqref{edf2} of Gauge field $A^\a$}  &(ii) $a_m+a_n\geq0$\\[0.2cm]
 & (iii) $b_m+b_n>0$\\
\hline
\multirow{3}{*}{Equation for $\phi_{\sss_m}$} & (iv) $b_m-a_m+1\geq0$ \\
    & (v) $a_n-a_m+1\geq0$ \\
    & (vi) $b_n-a_m+2\geq0$\\
   \hline
   \multirow{3}{*}{Equation for $\chi_{\sss_m}$} & (vii) $a_m-b_m+1=0$ \\
   & (viii) $b_n-b_m+1\geq0$\\
   & (ix) $a_n-b_m+2\geq0$\\
 \hline
\end{tabular}
\caption{Constraints on $a_m,b_m$ in the $J=0$ sector.}
\label{SU(N)2table}
\end{table}
The relevant equations that we need to look at are (ii), (v) and (vii) in Table[\ref{SU(N)2table}].
\bea{sun5}
&&\non(ii) : ~ a_m+a_n \geq 0,\\\non
&&(v) : ~ a_n-a_m+1 \geq 0,\\
&&(vii) : ~ a_m-b_m+1 = 0 \implies b_m= a_m+1.
\eea
Unlike $J\geq1$ sector, we  get no more constraints to specify $a_m$ . Hence there will be many  sectors describing $J=0$ case. At this point, we want to count how many sectors actually arise for the purely electric case of Carrollian $SU(N)$ Yang-Mills coupled to fermions considering the constraints \eqref{sun5}.\\

\noindent We can start with $SU(2)$ YM coupled to fermions case. We can see  ($a_1,a_2$) can have  5 distinct values taking in consideration the exchange symmetry between $a_1,a_2$. The free sectors which do not contain any interaction terms between the fermions and the gauge fields are also accounted for in this counting of 5 different sectors for $SU(2)$ YM coupled to fermions case. Similarly for $SU(3)$ coupled with fermions case, the set of ($a_1,a_2,a_3$) can take 11 distinct values (including the free sectors). The number of distinct values of $a_m$'s for $SU(4)$ and $SU(5)$ case including the free sectors are 17 and 26 respectively. So,  for $SU(N)$ coupled with fermions case, we can say that there are $(N^2+1)$ different sectors in the purely electric limit including the free sectors. For convenience we are showing the allowed values of $a_m$'s in $SU(N)$ Yang-Mills theory with fermions in Table[\ref{sunbigtable}]. Any other combination of these $a_m$'s, which obeys the constraints \eqref{sun5}, gives back no new sector.   
\begin{table}[ht]
\centering
  \begin{tabular}{| c | c c | c c c | c c c c | c c c c c |}
   \hline
& \multicolumn{2}{|c|}{$SU(2)$} & \multicolumn{3}{|c|}{$SU(3)$} & \multicolumn{4}{|c|}{$SU(4)$} & \multicolumn{5}{|c|}{ $SU(5)$} \\ 
& \multicolumn{2}{|c|}{+ fermion} & \multicolumn{3}{|c|}{+ fermion} & \multicolumn{4}{|c|}{+ fermion} & \multicolumn{5}{|c|}{ + fermion} \\ 
& \multicolumn{2}{|c|}{ $(a_1,a_2)$}& \multicolumn{3}{|c|}{ $(a_1,a_2,a_3)$} & \multicolumn{4}{|c|}{ $(a_1,a_2,a_3,a_4)$} & \multicolumn{5}{|c|}{ $(a_1,a_2,a_3,a_4,a_5)$}\\
\hline
1. & 0&0 &  0 & 0 & 0 &  0&0&0&0  &0&0&0&0&0 \\ \cline{2-15}

2. & 0&$\frac{1}{2}$    &    0 & 0& $\frac{1}{2}$ & 0&0&0&$\frac{1}{2}$ & 0&0&0&0&$\frac{1}{2}$ \\ \cline{2-3}
3. & $\frac{1}{2}$& $\frac{1}{2}$                  &    0 & $\frac{1}{2}$& $\frac{1}{2}$ & 0&0&$\frac{1}{2}$&$\frac{1}{2}$  & 0&0&0&$\frac{1}{2}$&$\frac{1}{2}$\\ \cline{2-3}
4.& 0&1   &  $\frac{1}{2}$& $\frac{1}{2}$& $\frac{1}{2}$ & 0&$\frac{1}{2}$&$\frac{1}{2}$&$\frac{1}{2}$   & 0&0&$\frac{1}{2}$&$\frac{1}{2}$&$\frac{1}{2}$\\ \cline{4-6}\cline{2-3}
5. & $\frac{1}{2}$ & $\frac{3}{2}$    & 0& 0& 1        & $\frac{1}{2}$&$\frac{1}{2}$&$\frac{1}{2}$&$\frac{1}{2}$ & 0&$\frac{1}{2}$&$\frac{1}{2}$&$\frac{1}{2}$&$\frac{1}{2}$\\\cline{2-3} \cline{7-10}
6. &  &  & 0 &1 & 1    & 0&0&0&1  & $\frac{1}{2}$&$\frac{1}{2}$&$\frac{1}{2}$&$\frac{1}{2}$&$\frac{1}{2}$\\ \cline{4-6} \cline{11-15}
7.& & &  $\frac{1}{2}$ & $\frac{1}{2}$ & $\frac{3}{2}$    & 0&0&1&1   & 0& 0&0&0&1\\
                               
8. &  &  & $\frac{1}{2}$ & $\frac{3}{2}$ & $\frac{3}{2}$  & 0&1&1&1  & 0& 0&0&1&1\\  \cline{4-6}\cline{7-10}
9. &  &  & $\frac{1}{2}$ & $1$ & $\frac{3}{2}$  & 0&0&$\frac{1}{2}$&1 &  0&0&1&1&1\\
10. &  &  & 0 & $\frac{1}{2}$ & 1  & 0&$\frac{1}{2}$&$\frac{1}{2}$&1 &  0&1&1&1&1\\ \cline{4-6} \cline{11-15}
11. &  &  &  & & & 0 &$\frac{1}{2}$ &1 & 1   & 0&0&0&$\frac{1}{2}$&1\\ \cline{7-10}
12. &  &  &  & & & $\frac{1}{2}$ &$\frac{1}{2}$&$\frac{1}{2}$ & $\frac{3}{2}$ & 0&0&$\frac{1}{2}$&$\frac{1}{2}$&1\\
13. &  &  &  & & & $\frac{1}{2}$ &$\frac{1}{2}$&$\frac{3}{2}$ & $\frac{3}{2}$ & 0&$\frac{1}{2}$&$\frac{1}{2}$&$\frac{1}{2}$&1\\
14. &   &  &  & & & $\frac{1}{2}$ &$\frac{3}{2}$&$\frac{3}{2}$ & $\frac{3}{2}$   & 0&0&$\frac{1}{2}$&1&1\\ \cline{7-10}
15. & &  &  & & & $\frac{1}{2}$ &$\frac{3}{2}$& 1 & 1    & 0&$\frac{1}{2}$&1&1&1\\ 
16. & &  &  & & & $\frac{1}{2}$ &$\frac{3}{2}$& $\frac{3}{2}$ & 1  & 0& $\frac{1}{2}$&$\frac{1}{2}$&1&1  \\ \cline{11-15}
17. & &  &  & & & $\frac{1}{2}$ &$\frac{1}{2}$& 1 & $\frac{3}{2}$  & $\frac{1}{2}$ &$\frac{1}{2}$&$\frac{1}{2}$&$\frac{1}{2}$  & $\frac{3}{2}$\\  \cline{7-10}
18. & & & & & & & & & & $\frac{1}{2}$ &$\frac{1}{2}$ & $\frac{1}{2}$  & $\frac{3}{2}$ &$\frac{3}{2}$\\
19. & & & & & & & & & & $\frac{1}{2}$ &$\frac{1}{2}$ & $\frac{3}{2}$  & $\frac{3}{2}$ &$\frac{3}{2}$\\
20. & & & & & & & & & & $\frac{1}{2}$ &$\frac{3}{2}$ & $\frac{3}{2}$  & $\frac{3}{2}$ &$\frac{3}{2}$\\ \cline{11-15}
21. & & & & & & & & & & $\frac{1}{2}$ &$\frac{1}{2}$ & $\frac{1}{2}$  & 1 &$\frac{3}{2}$\\
22. & & & & & & & & & & $\frac{1}{2}$ &$\frac{1}{2}$ & 1  & 1 &$\frac{3}{2}$\\
23. & & & & & & & & & & $\frac{1}{2}$ &1 & 1  & 1 &$\frac{3}{2}$\\
24. & & & & & & & & & & $\frac{1}{2}$ &1 & 1  & $\frac{3}{2}$ &$\frac{3}{2}$\\
25. & & & & & & & & & & $\frac{1}{2}$ &1 &  $\frac{3}{2}$ & $\frac{3}{2}$ &$\frac{3}{2}$\\
26. & & & & & & & & & & $\frac{1}{2}$ &  $\frac{1}{2}$  &1& $\frac{3}{2}$ &$\frac{3}{2}$\\
\hline
\end{tabular}
\caption{Possible $a_m$ values for different dimensional Carrollian $SU(N)$ Yang-Mills with fermions.}
\label{sunbigtable}
\end{table}

\subsubsection*{Gauge Transformation for Carrollian $SU(N)$ Yang-Mills with Fermions} 
Let us consider the transformations of gauge fields and fermions under $SU(N)$ gauge group:
\bea{}\label{gauge1} \delta A_{\mu}^{a}=\frac{1}{g}\p_{\mu} \Theta^{a} +f^{a}_{~bc}A^{b}_{\mu}\Theta^{c},~\delta \phi_{m}=i\Theta^{a}T^{a}_{mn}\phi_{n},~\delta \chi_{m}=i\Theta^{a}T^{a}_{mn}\chi_{n}. \eea
where $\Theta$ is the gauge parameter. We will now see gauge transformations in ultra-relativistic limit of $SU(N)$ Yang-Mills theory. We scale the gauge transformation parameter for the electric and magnetic legs as 
\begin{equation}\label{thetae1}
\Theta^\a \to \e^p \Theta^\a,~~\Theta^I \to \e^q \Theta^I .
\end{equation}
We would be plugging \eqref{thetae1} into \eqref{gauge1} along with the usual ultra-relativistic scaling of spacetime, gauge fields and fermions. Our aim is to find the value of $p,q$ which keep \eqref{gauge1} finite in the limit $\e\to0$. They are given by
\bea{} p=1,q=2,\,\,\Theta^{\a} \to \e \Theta^{\a},\,\,\Theta^I \to \e^2 \Theta^I.\eea
Thus we arrive at the following gauge transformation:
\bea{}\label{gasun}
&&\non \delta A_t^{\a}=\frac{1}{g}\p_t\Theta^{\a},~
\delta A_i^{\a}=\frac{1}{g}\p_i\Theta^{\a}+f^{\a}_{~I\b}A^{I}_{i}\Theta^{\b},\\
&&\non \delta A_t^I=\frac{1}{g}\p_t\Theta^I+f^I_{~\a \b}A_t^\a \Theta^\b,~\delta A_i^I=0,\\
&&\non \delta \phi_{\sss_m}= i \e^{a_{\sss_n}-a_{\sss_m}+1}\Theta^{\a} T^{\a}_{\sss_{mn}}\phi_{\sss_n}+ i \e^{a_{\sss_n}-a_{\sss_m}+2}\Theta^{I} T^{I}_{\sss_{mn}}\phi_{\sss_n},\\
&& \delta \chi_{\sss_m}= i \e^{b_{\sss_n}-b_{\sss_m}+1}\Theta^{\a} T^{\a}_{\sss_{mn}}\chi_{\sss_n}+ i \e^{b_{\sss_n}-b_{\sss_m}+2}\Theta^{I} T^{I}_{\sss_{mn}}\chi_{\sss_n}.
\eea
Choosing a particular representation $(a_m=0,~b_m=1)$ for all $m$'s, we rewrite the last two equations of \eqref{gasun} as
\be{gaugefersun}
\delta \phi_{\sss_m}= 0,~~~
 \delta \chi_{\sss_m}=0.
\ee
We can see that the above equations \eqref{gasun} keep the equations of motion for Carrollian Yang-Mills with fermions  invariant. \\
It is interesting to note here that \eqref{gaugefersun} holds for every sector except for the purely electric sector ($J=0$). The reason lies in the fact that $(a_n,\:b_n)$'s are all same for the sectors with $J\neq0$. For purely electric sector the last two equations in \eqref{gasun} give the gauge transformation for the spinors.

\subsubsection*{Checking symmetries of Carrollian Yang-Mills with Fermions: } 

We are ready to analyse the symmetries of the equations of motion of Carrollian Yang-Mills theory with fermions. We will be choosing an arbitrary value of $\{a_m,b_m\}$ in each of the sectors to find the invariance under conformal Carrollian generators.

 We will start with the sector where the number of magnetic legs $J=0$ and $\{a_m=0,b_m=1\}$ for all $m$. The equations of motion in this sector are
\bes
\bea{}&&
\p_t(\p_t A^{\a}_j-\p_j A_t^{\a})=0,~
 \partial_i(\p_i A_t^\a-\p_t A^{\a}_i)
-g \phi^{\dagger}_{\sss_m} T^\a_{\sss_{mn}}\,\phi_{\sss_n}=0 ,
\\&&
i\p_t \phi_m=0,~
i\p_t \chi_m+i\sigma_i \p_i \phi_m  =0.
\eea 
\ees  
Before finding the invariance, we write the values of the constants in representation theory. We should keep in mind that, here, all the gauge fields scale electrically.
\bea{} 
\i\{\underbrace{\Delta=\frac{3}{2},f=0,f'=-\frac{1}{2}}_{\text{Fermionic Field}},\underbrace{\Delta^\prime=1,a=0,b=1}_
{\text{Gauge Field}}\i\}.
\eea
The invariance under scale transformation are given as
\bes
\bea{}&&\non [D, \partial_i(\p_i A_t^\a-\p_t A^{\a}_i)
-g \phi^{\dagger}_{\sss_m} T^\a_{\sss_{mn}}\,\phi_{\sss_n}]=(\Delta' -1)(\p_i \p_i A^{\a}_{t}-\p_i\p_t A^{\a}_{i})\non\\ 
&&\hspace{6cm} -(2\Delta -3)g\phi^{\dagger}_{m}T^{\a}_{mn}\phi_{n},\\&&[D,\p_t(\p_t A^{\a}_j-\p_j A_t^\a)]=0,~
[D,i\p_t \phi_m]=0,~[D,i\p_t \chi_m+i\sigma_i \p_i \phi_m]= 0.~~~~~~~\eea\ees
Under special conformal transformation, 
\bes
\bea{}&&\non [K_l, \partial_i(\p_i A_t^\a-\p_t A^{\a}_i)
-g \phi^{\dagger}_{\sss_m} T^\a_{\sss_{mn}}\,\phi_{\sss_n}]=(4\Delta' -2\delta_{ii}+2)\p_l A_t^\a \non\\&&\hspace{6.5cm}+(2\delta_{ii}-2\Delta' -4)\p_t A_l^\a \non\\&&\hspace{6.5cm}+(2\Delta'-2)x_l [\p_{i}\p_{i}A_t^\a -\p_{t}\p_{i}A_i^\a]\non\\&&\hspace{6.5cm}-(4\Delta -6)x_l [g \phi^{\dagger}_{\sss_m} T^\a_{\sss_{mn}}\,\phi_{\sss_n}],\\&&[K_l,\p_t(\p_t A^{\a}_j-\p_j A_t^\a)]=-2(\Delta' -1)\delta_{lj}\p_t A^{\a}_{t},~
[K_l,i\p_t \phi_m]=0,\\&&[K_l,i\p_t \chi_m+i\sigma_i \p_i \phi_m]= i\sigma_l (2\Delta -3)\phi_m.\eea\ees
Under $M^{m_1,m_2,m_3}$, we get
\bea{}&&\non [M^{m_1,m_2,m_3}, \partial_i(\p_i A_t^\a-\p_t A^{\a}_i)
-g \phi^{\dagger}_{\sss_m} T^\a_{\sss_{mn}}\,\phi_{\sss_n}]=0\non\\ 
&&[M^{m_1,m_2,m_3},\p_t(\p_t A^{\a}_j-\p_j A_t^\a)]=0,~
[M^{m_1,m_2,m_3},i\p_t \phi_m]=0,\non\\ 
&&[M^{m_1,m_2,m_3},i\p_t \chi_m+i\sigma_i \p_i \phi_m]= 0.~~~~~~~\eea
Hence, all the above mentioned equations in $J=0$ sector, have infinite conformal Carrollian symmetry in $d=4$.

Next, we will examine the sections where $J=1$ and $J>1$. We are choosing $\{a_m=0,b_m=1\}$ for all $m$.
The relevant equations in these two  representative sectors are,
\bes{}
\bea{}
&&\hspace{-1cm}\mathcal{A}=\p_i \p_t A_i^I+gf^{I}_{~JK} A_i^J  \p_t  A^{K}_i=0,~\mathcal{B}= \p_t \p_t A^{I}_j=0,\\ \non\\
&&\hspace{-1cm}\mathcal{C}=\p_t(\p_t A^{\a}_j-\p_j A^\a_t+gf^{\a}_{~\b I}A^{\b}_t A^{I}_j) +gf^{\a}_{~\b I}A^{\b}_t \p_t A^{I}_j=0,\\ \non\\
&&\hspace{-1cm} \mathcal{E}=g f^{\a}_{~IJ}A^I_i \p_t A^{J}_i=0,\\ \non\\
&&\hspace{-1cm}\mathcal{F}= \p_i (\p_i A^{\a}_t - \p_t A^{\a}_i+g f^{\a}_{~I\b} A^{I}_i A^{\b}_t )- g f^{\a}_{~\b I} A^\b_i \p_t A^{I}_i+gf^{\a}_{~IJ}A^{I}_{i}(gf^{J}_{~K\rho}A^{K}_iA^{\rho}_{t})\non\\
&&+gf^{\a}_{~I\b} A_i^I (\p_i A^{\b}_t-\p_t A^{\b}_i+gf^{\b}_{~K\rho}A^{K}_iA^{\rho}_t)-g\phi^{\dagger}_m T^{\a}_{mn}\phi_n=0.\\ \non\\
&&\hspace{-1cm}\mathcal{G}=i\p_t \phi_m=0,~\mathcal {H}=i\p_t \chi_m+i\sigma_{i}\p_i \phi_m+g\sigma_{i}T^{I}_{mn} A^{I}_{i}\phi_n=0.~~~
\eea\ees
 The values of constants in representation theory are given by,
\bea{} 
\i\{\underbrace{\Delta=\frac{3}{2},f=0,f'=-\frac{1}{2}}_{\text{Fermionic Field}},\underbrace{\Delta^\prime=1,a=0,b=1}_
{\text{Gauge Field} ~A^\a},\underbrace{\Delta^\prime=1,a=1,b=0}_
{\text{Gauge Field}~ A^I}\i\}.
\eea
The invariance under scale transformation is 
\bea{}&&\non\hspace{-1.3cm}
[D, \mathcal{A}]=(\Delta'-1)g f^{I}_{~JK} A_i^J  \p_t  A^{K}_i,~[D,\mathcal{B}]=0,
\\\non
&&\hspace{-1.3cm}
[D,\mathcal{C}]=(\Delta' -1)gf^{\a}_{~\b I}[(\p_t A^{\b}_{t})A^{I}_j+2A^{\b}_{t}(\p_t A^{I}_j)],
 \\
&&\hspace{-1.3cm}
[D, \mathcal{E}]=0,~
[D, \mathcal{F}]= 0,~[D,\mathcal{G}] =0,~
[D,\mathcal{H}] =(\Delta'-1)g\sigma_{i}T^{I}_{mn} A^{I}_{i}\phi_n.
\eea 
Under special conformal transformation, we have
\bea{}&&\hspace{-1cm}\non
[K_l,\mathcal A]=(2\Delta' +4 -2\delta_{ii})\p_t A^{I}_{l}+2(\Delta'-1)x_l (gf^{I}_{~JK} A_i^J  \p_t  A^{K}_i),\\&&\hspace{-1cm}\non
[K_l,\mathcal C] = -(2\Delta' -2)[\delta_{lj}\p_t A^{\a}_{t}-x_l f^{\a}_{~\b I}g\{(\p_t A^{\b}_{t})A^{I}_j+2A^{\b}_{t}(\p_t A^{I}_j)\}],
\\&&\hspace{-1cm}\non
[K_l,\mathcal B]=0,~[K_l,\mathcal E]=0,~ 
[K_l,\mathcal G] =0,\\&&\hspace{-1cm}
[K_l,\mathcal H] =(2\Delta -3)i\sigma_l \phi_m +2(\Delta'-1)x_l (g\sigma_i T^{I}_{mn}A^{I}_{i}\phi_n).
\eea 
The invariance of the equations also hold for the other finite conformal Carrollian  generators.
Under infinite supertranslations, we have
\bea{}&&\hspace{-1cm}\non
[M^{m_1,m_2,m_3},\mathcal A]=0,[M^{m_1,m_2,m_3},\mathcal B]=0,~[M^{m_1,m_2,m_3},\mathcal C] = 0,
\\&&\hspace{-1cm}
[M^{m_1,m_2,m_3},\mathcal E]=0,~[M^{m_1,m_2,m_3},\mathcal F] =0,~ 
[M^{m_1,m_2,m_3},\mathcal G] =0,\\&&\hspace{-1cm}\non
[M^{m_1,m_2,m_3},\mathcal H] =i \Big({\sigma_i\over 2}\Big)\p_i(x^{m_1}y^{m_2}z^{m_3})\p_t \phi_m=0.
\eea 
The above analysis shows that both the $J\geq1$ sectors have finite as well as infinite conformal Carrollian symmetries in $d=4$.

\bigskip

\subsubsection*{Brief summary of the current section} Let us summarise the results that we have obtained in this section. We started out with a brief review on $SU(N)$ Carrollian Yang-Mills theory. We explained the scaling of gauge fields $(A^a_t,A^a_i)$ in detail and wrote down the equations of motion. These equations were invariant under conformal Carrollian generators including the infinite supertranslations in $d=4$. For further details about the invariance of these equations, the reader is referred to \cite{Bagchi:2016bcd}. 
One of the upshots of the analysis was that the number of ``nice" sectors with EOM having kinetic pieces, boiled down to just 3, viz. the purely electric and magnetic sectors, and one mixed sector with one magnetic leg. 

Then, we moved on to $SU(N)$ Carrollian Yang-Mills coupled with fermions. The scaling of gauge fields gave rise to $(\mathfrak D+1)$ sectors, of which only 3 were ``nice", as just mentioned.  For the sake of completeness, we did not discard the other sectors outright. We arranged the $(\mathfrak D+1)$ sectors depending on the number of magnetic legs, denoted by $J$. For $J=1$ and $J>1$, the equations $(\mathcal{A,B,C,G,H})$ (see eq.\eqref{mtA} - \eqref{mtH}) were held perfectly. For $J>1$, equation $\mathcal{E}$ \eqref{zeroeq} was true. This was the problematic equation in the pure YM case, which contained only interaction pieces. So in principle, all these sectors with $J>1$, except the purely magnetic sector, should be discarded on grounds of having a non-trivial realisation of this equation.  For $J=1$, we observed that the equation $\mathcal{E}$ did not hold and we had to consider the next order in $\epsilon$,  captured by equation $\mathcal{F}$ \eqref{fff}. Each of the $J=1$ and $J>1$ sectors  had two different subsectors  within them (see eq. \eqref{jg1} - \eqref{je2}).

We also considered $J=0$ case,  the purely electrical sector. The absence of the constraint $a_n=a_m$ for all $m,n$'s into the scaling of fermions (that appeared in the previous cases, see eq. \eqref{sun2}) resulted in a more diverse class of subsectors (see Table[\ref{sunbigtable}]).
   
We looked at the gauge transformations and the invariance of equations of motion in each subsector. We took a particular representation for each of the subsectors in  $J=0,\geq1$ cases. We investigated the symmetries of the Carrollian YM theory coupled to fermions at the level of equations of motion. Again, like in the previous sections, we found the emergence of finite as well as infinite conformal Carrollian symmetries $d=4$ in all of these subsectors. This emergence of infinite enhancement of conformal Carrollian symmetries in $d=4$, thus, seems to be a generic feature of all field theories that arise as a ultra-relativistic limit of classically conformally invariant theories in $d=4$.

\clearpage

\section{Conclusions and Future Directions}
In this paper, we have systematically constructed conformal Carrollian field theories in $d=4$ by following the ultra-relativistic limit of relativistic CFTs in $d=4$. We have started with free scalars, fermions and reviewed previous constructions of Carrollian gauge theories. Then we have constructed arbitrary interacting theories by adding matter to general Carrollian gauge theories. The intricacies of the limit resulted in a multitude of different sectors of these gauge field theories. We showed that in all these sectors the equations of motion possess an infinite dimensional symmetry. We again stress here that this enhancement of symmetries in conformal field theories in the ultra-relativistic limit seems to be a generic feature of this limit. We also saw that algebraically the process of infinite extension for $d>4$ was very similar to $d=4$. Taking this analogy further, we propose that if one starts with any relativistic conformal field theory in any dimension and constructs its Carrollian version by taking $c\to 0$ limit, the equations of motion of the ultra-relativistic theory so constructed would be invariant under the infinite conformal Carrollian  algebra. 

\medskip

Our analysis in this paper, as stated earlier, has been a direct follow-up to earlier work \cite{Bagchi:2016bcd}. The underlying mathematical formulation of the investigation of symmetries from the point of view of equations of motion actually is very similar in nature to the investigations of non-relativistic systems and the emergence of infinite Galilean conformal symmetry in theories derived as a limit of relativistic conformal field theories \cite{Bagchi:2017yvj,Bagchi:2015qcw,Bagchi:2014ysa}. Given the similarity of these analyses, it is important to compare and contrast the two different singular limits of gauge theories. In \cite{Bagchi:2017yvj}, the non-relativistic counterpart of the field theories studied in the current paper were constructed, namely Galilean scalar, fermions, Yukawa theory and interacting theories like electrodynamics and $SU(2)$ Yang-Mills coupled to matter.  This analysis has been extended for the case of $SU(N)$ Yang-Mills coupled to matter in appendix \ref{Galilean $SU(N)$ Yang-Mills Theory with matter}, which also contains a brief introduction to aspects of the non-relativistic symmetries for the uninitiated reader. The important question of the connection these two opposite limits (Galilean and Carrollian) in terms of their similarities and differences is discussed in detail in Appendix \ref{Comparison between Galilean and conformal Carrollian  field theories}.

\medskip

The infinite enhancement of symmetries in the ultra-relativistic limit, that has been the central theme of our present paper, seems intimately related to the geometry of the spacetime that these field theories live on. If we start with relativistic CFTs that live in flat spacetimes, the Carrollian limit leads to a degeneration of the metric structure of the Minkowskian manifold:
\be{}
ds^2 \to - \e^2 dt^2 + dx_i^2; \quad {\rm I\!R}^{1,d} \to {\rm I\!R}^d\times {\rm I\!R}_t  
\ee
The relativistic conformal structure is more restrictive as it involves space and time on the same footing. 
In the ultra-relativistic case, the (pseudo) Riemannian structure (${\rm I\!R}^{1,d}$) turns into a fibre bundle structure with a ${\rm I\!R}^d$ base and temporal one-dimensional ${\rm I\!R}_t$ fibres. Now there is more ``wiggle-room" and the conformal structure gets split into that of the base and the fibre. This degeneration of the spacetime metric into a purely spatial part (for the base) and a purely temporal part (for the fibre) seems to be at the heart of these infinite dimensional enhancements in the ultra-relativistic limit. This also should be true in the non-relativistic limit, where similar infinite enhancements are noticed. We wish to investigate these geometrical aspects involving Carrollian and its dual Newton-Cartan structures more carefully in the near future. 

\medskip

As we stated in the introduction, the ultimate aim of this programme is to make connections with holography in asymptotically flat spacetimes. As a specific case, we would like to understand what happens to the dual field theory when we take a flat limit on the best known example of the AdS/CFT duality, viz. the duality between type IIB superstring theory on AdS$_5 \times$S$^5$ and $\mathcal{N}=4$ Supersymmetric Yang-Mills (SYM) theory in $d=4$. In the bulk, the radii of the AdS$_5$ and S$^5$ are equal and hence taking a limit on one induces a limit on the other, thereby giving us type IIB superstring theory on a 10d flat spacetime. On the boundary, the limit would lead to a Carrollian version of $\mathcal{N}=4$ SYM. In future work, we would attempt an understanding of SYM theories with various supersymmetry building up to this example. The analysis done in this paper would provide the basic ingredients for this investigation. It seems very unclear at present what these various sectors arising in the Carrollian limit would correspond to in the holographic setting. The infinite dimensional emergent symmetry, on the other hand, has tantalising hints of a new integrable sector in the parent gauge theories and relatedly holography. 

\medskip

One of the stumbling blocks of our analysis in this paper, and our programme as a whole, has been the lack of an action principal. For the case of Carrollian electrodynamics, this has been partially addressed and one has an action for the Electric sector \cite{Basu:2018dub}, and the same symmetries have been understood by a rigorous canonical analysis {\footnote{For another take on action principles and Carrollian theories, the reader is pointed to \cite{Bergshoeff:2017btm}.}}. In ongoing work \cite{BBN}, it has been possible to show that one can have a similar action principal for scalars coupled to Carrollian electrodynamics for a particular sector. Generalisation of this to the other possible sectors and then to the case of Yang-Mills theories and SYM is the obvious goal of this line of work. With an action, quantities of interest like correlation functions can be derived by functional methods and this would also pave the way to a possible quantisation of the Carrollian theories and investigations into anomalies to understand whether the infinite dimensional symmetries survive in the quantum regime.

\bigskip

\section*{Acknowledgements}
Discussions with Rudranil Basu, Joydeep Chakrabortty, Shankhadeep Chakrabortty, Daniel Grumiller are gratefully acknowledged. AB also acknowledges the warm hospitality of the Department of Mathematics and Grey College at Durham University and the financial support of the Institute of Advanced Study, Durham through a Senior Research Fellowship (COFUNDed between Durham University and the European Union under grant agreement number 609412), during the final stages of this work. 

This work is also partially supported by the following grants: DST-INSPIRE faculty award, SERB Early Career Research Award (ECR/2017/000873), DST-Max Planck mobility award, DST-BMWF India-Austria bilateral grant, Royal Society International Exchange grant, SERB extra-mural grant (EMR/2016/008037). 
 
\bigskip 
\bigskip

\appendix
\section*{Appendices}

\section{Carrollian $SU(2)$ Yang-Mills with Fermions}
\label{Carrollian $SU(2)$ Yang-Mills with Fermions}
In the main body of the paper, we have dealt with a general $SU(N)$ Carrollian Yang-Mills theory coupled to fermionic matter. It may be of interested for the reader to understand the elementary details and hence in this appendix, we present the complete details of the $SU(2)$ Carrollian theory with fermions. 

\medskip

\noindent In $SU(2)$ Yang-Mills with fermions, we have three colour indices for the gauge fields and a doublet of fermions  in the fundamental representation. The relativistic equations of motion are given by \eqref{gsymeomc}. In the Carrollian limit, we scale the fermions as,
\begin{eqnarray}\label{su2fs1}
\phi_{\sss_1}\to\epsilon^{\alpha_1}\phi_{\sss_1}, \, \, \, \chi_{\sss_1}\to\epsilon^{\beta_1}\chi_{\sss_1},~~
\phi_{\sss_2}\to\epsilon^{\alpha_2}\phi_{\sss_2}, \, \, \, \chi_{\sss_2}\to\epsilon^{\beta_2}\chi_{\sss_2}.
\end{eqnarray}
In \cite{Bagchi:2016bcd}, the detailed analysis of Carrollian $SU(2)$ Yang-Mills is carried out. There are four different sectors in case of $SU(2)$ Yang Mills: EEE (purely Electric), MMM (purely Magnetic), EEM and EMM (2 mixed sectors). Following the same trail, in case of $SU(2)$ Yang-Mills coupled to fermions, we scale the gauge fields in the similar way. Let us discuss the different sectors in details below.

\subsection{EEE sector}
\begin{figure}[t]
\centering
\includegraphics[width=8cm]{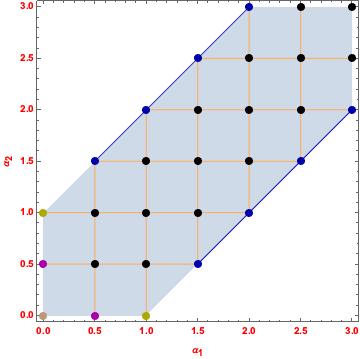}
\caption{Allowed values of $(\a_1,\a_2)$ in EEE sector.}
\floatfoot{\small {{\bf{Key:}} Points in black: free Carrollian theories (Case 2), Coloured points: non-trivial interacting sectors.}}
\label{EEEplot}
\end{figure}

In this sector, all  gauge fields are scaled electrically. The scaling is given as
\begin{eqnarray}
A_i^a\to \epsilon A_i^a,\,\, A_t^a\to A_t^a,\phi_{\sss_m}\to\epsilon^{\alpha_m}\phi_{\sss_m}, \, \chi_{\sss_m}\to\epsilon^{\beta_m}\chi_{\sss_m}
\end{eqnarray}
where $m=1,2$. The fermions are scaled as in \eqref{su2fs1}. The equations of motion in this sector are given as
\bes
\bea{}&&\hspace{-1cm}
\p_t(\p_t A^{a}_j-\p_j A_t^a)-g[\epsilon^{\alpha_m+\beta_n+1} \phi^{\dagger}_{\sss_m}\sigma_j T^{a}_{\sss_{mn}}\chi_{\sss_n}
+\epsilon^{\b_m+\a_n+1}\chi^{\dagger}_{\sss_m}\sigma_j T^a_{\sss_{mn}}\phi_{\sss_n}]=0,
 \\&&\hspace{-1cm}
 \partial_i(\p_i A_t^a-\p_t A^{a}_i)
-g[\epsilon^{\alpha_m+\alpha_n} \phi^{\dagger}_{\sss_m} T^a_{\sss_{mn}}\,\phi_{\sss_n}
+\e^{\beta_{m}+\beta_n} \chi^{\dagger}_{\sss_m}\,T^a_{\sss{mn}}\,\chi_{\sss_n}]=0 ,
\\&&\hspace{-1cm}
i\p_t \phi_m + i\e^{\b_m -\a_m +1}\sigma_i \p_i \chi_m+ \e^{\a_n-\a_m+1} gT^{a}_{mn}A^{a}_{t}\phi_n + \e^{\b_n -\a_m +2} g\sigma_i T^{a}_{mn}A^{a}_{i} \chi_n =0,
 \\&&\hspace{-1cm}
i\p_t \chi_m + i\e^{\a_m -\b_m +1}\sigma_i \p_i \phi_m+ \e^{\b_n-\b_m+1} gT^{a}_{mn}A^{a}_{t}\chi_n + \e^{\a_n -\b_m +2} g\sigma_i T^{a}_{mn}A^{a}_{i} \phi_n =0.
\eea 
\ees
The above equations must reduce to free equations of a particular field in absence of rest of the fields. Also, if we replace $T^a$ by an identity matrix and set the structure constant to be zero, we should  get back the equations of $U(1)$ Carrollian electrodynamics coupled to fermions. For example, suppose we are looking at the limit EMM of the $SU(2)$ gauge fields coupled to fermions. In this sector $A^1$ scales electrically and $A^2, A^3$ scales magnetically. So we demand that if we replace $T_a$ by an identity matrix, the $A^1$ equation must reduce to \eqref{c1efed} or \eqref{c2efed} while the rest of the gauge fields must follow \eqref{c1mfed} or \eqref{c2mfed}. The same goes for the fermions as well. If we turn off the gauge fields in the equations obtained in EMM case, we must get back the free fermion equations. If we keep the gauge field, replace $T_a$ by an identity matrix, we should get the fermion equation from either \eqref{c1efed},\eqref{c2efed}  or \eqref{c1mfed}, \eqref{c2mfed}(depending on how the particular gauge field coupled to the fermion scales).  The above conditions impose constraints on the constants $\{\a_m,\b_m\}$. Please see Table[\ref{EEEtable}].
\begin{table}[t]
\centering
    \begin{tabular}{ | l | p{5cm} |  p{5cm} |}
   \hline
\multicolumn{3}{ |c| }{EEE sector} \\
\hline
Free equation  & $U(1)$+Fermions & Constraints\\ [0.2cm]
\hline
$\p_t(\p_t A_i^a-\partial_i A_t^a)=0$ & $\p_t(\p_t A_i-\partial_i A_t)=0$  &(i) $\a_n+\b_m+1>0$\\[0.2cm]
   \hline
\multirow{2}{*}{$\partial_i(\partial_i A_t^a-\p_t A_i^a)=0$} & $\partial_i(\partial_i A_t-\p_t A_i)- e\phi^{\dagger}\phi=0$,\: &(ii) $\alpha_n+\a_m\geq0$\\
   & $\partial_i(\partial_i A_t-\p_t A_i)=0$  &  (iii) $\b_n+\b_m>0$\\[0.2cm]
 \hline
   \multirow{3}{*}{$i\p_t \phi=0$} & \multirow{3}{*}{$i \p_t \phi=0$} & (iv) $\b_m-\a_m+1>0$,\\
   & & (v) $\b_n-\a_m+2\geq0$,$(n \neq m)$, \\
   & & (vi) $\a_n-\a_m+1\geq0$\\[0.2cm]
 \hline
   \multirow{3}{*}{$i\p_t \chi+i\sigma_i \partial_i \phi=0$} & \multirow{3}{*}{$i\p_t \chi+i\sigma_i \partial_i \phi=0$} &  (vii) $\a_m-\b_m+1=0$,\\
   & &(viii) $\b_n-\b_m+1\geq0$,\\
   & &(ix) $\a_n-\b_m+2>0$\\[0.2cm]
   \hline
\end{tabular}\caption{Constraints on $\a_m,\b_m$ in EEE sector.}
\label{EEEtable}
\end{table}
There is no equality in some of the constraints in the Table[\ref{EEEtable}] to reduce it to Carrollian electrodynamics with fermions case.
Constraint (ii),(vi) and (vii) gives
\bes \label{coneee}
\begin{eqnarray}
\a_1\geq0,\,\, \a_2\geq0,\\
-1\leq \a_2-\a_1\leq 1,\\
\a_m=\b_m-1.
\end{eqnarray} 
\ees
The rest of the constraints give no new information. In Mathematica, we can do a Regionplot[{\ref{EEEplot}}] to plot \eqref{coneee} and find the possible values of $\a_1,\a_2$.
Each intersecting point in Regionplot[{\ref{EEEplot}}]  corresponds to a sector in the EEE limit. Some of the possible values are:
\begin{equation}
(\a_1,\a_2)=(0,0),(0,\frac{1}{2}),(0,1),(\frac{1}{2},1),(\frac{3}{2},\frac{1}{2}),(\frac{1}{2},\frac{1}{2}),(\frac{1}{2},0)\dots
\end{equation}
We will now discuss different sectors within EEE limit. All the intersecting points in  Plot[\ref{EEEplot}] do not give unique sectors. We are enlisting below only the different sectors.
\begin{itemize}
\item Case 1: ($\a_1=0,\a_2=0,\b_1=1,\b_2=1$):
\bes
\bea{}&&
\p_t(\p_t A^{a}_j-\p_j A_t^a)=0,~
 \partial_i(\p_i A_t^a-\p_t A^{a}_i)
-g \phi^{\dagger}_{\sss_m} T^a_{\sss_{mn}}\,\phi_{\sss_n}=0 ,
\\&&
i\p_t \phi_m=0,~
i\p_t \chi_m+i\sigma_i \p_i \phi_m  =0.
\eea 
\ees
\item Case 2: ($\a_1=\frac{1}{2},\a_2=\frac{1}{2},\b_1=\frac{3}{2},\b_2=\frac{3}{2}$): (Reproduces free sector)
\bes
\bea{}&&
\p_t(\p_t A^{a}_j-\p_j A_t^a)=0,~
 \partial_i(\p_i A_t^a-\p_t A^{a}_i)=0 ,
\\&&
i\p_t \phi_m =0,~
i\p_t \chi_m +i\sigma_i \p_i \phi_m =0.
\eea 
\ees
\item Case 3: ($\a_1=0,\a_2=\frac{1}{2},\b_1=1,\b_2=\frac{3}{2}$):
\bes
\bea{}&&\hspace{-1cm}
\p_t(\p_t A^{a}_j-\p_j A_t^a)=0,~
 \partial_i(\p_i A_t^a-\p_t A^{a}_i)-g \phi^{\dagger}_{\sss_1} T^a_{\sss_{11}}\,\phi_{\sss_1}=0 ,
\\&&\hspace{-1cm}
i\p_t \phi_m =0,~
i\p_t \chi_m +i\sigma_i \p_i \phi_m =0.
\eea 
\ees

\item Case 4: ($\a_1=0,\a_2=1,\b_1=1,\b_2=2$):
\bes
\bea{}&&\hspace{-1cm}
\p_t(\p_t A^{a}_j-\p_j A_t^a)=0,~
 \partial_i(\p_i A_t^a-\p_t A^{a}_i)-g \phi^{\dagger}_{\sss_1} T^a_{\sss_{11}}\,\phi_{\sss_1}=0 ,
\\&&\hspace{-1cm}
i\p_t \phi_1 =0,~i\p_t \phi_2 +gT^{a}_{21}A^{a}_{t}\phi_1=0,~
i\p_t \chi_1 +i\sigma_i \p_i \phi_1 =0,\\&&\hspace{-1cm}
i\p_t\chi_2 +gT^{a}_{21}A^{a}_{t}\chi_1 +i\sigma_{i}\p_i \phi_2 +g\sigma_{i}T^{a}_{21}A^{a}_{i}\phi_1=0.
\eea 
\ees
\item Case 5: ($\a_1=\frac{1}{2},\a_2=\frac{3}{2},\b_1=\frac{3}{2},\b_2=\frac{5}{2}$):
\bes
\bea{}&&\hspace{-1cm}
\p_t(\p_t A^{a}_j-\p_j A_t^a)=0,~
 \partial_i(\p_i A_t^a-\p_t A^{a}_i)=0 ,
\\&&\hspace{-1cm}
i\p_t \phi_1 =0,~i\p_t \phi_2 +gT^{a}_{21}A^{a}_{t}\phi_1=0,~
i\p_t \chi_1 +i\sigma_i \p_i \phi_1 =0,\\&&\hspace{-1cm}
i\p_t\chi_2 +gT^{a}_{21}A^{a}_{t}\chi_1 +i\sigma_{i}\p_i \phi_2 +g\sigma_{i}T^{a}_{21}A^{a}_{i}\phi_1=0.
\eea 
\ees
\end{itemize}
We do not need to consider any other values of $\a_1,\a_2$ as we  get no new sector out of it. For the higher $\a_{\sss_m}$'s the interaction part either drops off or we get back the same equations from the above mentioned cases.  In Plot[\ref{EEEplot}], the dots in black represent the sectors where there is no interaction term between fermions and gauge fields. The coloured dots represent the distinct non trivial sections. The different values of $\a_1,\a_2$ returning same sectors are connected through lines in the same figure.  So for the EEE sector we get five different sectors (including the free sectors). 
\subsubsection*{Gauge Transformation for EEE sector} 
Let us consider relativistic gauge transformations of  gauge  and fermions fields as
\bea{}\label{gauge} \delta A_{\mu}^{a}=\frac{1}{g}\p_{\mu} \Theta^{a} +f^{a}_{~bc}A^{b}_{\mu}\Theta^{c},~\delta \phi_{m}=i\Theta^{a}T^{a}_{mn}\phi_{n},~\delta \chi_{m}=i\Theta^{a}T^{a}_{mn}\chi_{n}. \eea
We will now see gauge transformations in this sector. We scale the gauge transformation parameter for  EEE sector by
\begin{equation}\label{thetae}
\Theta^a \to \e^p \Theta^a.
\end{equation}
We would be plugging \eqref{thetae} into \eqref{gauge} along with the usual ultra-relativistic scaling of spacetime, gauge fields and fermions. Our aim is to find the value of $p$ which keeps \eqref{gauge} finite in the limit $\e\to0$. It is given by 
\bea{} p  =1,~~\Theta^a \to \e \Theta^a.\eea
 Thus we arrive at the following gauge transformation for EEE sector of Carrollian $SU(2)$ Yang Mills coupled to fermions:
\bea{} \label{gaeee}
&&\non \delta A_t^a =\frac{1}{g}\p_t\Theta^a,~~\delta A_i^a=\frac{1}{g}\p_i\Theta^a,\\&&
\delta \phi_{m}=  \e^{\a_{\sss_n}-\a_{\sss_m}+1}i\Theta^a T^a_{\sss_{mn}}\phi_n,~~\delta \chi_{m}= \e^{\b_{\sss_n}-\b_{\sss_m}+1}i\Theta^a T^a_{\sss_{mn}}\chi_n.
\eea
To see the invariance under these transformation, we have to take one set of values of the coefficients $(\a_m,\b_m)$ along with $\epsilon \rightarrow 0$ limit and plug them into \eqref{gaeee}.

\subsubsection*{Checking symmetries of EEE sector } 
Next, we will find the invariance of the equations of motion. For that we will choose a representative sector, that is $\a_1=0,\a_2=0,\b_1=1,\b_2=1$. The equations are
\bes
\bea{}&&
\p_t(\p_t A^{a}_j-\p_j A_t^a)=0,~
 \partial_i(\p_i A_t^a-\p_t A^{a}_i)
-g \phi^{\dagger}_{\sss_m} T^a_{\sss_{mn}}\,\phi_{\sss_n}=0 ,
\\&&
i\p_t \phi_m=0,~
i\p_t \chi_m+i\sigma_i \p_i \phi_m  =0.
\eea 
\ees
Before checking invariance under Carrollian symmetry, we remind the readers, the values of constants in  representation thoery:
\bea{} 
\i\{\underbrace{\Delta=\frac{3}{2},f=0,f'=-\frac{1}{2}}_{\text{Fermionic Field}},\underbrace{\Delta^\prime=1,a=0,b=1}_
{\text{Gauge Field}}\i\}.
\eea
Under scale transformation,
\bea{}&&\non [D, \partial_i(\p_i A_t^a-\p_t A^{a}_i)
-g \phi^{\dagger}_{\sss_m} T^a_{\sss_{mn}}\,\phi_{\sss_n}]=(\Delta' -1)(\p_i \p_i A^{a}_{t}-\p_i\p_t A^{a}_{i})-(2\Delta -3)g\phi^{\dagger}_{m}T^{a}_{mn}\phi_{n},\\&&[D,\p_t(\p_t A^{a}_j-\p_j A_t^a)]=0,~
[D,i\p_t \phi_m]=0,~[D,i\p_t \chi_m+i\sigma_i \p_i \phi_m]= 0.\eea
Under special conformal transformation, 
\bes
\bea{}&&\non [K_l, \partial_i(\p_i A_t^a-\p_t A^{a}_i)
-g \phi^{\dagger}_{\sss_m} T^a_{\sss_{mn}}\,\phi_{\sss_n}]=(4\Delta' -2\delta_{ii}+2)\p_l A_t^a +(2\delta_{ii}-2\Delta' -4)\p_t A_l^a \non\\&&\hspace{6.5cm}+(2\Delta'-2)x_l [\p_{i}\p_{i}A_t^a -\p_{t}\p_{i}A_i^a]\non\\&&\hspace{6.5cm}-(4\Delta -6)x_l [g \phi^{\dagger}_{\sss_m} T^a_{\sss_{mn}}\,\phi_{\sss_n}],\\&&[K_l,\p_t(\p_t A^{a}_j-\p_j A_t^a)]=-2(\Delta' -1)\delta_{lj}\p_t A^{a}_{t},~
[K_l,i\p_t \phi_m]=0,\\&&[K_l,i\p_t \chi_m+i\sigma_i \p_i \phi_m]= i\sigma_l (2\Delta -3)\phi_m.\eea\ees
Under infinite supertranslations,
\bea{}&&\non [M^{m_1,m_2,m_3}, \partial_i(\p_i A_t^a-\p_t A^{a}_i)
-g \phi^{\dagger}_{\sss_m} T^\a_{\sss_{mn}}\,\phi_{\sss_n}]= -\p_i(x^{m_1}y^{m_2}z^{m_3})\p_t(\p_t A^{a}_j-\p_j A_t^a)=0\non\\ 
&&[M^{m_1,m_2,m_3},\p_t(\p_t A^{a}_j-\p_j A_t^a)]=0,~
[M^{m_1,m_2,m_3},i\p_t \phi_m]=0,\non\\ 
&&[M^{m_1,m_2,m_3},i\p_t \chi_m+i\sigma_i \p_i \phi_m]= 0.~~~~~~~\eea
The equations are invariant under infinite conformal Carrollian symmetries in $d=4$.

\subsection{EEM sector} 
\begin{figure}[t]
\centering
\includegraphics[width=8cm]{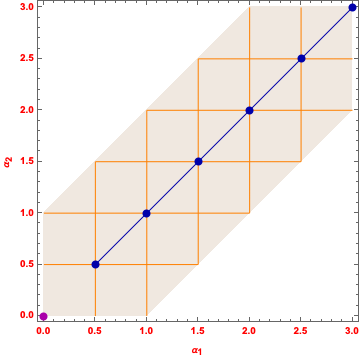}
\caption{Allowed region for EEM sector.}
\floatfoot{\small {{\bf{Key:}}  Point in magenta: Case 1, Points in blue (same sectors are connected by a line): Case 2. Shaded region describe equations \eqref{coneem}.}}
\label{EEMplot}
\end{figure}

We would be looking at the two mixed sectors. First, we are discussing EEM, where two of the gauge fields scale electrically and the other one scales magnetically. The scaling in this limit is given by
\bea{}\hspace{-0.5cm}
A_t^{1,2}\to  A_t^{1,2},~ A_i^{1,2}\to \epsilon A_i^{1,2},~
A_t^{3}\to \epsilon A_t^3,\,\, A_i^3\to A_i^3,~
\phi_{\sss_m} \to \e^{\a_m}\phi_{\sss_m},\,\,\chi_{\sss_m} \to \e^{\b_m}\chi_{\sss_m}.
\eea
The equations of motion are:
\bes
\bea{}
&&\hspace{-2cm}
\bullet\, \text{For gauge field } A^{1,2}: \non\\ &&
\hspace{-1cm}\p_t(\p_t A^{1}_j-\p_j A_t^1 +gA^{2}_{t}A^{3}_j)+gA^{2}_{t}\p_t A^{3}_j\non\\
&&\hspace{.35cm}-g[\epsilon^{\alpha_m+\beta_n+1} \phi^{\dagger}_{\sss_m}\sigma_j T^{1}_{\sss_{mn}}\chi_{\sss_n}
+\epsilon^{\b_m+\a_n+1}\chi^{\dagger}_{\sss_m}\sigma_j T^1_{\sss_{mn}}\phi_{\sss_n}]=0,\label{hgo3}
 \\&&\hspace{-1cm}
\p_t(\p_t A^{2}_j-\p_j A_t^2 -gA^{1}_{t}A^{3}_j)-gA^{1}_{t}\p_t A^{3}_j\non\\&&\hspace{.35cm}-g[\epsilon^{\alpha_m+\beta_n+1} \phi^{\dagger}_{\sss_m}\sigma_j T^{2}_{\sss_{mn}}\chi_{\sss_n}
+\epsilon^{\b_m+\a_n+1}\chi^{\dagger}_{\sss_m}\sigma_j T^2_{\sss_{mn}}\phi_{\sss_n}]=0,\label{hgo4}
 \\
 &&\hspace{-1cm}
 \partial_i(\p_i A_t^1-\p_t A^{1}_i-gA^{3}_{i}A^{2}_{t})-gA^{2}_{i}\p_t A^{3}_i-gA^{3}_i (\p_i A^{2}_{t} -\p_t A^{2}_i+gA^{3}_iA^{1}_{t})
\non\\&&\hspace{3.35cm}-g[\epsilon^{\alpha_m+\alpha_n} \phi^{\dagger}_{\sss_m} T^1_{\sss_{mn}}\,\phi_{\sss_n}
+\e^{\beta_{m}+\beta_n} \chi^{\dagger}_{\sss_m}\,T^1_{\sss{mn}}\,\chi_{\sss_n}]=0,\label{hgo1}
\\&&\hspace{-1cm}
 \partial_i(\p_i A_t^2-\p_t A^{2}_i+gA^{3}_{i}A^{1}_{t})+gA^{1}_{i}\p_t A^{3}_i+gA^{3}_i (\p_i A^{1}_{t} -\p_t A^{1}_i-gA^{3}_iA^{2}_{t})
\non\\&&\hspace{3.35cm}-g[\epsilon^{\alpha_m+\alpha_n} \phi^{\dagger}_{\sss_m} T^2_{\sss_{mn}}\,\phi_{\sss_n}
+\e^{\beta_{m}+\beta_n} \chi^{\dagger}_{\sss_m}\,T^2_{\sss{mn}}\,\chi_{\sss_n}]=0.\label{hgo2}
\eea
\ees
\bes
\bea{}
&&\hspace{-2cm}
\bullet\, \text{For gauge field } A^{3}: \non\\
&&
\p_t\p_t A^{3}_j-g[\epsilon^{\alpha_m+\beta_n+2} \phi^{\dagger}_{\sss_m}\sigma_j T^{3}_{\sss_{mn}}\chi_{\sss_n}
+\epsilon^{\b_m+\a_n+2}\chi^{\dagger}_{\sss_m}\sigma_j T^3_{\sss_{mn}}\phi_{\sss_n}]=0,
 \\&& 
 \partial_i \p_t A^{3}_i
+g[\epsilon^{\alpha_m+\alpha_n +1} \phi^{\dagger}_{\sss_m} T^3_{\sss_{mn}}\,\phi_{\sss_n}
+\e^{\beta_{m}+\beta_n +1} \chi^{\dagger}_{\sss_m}\,T^3_{\sss{mn}}\,\chi_{\sss_n}]=0.
\eea\ees
\bea{}
&&\hspace{-1cm}
\bullet\, \text{For spinors } \phi_{\sss m}, \chi_{\sss m}: \non\\
&&
i\p_t \phi_m+\e^{\b_m-\a_m+1}i\sigma_i \p_i \chi_m+\e^{\a_n-\a_m+1} gT^{1}_{mn}A^{1}_{t}\phi_n +\e^{\a_n-\a_m+1} gT^{2}_{mn}A^{2}_{t}\phi_n\non\\
&&+\e^{\a_n-\a_m+2} gT^{3}_{mn}A^{3}_{t}\phi_n+\e^{\b_n -\a_m +2} g\sigma_i T^{1}_{mn}A^{1}_{i} \chi_n\non\\&&+\e^{\b_n -\a_m+2 } g\sigma_i T^{2}_{mn}A^{2}_{i} \chi_n+\e^{\b_n -\a_m+1} g\sigma_i T^{3}_{mn}A^{3}_{i} \chi_n =0,
\eea
\bea{}
 &&
i\p_t \chi_m+\e^{\a_m-\b_m+1}i\sigma_i \p_i \phi_m+\e^{\b_n-\b_m+1} gT^{1}_{mn}A^{1}_{t}\chi_n +\e^{\b_n-\b_m+1} gT^{2}_{mn}A^{2}_{t}\chi_n\non\\
&&+\e^{\b_n-\b_m+2} gT^{3}_{mn}A^{3}_{t}\chi_n+\e^{\a_n -\b_m +2} g\sigma_i T^{1}_{mn}A^{1}_{i} \phi_n\non\\
&&+\e^{\a_n -\b_m+2 } g\sigma_i T^{2}_{mn}A^{2}_{i} \phi_n+\e^{\a_n -\b_m+1} g\sigma_i T^{3}_{mn}A^{3}_{i} \phi_n =0.
\eea
Following the same reasoning mentioned above we can find the constraints on $(\a_m,\b_m)$ in Table[\ref{EEMtable}]. \\
\begin{table}[t]
\centering
    \begin{tabular}{| l |  p{5cm} |}
    \hline
\multicolumn{2}{ |c| }{EEM sector} \\
\hline
  From equation & Constraints\\ [0.2cm]
\hline
\multirow{2}{*}{Equations (\ref{hgo1}-\ref{hgo2}) of Gauge field $A^{1,2}$}  &(i) $\a_n+\a_m\geq0$\\[0.2cm]
 & (ii )$\b_n+\b_m>0$\\
   \hline
   Equations (\ref{hgo3}-\ref{hgo4}) of Gauge field $A^{1,2}$ &(iii) $\a_n+\b_m+1>0$\\[0.2cm]
\hline
\multirow{3}{*}{Second equation of Gauge field  $A^{3}$}  &(iv) $\a_n+\a_m+1\geq0$\\[0.2cm]
 & (v)$\b_n+\b_m+1>0$\\
 \hline
First equation of Gauge field  $A^{3}$  &  (vi)$\a_n+\b_m+2>0$\\[0.2cm]
\hline
\multirow{3}{*}{Equation for $\phi_{\sss_m}$} & (vii) $\b_m-\a_m+1>0$,\\[0.2cm]
    & (viii) $\a_n-\a_m+1\geq0$ \\
    & (ix) $\b_n-\a_m+1>0$\\
   \hline
   \multirow{3}{*}{Equation for $\chi_{\sss_m}$}&  (x) $\a_m-\b_m+1=0$\\
   & (xi) $\b_n-\b_m+1\geq0$\\
   & (xii)$\a_n-\b_m+1\geq0$\\
 \hline
\end{tabular}
\caption{Constraints on $\a_n,\b_n$ in EEM sector.}
\label{EEMtable}
\end{table}
\\
From constraints (i),(viii) and (x) we find:
\bes \label{coneem}
\begin{eqnarray}
\a_1 \geq 0, \, \a_2 \geq 0,\\
-1\leq \a_2 -\a_1 \leq 1,\\
\a_m=\b_m-1.
\end{eqnarray}
\ees
We can plot  the region described by \eqref{coneem} in Fig[\ref{EEMplot}].
However from constraint (xii),
\begin{eqnarray} \label{con2eem} 
 \a_n-\b_m+1\geq0  \implies \a_1=\a_2.
\end{eqnarray}
The rest of the constraints trivially satisfy these four major constraints.
 Hence imposing \eqref{con2eem}, we get only a subset of the intersecting points from the plot above. The allowed values  are $$(\a_1,\a_2)=(0,0),\:(\frac{1}{2},\frac{1}{2}),\:(1,1)\dots$$ From the allowed values of $(\a_1,\a_2)$ we can construct only two different sectors.\\
\begin{itemize}
\item Case 1: ( $\a_1=0,\a_2=0,\b_1=1,\b_2=1$):
\bes
\bea{}&&\hspace{-1cm}
\p_t(\p_t A^{1}_j-\p_j A_t^1 +gA^{2}_{t}A^{3}_j)+gA^{2}_{t}\p_t A^{3}_j=0,
 \\&&\hspace{-1cm}
\p_t(\p_t A^{2}_j-\p_j A_t^2 -gA^{1}_{t}A^{3}_j)-gA^{1}_{t}\p_t A^{3}_j=0,
 \\&&\hspace{-1cm}
\p_t\p_t A^{3}_j=0,~~
 \partial_i \p_t A^{3}_i=0,\\\non
\\&&\hspace{-1cm}
 \partial_i(\p_i A_t^1-\p_t A^{1}_i-gA^{3}_{i}A^{2}_{t})-gA^{2}_{i}\p_t A^{3}_i-gA^{3}_i (\p_i A^{2}_{t} -\p_t A^{2}_i+gA^{3}_iA^{1}_{t})
\non\\&&\hspace{8cm}-g\phi^{\dagger}_{\sss_m} T^1_{\sss_{mn}}\,\phi_{\sss_n}=0 ,
\\&&\hspace{-1cm}
 \partial_i(\p_i A_t^2-\p_t A^{2}_i+gA^{3}_{i}A^{1}_{t})+gA^{1}_{i}\p_t A^{3}_i+gA^{3}_i (\p_i A^{1}_{t} -\p_t A^{1}_i-gA^{3}_iA^{2}_{t})
\non\\&&\hspace{8cm}-g\phi^{\dagger}_{\sss_m} T^2_{\sss_{mn}}\,\phi_{\sss_n}=0,\\\non
\\&&\hspace{-1cm}
i\p_t \phi_m  =0,~
i\p_t \chi_m +i\sigma_i \p_i \phi_m +g\sigma_i T^{3}_{mn}A^{3}_{i} \phi_n =0.
\eea \ees

\item Case 2: ($\a_1 >0,\a_2 >0,\b_1=\a_1+1,\b_2=\a_2+1$):
\bes
\bea{}&&\hspace{-1cm}
\p_t(\p_t A^{1}_j-\p_j A_t^1 +gA^{2}_{t}A^{3}_j)+gA^{2}_{t}\p_t A^{3}_j=0,
 \\&&\hspace{-1cm}
\p_t(\p_t A^{2}_j-\p_j A_t^2 -gA^{1}_{t}A^{3}_j)-gA^{1}_{t}\p_t A^{3}_j=0,
 \\&&\hspace{-1cm}
\p_t\p_t A^{3}_j=0,~~
 \partial_i \p_t A^{3}_i=0,\\\non
\\&&\hspace{-1cm}
 \partial_i(\p_i A_t^1-\p_t A^{1}_i-gA^{3}_{i}A^{2}_{t})-gA^{2}_{i}\p_t A^{3}_i-gA^{3}_i (\p_i A^{2}_{t} -\p_t A^{2}_i+gA^{3}_iA^{1}_{t})=0 ,
\\&&\hspace{-1cm}
 \partial_i(\p_i A_t^2-\p_t A^{2}_i+gA^{3}_{i}A^{1}_{t})+gA^{1}_{i}\p_t A^{3}_i+gA^{3}_i (\p_i A^{1}_{t} -\p_t A^{1}_i-gA^{3}_iA^{2}_{t})=0,\\\non
\\&&\hspace{-1cm}
i\p_t \phi_m =0,~
i\p_t \chi_m +i\sigma_i \p_i \phi_m  +g\sigma_i T^{3}_{mn}A^{3}_{i} \phi_n =0.
\eea \ees
\end{itemize}
In Case 2, as the last equation is different, this sector does not fully reduce to the free equations of $SU(2)$ Yang-Mills theory and free fermions. In Fig[\ref{EEMplot}], we showed the Case 1 in magenta and the rest of the sectors in blue.  
\subsubsection*{Gauge Transformation for EEM sector} For the EEM sector, we scale the gauge transformation parameter as,
\begin{equation}\label{thetaeem}
\Theta^{1,2} \to \e^p \Theta^{1,2},\,\,\Theta^{3} \to \e^q \Theta^{3}.
\end{equation}
We want to find $p,q$  in the same way mentioned before. Now we scale $\Theta^a$'s differently for $A^{1,2}$ and $A^3$. The values comes out to be
\bea{} p=1,q=2,\,\,\Theta^{1,2} \to \e \Theta^{1,2},\,\,\Theta^3 \to \e^2 \Theta^3 \eea
The gauge transformation in this sector is given by
\bea{}\label{gaeem}
&&\non \delta A_t^{1}=\frac{1}{g}\p_t\Theta^{1},~\delta A_i^{1}=\frac{1}{g}\p_i\Theta^{1}-A^{3}_{i}\Theta^2,\\&&
 \delta A_t^{2}=\frac{1}{g}\p_t\Theta^{2},~\delta A_i^{2}=\frac{1}{g}\p_i\Theta^{2}+A^{3}_{i}\Theta^1,\non\\&&
 \delta A_t^3=\frac{1}{g}\p_t\Theta^3+A_t^1\Theta^2-A_t^2\Theta^1,~\delta A_i^3=0,\non\\
&&\non \delta \phi_{\sss_m}= \e^{\a_{\sss_n}-\a_{\sss_m}+1}[i\Theta^{1} T^{1}_{\sss_{mn}}\phi_{\sss_n}+i\Theta^{2} T^{2}_{\sss_{mn}}\phi_{\sss_n}]+ \e^{\a_{\sss_n}-\a_{\sss_m}+2}i\Theta^{3} T^{3}_{\sss_{mn}}\phi_{\sss_n},\\
&&\delta \chi_{\sss_m}= \e^{\b_{\sss_n}-\b_{\sss_m}+1}[i\Theta^{1} T^{1}_{\sss_{mn}}\chi_{\sss_n}+i\Theta^{2} T^{2}_{\sss_{mn}}\chi_{\sss_n}]+ \e^{\b_{\sss_n}-\b_{\sss_m}+2}i\Theta^3 T^3_{\sss_{mn}}\chi_{\sss_n}.
\eea
The invariance of each subsector under these transformations can be seen when we take one set of values of the coefficients $(\a_m,\b_m)$ along with $\epsilon \rightarrow 0$ limit and plug them into \eqref{gaeem}.

\subsubsection*{Checking symmetries of EEM sector} 
The representative limit for this sector is taken as $\a_1=0,\a_2=0,\b_1=1,\b_2=1$. The equations are
\bes
\bea{}&&\hspace{-1cm}
\p_t(\p_t A^{1}_j-\p_j A_t^1 +gA^{2}_{t}A^{3}_j)+gA^{2}_{t}\p_t A^{3}_j=0,
 \\&&\hspace{-1cm}\label{eem11s}
\p_t(\p_t A^{2}_j-\p_j A_t^2 -gA^{1}_{t}A^{3}_j)-gA^{1}_{t}\p_t A^{3}_j=0,
 \\&&\hspace{-1cm}
\p_t\p_t A^{3}_j=0,~~
 \partial_i \p_t A^{3}_i=0,\\\non
\\&&\hspace{-1cm}\label{eem12s}
 \partial_i(\p_i A_t^1-\p_t A^{1}_i-gA^{3}_{i}A^{2}_{t})-gA^{2}_{i}\p_t A^{3}_i-gA^{3}_i (\p_i A^{2}_{t} -\p_t A^{2}_i+gA^{3}_iA^{1}_{t})
\non\\&&\hspace{8cm}-g\phi^{\dagger}_{\sss_m} T^1_{\sss_{mn}}\,\phi_{\sss_n}=0 ,
\\&&\hspace{-1cm}\label{eem10s}
 \partial_i(\p_i A_t^2-\p_t A^{2}_i+gA^{3}_{i}A^{1}_{t})+gA^{1}_{i}\p_t A^{3}_i+gA^{3}_i (\p_i A^{1}_{t} -\p_t A^{1}_i-gA^{3}_iA^{2}_{t})
\non\\&&\hspace{8cm}-g\phi^{\dagger}_{\sss_m} T^2_{\sss_{mn}}\,\phi_{\sss_n}=0,\\\non
\\&&\hspace{-1cm}
i\p_t \phi_m  =0,~
i\p_t \chi_m +i\sigma_i \p_i \phi_m +g\sigma_i T^{3}_{mn}A^{3}_{i} \phi_n =0.
\eea \ees
The values of the constants are given as
\bea{}
\i\{\underbrace{\Delta=\frac{3}{2},f=0,f'=-\frac{1}{2}}_{\text{Fermionic Field}},\underbrace{\Delta^\prime=1,a_{1,2}=0,b_{1,2}=1,a_{3}=1,b_{3}=0}_
{\text{Gauge Field}}\i\}.\eea
The invariance under scale transformation becomes
\bes
\bea{}&&\hspace{-1.3cm}
[D,\p_t(\p_t A^{1}_j-\p_j A_t^1 +gA^{2}_{t}A^{3}_j)+gA^{2}_{t}\p_t A^{3}_j]=(\Delta' -1)g[(\p_t A^{2}_{t})A^{3}_j+2A^{2}_{t}(\p_t A^{3}_j)],
 \\&&\hspace{-1.3cm}
[D,\p_t\p_t A^{3}_j]=0,~~
[D, \partial_i \p_t A^{3}_i]=0,~[D,i\p_t \phi_m]  =0,\\\non
\\&&\hspace{-1.3cm}\non
[D, \eqref{eem12s}]=(\Delta' -1)(\p_i \p_i A^{1}_{t} -\p_i \p_t A^{1}_i) -(2\Delta' -2)g[(\p_i A^{3}_i)A^{2}_{t}+2A^{3}_i(\p_i A^{2}_{t}) \\&&\non\hspace{1cm}+A^{2}_{i}(\p_t A^{3}_i)-A^{3}_{i}(\p_t A^{2}_i)]-(3\Delta' -3)(gA^{3}_{i} A^{3}_iA^{1}_{t}) \\&&\hspace{1cm} -(2\Delta -3)(g\phi^{\dagger}_m T^{1}_{mn}\phi_n) ,
\\&&\hspace{-1.3cm}
[D,i\p_t \chi_m +i\sigma_i \p_i \phi_m +g\sigma_i T^{3}_{mn}A^{3}_{i} \phi_n] =(\Delta'-1)g\sigma_{i}T^{3}_{mn}A^{3}_{i}\phi_n.
\eea \ees
Under $K_l$, we have
\bes
\bea{}&&\hspace{-1.3cm}
[K_l,\p_t(\p_t A^{1}_j-\p_j A_t^1 +gA^{2}_{t}A^{3}_j)+gA^{2}_{t}\p_t A^{3}_j]=-(2\Delta' -2)[\delta_{lj}\p_t A^{1}_{t}-x_l g(\p_t A^{2}_{t})A^{3}_j\non\\&&\hspace{6.7cm}-2x_l gA^{2}_{t}(\p_t A^{3}_j)],
 \\&&\hspace{-1.3cm}
[K_l,\p_t\p_t A^{3}_j]=0,~~
[K_l, \partial_i \p_t A^{3}_i]=(2\Delta' +4-2\delta_{ii})\p_t A^{3}_{l},~[K_l ,i\p_t \phi_m]  =0,\\\non
&&\hspace{-1.3cm}\non
[K_l,i\p_t \chi_m +i\sigma_i \p_i \phi_m +g\sigma_i T^{3}_{mn}A^{3}_{i} \phi_n] =(2\Delta -3)i\sigma_l \phi_m+(2\Delta'-2)x_l (g\sigma_{i}T^{3}_{mn}A^{3}_{i}\phi_n).
\eea \ees
The equations are invariant under $M^{m_1,m_2,m_3}$. Similarly, the invariance of \eqref{eem11s} and \eqref{eem10s} can also be checked respectively.

\subsection{EMM sector}
\begin{figure}[t]
\centering
\includegraphics[width=8cm]{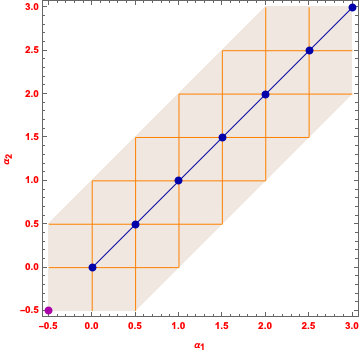}
\caption{Allowed region for EMM sector.}
\floatfoot{\small {{\bf{Key:}}  Point in magenta: Case 1, Points in blue (same sectors are connected by a line): Case 2. Shaded region is described by  \eqref{conemm}.}}
\label{EMMplot}
\end{figure}

We are looking at the other mixed sector EMM. Here, two of the gauge fields scale magnetically and the other one scales electrically.
\bea{}\hspace{-0.5cm}
A_t^{1}\to  A_t^{1},~ A_i^{1}\to \epsilon A_i^{1},~
A_t^{2,3}\to \epsilon A_t^{2,3},\,\, A_i^{2,3}\to A_i^{2,3},~
\phi_{\sss_m} \to \e^{\a_m}\phi_{\sss_m},\,\,\chi_{\sss_m} \to \e^{\b_m}\chi_{\sss_m}.
\eea
The equations of motion are
\bes
\bea{}\non
&&\hspace{-2cm}\bullet \text{For gauge field } A^{1}:\\
&&\hspace{-1cm}
\p_t(\p_t A^{1}_j-\p_j A_t^1)-g[\epsilon^{\alpha_m+\beta_n+1} \phi^{\dagger}_{\sss_m}\sigma_j T^{1}_{\sss_{mn}}\chi_{\sss_n}
+\epsilon^{\b_m+\a_n+1}\chi^{\dagger}_{\sss_m}\sigma_j T^1_{\sss_{mn}}\phi_{\sss_n}]=0,\label{hgd1}
\\&&\hspace{-1cm}
  g(A^2_i \p_t A^{3}_i-A^3_i \p_t A^{2}_i)
+g[\epsilon^{\alpha_m+\alpha_n+1} \phi^{\dagger}_{\sss_m} T^1_{\sss_{mn}}\,\phi_{\sss_n}
+\e^{\beta_{m}+\beta_n+1} \chi^{\dagger}_{\sss_m}\,T^1_{\sss{mn}}\,\chi_{\sss_n}]=0 \label{hgd2},\\ \non
\\
&&\hspace{-2cm}\bullet \text{For gauge field } A^{2,3}:\non
 \\&&\hspace{-1cm}
\p_t\p_t A^{(2,3)}_j-g[\epsilon^{\alpha_m+\beta_n+2} \phi^{\dagger}_{\sss_m}\sigma_j T^{(2,3)}_{\sss_{mn}}\chi_{\sss_n}
+\epsilon^{\b_m+\a_n+2}\chi^{\dagger}_{\sss_m}\sigma_j T^{(2,3)}_{\sss_{mn}}\phi_{\sss_n}]=0,
 \\&&\hspace{-1cm}
 \partial_i \p_t A^{(2,3)}_i
+g[\epsilon^{\alpha_m+\alpha_n +1} \phi^{\dagger}_{\sss_m} T^{(2,3)}_{\sss_{mn}}\,\phi_{\sss_n}
+\e^{\beta_{m}+\beta_n +1} \chi^{\dagger}_{\sss_m}\,T^{(2,3)}_{\sss{mn}}\,\chi_{\sss_n}]=0,
\eea \ees
\bes
\bea{}
&&\hspace{-2cm}
\bullet\, \text{For spinors } \phi_{\sss m}, \chi_{\sss m}: \non\\
&&
i(\p_t \phi_m+\e^{\b_m-\a_m+1}\sigma_i \p_i \chi_m )+\e^{\a_n-\a_m+1} gT^{1}_{mn}A^{1}_{t}\phi_n +\e^{\a_n-\a_m+2} gT^{2}_{mn}A^{2}_{t}\phi_n\non\\
&&+\e^{\a_n-\a_m+2} gT^{3}_{mn}A^{3}_{t}\phi_n+\e^{\b_n -\a_m +2} g\sigma_i T^{1}_{mn}A^{1}_{i} \chi_n\non\\
&&+\e^{\b_n -\a_m+1 } g\sigma_i T^{2}_{mn}A^{2}_{i} \chi_n+\e^{\b_n -\a_m+1} g\sigma_i T^{3}_{mn}A^{3}_{i} \chi_n =0,\\\non
  \\&&
i(\p_t \chi_m+\e^{\a_m-\b_m+1}\sigma_i \p_i \phi_m )+\e^{\b_n-\b_m+1} gT^{1}_{mn}A^{1}_{t}\chi_n +\e^{\b_n-\b_m+1} gT^{2}_{mn}A^{2}_{t}\chi_n\non\\
&&+\e^{\b_n-\b_m+1} gT^{3}_{mn}A^{3}_{t}\chi_n+\e^{\a_n -\b_m +2} g\sigma_i T^{1}_{mn}A^{1}_{i} \phi_n\non\\
&&+\e^{\a_n -\b_m+1 } g\sigma_i T^{2}_{mn}A^{2}_{i} \phi_n+\e^{\a_n -\b_m+1} g\sigma_i T^{3}_{mn}A^{3}_{i} \phi_n =0.
\eea \ees
The constraints on $(\a_m,\b_m)$ are described in Table[\ref{EMMtable}].
\begin{table}[t]
\centering
    \begin{tabular}{| l |  p{5cm} |}
   
    \hline
\multicolumn{2}{ |c| }{EMM sector} \\
\hline
  From equation & Constraints\\ [0.2cm]
\hline
\multirow{2}{*}{Equation \eqref{hgd2} of Gauge field  $A^{1}$}  &(i) $\a_n+\a_m+1\geq0$\\[0.2cm]
 & (ii )$\b_n+\b_m+1>0$\\
   \hline
   Equation \eqref{hgd1} of Gauge field  $A^{1}$ &(iii) $\a_n+\b_m+1>0$ (no equality to reduce it to U(1))\\[0.2cm]
\hline
\multirow{3}{*}{Second equation of Gauge field  $A^{2,3}$}  &(iv) $\a_n+\a_m+1\geq0$\\[0.2cm]
 & (v)$\b_n+\b_m+1>0$\\
 \hline
First equation of Gauge field $A^{2,3}$  &  (vi)$\a_n+\b_m+2>0$\\[0.2cm]
\hline
\multirow{3}{*}{Equation for $\phi_{\sss_m}$} & (vii) $\b_m-\a_m+1>0$,\\[0.2cm]
    & (viii) $\a_n-\a_m+1\geq0$ \\
    & (ix) $\b_n-\a_m+1\geq0$\\
   \hline
   \multirow{3}{*}{Equation for $\chi_{\sss_m}$}&  (x) $\a_m-\b_m+1=0$\\
   & (xi) $\b_n-\b_m+1\geq0$\\
   & (xii) $\a_n-\b_m+1\geq0$\\
 \hline
\end{tabular}
\caption{Constraints on $\a_m,\b_m$ in EMM sector.}
\label{EMMtable}
\end{table}
From constraints (i),(viii) and (x) we find:
\bes \label{conemm}
\begin{eqnarray}
\a_1 \geq -\frac{1}{2}, \, \a_2 \geq -\frac{1}{2},\\
-1\leq \a_2 -\a_1 \leq 1,\\
\a_m=\b_m-1.
\end{eqnarray}
\ees
The region covered by the inequalities above is described by Plot[\ref{EMMplot}].
The last constraint (xii) adds another condition,
\begin{eqnarray}\label{con2emm}
\a_n-\b_m+1\geq0  \implies \a_1=\a_2.
\end{eqnarray}
The rest of the constraints from the table are satisfied trivially.
 Hence imposing \eqref{con2emm}, the allowed values are $$(\a_1,\a_2)=(-\frac{1}{2},-\frac{1}{2}).(0,0),\:(\frac{1}{2},\frac{1}{2}),\:(1,1)\dots$$ From the allowed values of $(\a_1,\a_2)$ we can construct only two different sectors.\\
\begin{itemize}
\item Case 1: \{$\a_1=-\frac{1}{2},\a_2=-\frac{1}{2},\b_1=\frac{1}{2},\b_2=\frac{1}{2}$\}:
\bes
\bea{}&&\hspace{-1cm}
\p_t(\p_t A^{1}_j-\p_j A_t^1)=0,~
\p_t\p_t A^{(2,3)}_j=0,
 \\&&\hspace{-1cm}
  g(A^2_i \p_t A^{3}_i-A^3_i \p_t A^{2}_i)
+g \phi^{\dagger}_{\sss_m} T^1_{\sss_{mn}}\,\phi_{\sss_n}=0 ,
\\&&\hspace{-1cm}
 \partial_i \p_t A^{(2,3)}_i
+g\phi^{\dagger}_{\sss_m} T^{(2,3)}_{\sss_{mn}}\,\phi_{\sss_n}=0,\\&&\hspace{-1cm}
i\p_t \phi_m =0,~
i\p_t \chi_m +i\sigma_i \p_i \phi_m + g\sigma_i T^{2}_{mn}A^{2}_{i} \phi_n + g\sigma_i T^{3}_{mn}A^{3}_{i} \phi_n =0.
\eea \ees

\item Case 2: \{$\a_1 > -\frac{1}{2},\a_2>-\frac{1}{2},\b_1=\a_1+1,\b_2=\a_2+1$\}:
\bes
\bea{}&&\hspace{-1cm}
\p_t(\p_t A^{1}_j-\p_j A_t^1)=0,~
\p_t\p_t A^{(2,3)}_j=0,
 \\&&\hspace{-1cm}
  g(A^2_i \p_t A^{3}_i-A^3_i \p_t A^{2}_i)=0 ,~
 \partial_i \p_t A^{(2,3)}_i=0,
\\&&\hspace{-1cm}
i\p_t \phi_m=0,~
i\p_t \chi_m +i\sigma_i \p_i \phi_m + g\sigma_i T^{2}_{mn}A^{2}_{i} \phi_n + g\sigma_i T^{3}_{mn}A^{3}_{i} \phi_n =0.
\eea \ees
\end{itemize}
\subsubsection*{Gauge Transformation for EMM sector} For the EMM sector, we scale the gauge transformation parameter as,
\begin{equation}\label{thetaeem}
\Theta^{1} \to \e^p \Theta^{1},\,\,\Theta^{2,3} \to \e^q \Theta^{2,3}.
\end{equation}
 We scale $\Theta^a$'s differently for $A^{1}$ and $A^{2,3}$. The values of $p,q$ comes out to be
\bea{}  p=1,q=2,\,\,\Theta^{1} \to \e \Theta^{1},\,\,\Theta^{2,3} \to \e^2 \Theta^{2,3}.\eea
The gauge transformations in this sector is given by
\bea{}\label{gaemm}
&&\non \delta A_t^{1}=\frac{1}{g}\p_t\Theta^{1},~\delta A_i^{1}=\frac{1}{g}\p_i\Theta^{1},~\delta A_t^{2,3}=\frac{1}{g}\p_t\Theta^{2,3},~
\delta A_i^{2,3}=0,\\
&&\non \delta \phi_{\sss_m}=  \e^{\a_{\sss_n}-\a_{\sss_m}+1}i\Theta^{1} T^{1}_{\sss_{mn}}\phi_{\sss_n}+  \e^{\a_{\sss_n}-\a_{\sss_m}+2}[i\Theta^{2} T^{2}_{\sss_{mn}}\phi_{\sss_n}+i\Theta^{3} T^{3}_{\sss_{mn}}\phi_{\sss_n}],\\
&&\delta \chi_{\sss_m}= \e^{\b_{\sss_n}-\b_{\sss_m}+1}i\Theta^{1} T^{1}_{\sss_{mn}}\chi_{\sss_n}+  \e^{\b_{\sss_n}-\b_{\sss_m}+2}[i\Theta^{2} T^{2}_{\sss_{mn}}\chi_{\sss_n}+i\Theta^{3} T^{3}_{\sss_{mn}}\chi_{\sss_n}].
\eea
The invariance under these transformations can be seen when we take one set of values of the coefficients $(\a_m,\b_m)$ along with $\epsilon \rightarrow 0$ limit and plug them into \eqref{gaemm}.
\subsubsection*{Checking symmetries of EMM sector} 
We will take $\a_1=0,\a_2=0,\b_1=1,\b_2=1$ as the representative limit for this sector. The equations are
 \bes
\bea{}&&
\p_t(\p_t A^{1}_j-\p_j A_t^1)=0,~
\p_t\p_t A^{(2,3)}_j=0,
 \\&&
  g(A^2_i \p_t A^{3}_i-A^3_i \p_t A^{2}_i)=0 ,~
 \partial_i \p_t A^{(2,3)}_i=0,
\\&&
i\p_t \phi_m=0,~
i\p_t \chi_m +i\sigma_i \p_i \phi_m + g\sigma_i T^{2}_{mn}A^{2}_{i} \phi_n + g\sigma_i T^{3}_{mn}A^{3}_{i} \phi_n =0.
\eea \ees
The values of the constants in the representation theory are
\bea{}
\i\{\underbrace{\Delta=\frac{3}{2},f=0,f'=-\frac{1}{2}}_{\text{Fermionic Field}},\underbrace{\Delta^\prime=1,a_{1}=0,b_{1}=1,a_{2,3}=1,b_{2,3}=0}_
{\text{Gauge Field}}\i\}.\eea
Since, most of the equations do not contain interaction pieces, the invariance under $D,K_l$ and $M$ are quite trivial to find. The equations come out to be invariant in $d=4$ case. However, in the EMM sector we do not get any kinetic term in the third equation for Case 1 and 2. Hence, we need to discard this entire EMM sector of Carrollian $SU(2)$ Yang-Mills with fermions.

\subsection{MMM sector}
\begin{figure}[t]
\centering
\includegraphics[width=8cm]{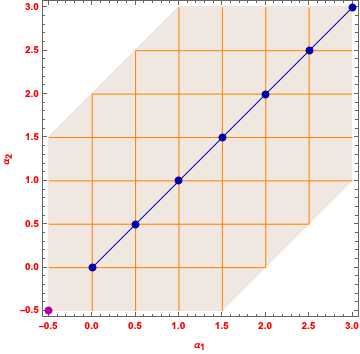}
\caption{Allowed region for MMM sector.}
\floatfoot{\small {{\bf{Key:}}  Point in magenta: Case 1, Points in blue (same sectors are connected by a line): Case 2. Shaded region is described  by \eqref{conmmm}.}}
\label{MMMplot}
\end{figure}

In this sector, all gauge fields are scaled magnetically. The scaling are given as
\begin{eqnarray}
A_i^a\to  A_i^a,\,\, A_t^a\to \epsilon A_t^a,~\phi_{\sss_m}\to\epsilon^{\alpha_m}\phi_{\sss_m}, \, \chi_{\sss_m}\to\epsilon^{\beta_m}\chi_{\sss_m}.
\end{eqnarray}
The equations of motion in this sector are given as
\bes
\bea{}&&\hspace{-1cm}
\p_t\p_t A^{a}_j-g[\epsilon^{\alpha_m+\beta_n+2} \phi^{\dagger}_{\sss_m}\sigma_j T^{a}_{\sss_{mn}}\chi_{\sss_n}
+\epsilon^{\b_m+\a_n+2}\chi^{\dagger}_{\sss_m}\sigma_j T^a_{\sss_{mn}}\phi_{\sss_n}]=0,
 \\&&\hspace{-1cm}
 \partial_i \p_t A^{a}_i+gf^{abc}A^{b}_{i}\p_t A^{c}_i
+g[\epsilon^{\alpha_m+\alpha_n +1} \phi^{\dagger}_{\sss_m} T^a_{\sss_{mn}}\,\phi_{\sss_n}
+\e^{\beta_{m}+\beta_n +1} \chi^{\dagger}_{\sss_m}\,T^a_{\sss{mn}}\,\chi_{\sss_n}]=0 ,
\\&&\hspace{-1cm}
i(\p_t \phi_m+\e^{\b_m-\a_m+1}\sigma_i \p_i \chi_m )+\e^{\a_n-\a_m+2} gT^{a}_{mn}A^{a}_{t}\phi_n+\e^{\b_n -\a_m+1} g\sigma_i T^{a}_{mn}A^{a}_{i} \chi_n =0,\\&&\hspace{-1cm}
i(\p_t \chi_m+\e^{\a_m-\b_m+1}\sigma_i \p_i \phi_m )+\e^{\b_n-\b_m+2} gT^{a}_{mn}A^{a}_{t}\chi_n+\e^{\a_n -\b_m+1} g\sigma_i T^{a}_{mn}A^{a}_{i} \phi_n =0.~~~
\eea 
\ees
Imposing the constraints  we can write down Table[{\ref{MMMtable}}] to find out the allowed values of $\alpha_m,\beta_m$ for the MMM sector.
\begin{table}[t]
\centering
    \begin{tabular}{ | l | p{5cm} |  p{5cm} |}
   
    \hline
\multicolumn{3}{ |c| }{MMM sector} \\
\hline
 Free equation  & $U(1)$+Fermions & Constraints\\ [0.2cm]
\hline
$\p_t \p_t A_i^a=0$ & $ \p_t \p_t A_i=0$  &(i) $\a_n+\b_m+2>0$\\[0.2cm]
   \hline
\multirow{2}{*}{$\p_i\p_t A_i^a+gf^{abc}A^{b}_i \p_tA_i^c=0$} & $\p_i\p_t A_i+ e\phi^{\dagger}\phi=0$  &(ii) $\alpha_n+\a_m+1\geq0$\\
   & $\p_i\p_t A_i=0$  &  (iii) $\b_n+\b_m+1>0$\\[0.2cm]
 \hline
   \multirow{3}{*}{$i\p_t \phi=0$} & \multirow{3}{*}{$i \p_t \phi=0$} & (iv) $\b_m-\a_m+1>0$,\\
   & & (v) $\b_n-\a_m+1\geq0$,$(n \neq m)$, \\
   & & (vi) $\a_n-\a_m+2\geq0$\\[0.2cm]
 \hline
   \multirow{3}{*}{$i\p_t \chi+i\sigma_i \partial_i\phi=0$} & \multirow{3}{*}{$i\p_t \chi+i\sigma_i \partial_i \phi+e\s_i A_i\phi=0$} &  (vii) $\a_m-\b_m+1=0$,\\
   & &(viii) $\b_n-\b_m+2\geq0$,\\
   & &(ix) $\a_n-\b_m+1\geq0,\, (n \neq m)$\\[0.2cm]
   \hline
\end{tabular}
\caption{Constraints on $\a_m,\b_m$ in MMM sector.}
\label{MMMtable}
\end{table}
From constraints (ii),(vi) and (vii) we find:
\bes \label{conmmm}
\begin{eqnarray}
\a_1 \geq -\frac{1}{2}, \, \a_2 \geq -\frac{1}{2},\\
-2\leq \a_2 -\a_1 \leq 2,\\
\a_m=\b_m-1.
\end{eqnarray} 
\ees
We can plot (\ref{MMMplot}) to describe the region contained   by \eqref{conmmm}.
However we have another constraint (ix),
\begin{eqnarray}\label{con2mmm}
\nonumber \a_n-\b_m+1\geq0 &\implies& \a_n-(\a_m+1)+1\geq 0 \implies \a_n-\a_m\geq 0,\,\,\, (for\: n \neq m)\\
&\implies& \a_1-\a_2\geq 0,\, \a_2-\a_1\geq 0.
\end{eqnarray} 
Eq \eqref{con2mmm} can hold if and only if 
$
\a_1=\a_2
$ which suggests $\b_1=\b_2$. The rest of the constraints trivially satisfy these four major constraints. Hence the allowed values for $(\a_1,\a_2)$ are written from the intersecting points in the plot above (also strictly imposing \eqref{con2mmm}):$$(\a_1,\a_2)=(-\frac{1}{2},-\frac{1}{2}),\:(0,0),\:(\frac{1}{2},\frac{1}{2}),\:(1,1)\dots$$ From the allowed values of $(\a_1,\a_2)$ we can construct only two different sectors.\
\begin{itemize}
\item Case 1: \{$\a_1=-\frac{1}{2},\a_2=-\frac{1}{2},\b_1=\frac{1}{2},\b_2=\frac{1}{2}$\}:
\bes
\bea{}&&\hspace{-1cm}
\p_t\p_t A^{a}_j=0,~
 \partial_i \p_t A^{a}_i+gf^{abc}A^{b}_{i}\p_t A^{c}_i
+g\phi^{\dagger}_{\sss_m} T^a_{\sss_{mn}}\,\phi_{\sss_n}=0 ,
\\&&\hspace{-1cm}
i\p_t \phi_m =0,~
i\p_t \chi_m+i\sigma_i \p_i \phi_m + g\sigma_i T^{a}_{mn}A^{a}_{i} \phi_n =0.
\eea 
\ees

\item Case 2: \{$\a_1> -\frac{1}{2},\a_2> -\frac{1}{2},\b_1=\a_1+1,\b_2=\a_2+1$\}: 
\bes
\bea{}&&\hspace{-1cm}
\p_t\p_t A^{a}_j=0,~
 \partial_i \p_t A^{a}_i+gf^{abc}A^{b}_{i}\p_t A^{c}_i=0,
\\&&\hspace{-1cm}
i\p_t \phi_m =0,~
i\p_t \chi_m+i\sigma_i \p_i \phi_m + g\sigma_i T^{a}_{mn}A^{a}_{i} \phi_n =0.
\eea 
\ees
\end{itemize}
\subsubsection*{Gauge Transformation for MMM sector} For the MMM sector we scale the gauge transformation parameter as,
\begin{equation}\label{thetam}
\Theta^a \to \e^q \Theta^a.
\end{equation}
We want to find the value of $q$ that keeps the gauge transformation \eqref{gauge} of fermions and gauge fields finite  along with the ultra-relativistic spacetime contraction. The value of $q$ comes out to be
\bea{} q=2,\,\,\Theta^a \to \e^2 \Theta^a. \eea
 We also scale the gauge fields magnetically and the fermions in the usual way. Thus we arrive at the gauge transformation of MMM limit of Carrollian $SU(2)$ Yang Mills coupled to fermions:
\bea{}\label{gammm}
&&\non \delta A_t^a=\frac{1}{g}\p_t\Theta^a,~
\delta A_i^a=0 ,\\
&&\delta \phi_{\sss_m}=\e^{\a_{\sss_n}-\a_{\sss_m}+2}i\Theta^a T^a_{\sss_{mn}}\phi_n,~
\delta \chi_{\sss_m}= \e^{\b_{\sss_n}-\b_{\sss_m}+2}i\Theta^a T^a_{\sss_{mn}}\chi_n .
\eea
When we take one set of values of the coefficients $(\a_m,\b_m)$ along with $\epsilon \rightarrow 0$ limit and plug them into \eqref{gammm}, the equations remain invariant in that subsector.
\subsubsection*{Checking symmetries of MMM sector} 
The representative limit for this sector is chosen to be $\a_1=0,\a_2=0,\b_1=1,\b_2=1$. The equations are
\bes
\bea{}&&
\p_t\p_t A^{a}_j=0,~
 \partial_i \p_t A^{a}_i+gf^{abc}A^{b}_{i}\p_t A^{c}_i=0,
\\&&
i\p_t \phi_m =0,~
i\p_t \chi_m+i\sigma_i \p_i \phi_m + g\sigma_i T^{a}_{mn}A^{a}_{i} \phi_n =0.
\eea 
\ees
The values of the constants are
\bea{} 
\i\{\underbrace{\Delta=\frac{3}{2},f=0,f'=-\frac{1}{2}}_{\text{Fermionic Field}},\underbrace{\Delta^\prime=1,a=1,b=0}_
{\text{Gauge Field}}\i\}.
\eea
The invariance under scale transformation is 
\bes
\bea{}&&\hspace{-1cm}
[D,\p_t\p_t A^{a}_j]=0,~
[D, \partial_i \p_t A^{a}_i+gf^{abc}A^{b}_{i}\p_t A^{c}_i]=(\Delta'-1)g f^{abc}A^{b}_{i}\p_t A^{c}_i,
\\&&\hspace{-1cm}
[D,i\p_t \phi_m] =0,~
[D,i\p_t \chi_m+i\sigma_i \p_i \phi_m + g\sigma_i T^{a}_{mn}A^{a}_{i} \phi_n] =(\Delta'-1)g\sigma_i T^{a}_{mn}A^{a}_{i}\phi_n.
\eea 
\ees
Under SCT, we have
\bea{}&&\hspace{-1cm}\non
[K_l, \partial_i \p_t A^{a}_i+gf^{abc}A^{b}_{i}\p_t A^{c}_i]=(2\Delta' +4 -2\delta_{ii})\p_t A^{a}_{l}+2(\Delta'-1)x_l (g f^{abc}A^{b}_{i}\p_t A^{c}_i),
\\&&\hspace{-1cm}\non
[K_l,i\p_t \chi_m+i\sigma_i \p_i \phi_m + g\sigma_i T^{a}_{mn}A^{a}_{i} \phi_n] =(2\Delta -3)i\sigma_l \phi_m +2(\Delta'-1)x_l (g\sigma_i T^{a}_{mn}A^{a}_{i}\phi_n),\\&&\hspace{-1cm}
[K_l,i\p_t \phi_m] =0,~[K_l,\p_t\p_t A^{a}_j]=0.
\eea 
The equations are also invariant under $M^{m_1,m_2,m_3}$.
\bea{}&&\hspace{-1cm}\non
[M^{m_1,m_2,m_3}, \partial_i \p_t A^{a}_i+gf^{abc}A^{b}_{i}\p_t A^{c}_i]=0,
\\&&\hspace{-1cm}\non
[M^{m_1,m_2,m_3},i\p_t \chi_m+i\sigma_i \p_i \phi_m + g\sigma_i T^{a}_{mn}A^{a}_{i} \phi_n] =i \Big({\sigma_i\over 2}\Big)\p_i(x^{m_1}y^{m_2}z^{m_3})\p_t \phi=0,\\&&\hspace{-1cm}
[M^{m_1,m_2,m_3},i\p_t \phi_m] =0,~[M^{m_1,m_2,m_3},\p_t\p_t A^{a}_j]=0.
\eea

\newpage

\section{Galilean $SU(N)$ Yang-Mills Theory with matter} 
\label{Galilean $SU(N)$ Yang-Mills Theory with matter}
In this section we extend the analysis of \cite{Bagchi:2017yvj}, to our non-relativistic counterpart of $SU(N)$ Yang-Mills coupled to matter. It is a further generalisation of Galilean $SU(2)$ Yang-Mills with matter described in \cite{Bagchi:2017yvj}. 




\subsection*{Galilean Conformal Algebra}
\noindent We start with a brief review on the Galilean Conformal Algebra (GCA). GCA is obtained from conformal algebra by performing an Inonu-Wigner contraction. In the process of going to the Galilean framework, the underlying Lorentz symmetry breaks down. The spacetime scale differently, as
\be{ged} x_i \rightarrow \e x_i,~t \rightarrow t .\ee
The generators of GCA can be found by imposing the limit \eqref{ged} on relativistic conformal algebra. The generators in this limit emerge as 
\bea{} L^{(n)}=-t^{n+1}\p_t - (n+1)t^n x_i \p_i,~M^{(n)}_{i}=t^{n+1}\p_i ~\text{for}~n=0,\pm 1,~J_{ij}= -x_{[i}\p_{j]}.
\eea
where $L^{(-1,0,1)}=H,D,K$ ($H, D$ and $K$ are the Galilean Hamiltonian, dilatation and temporal SCT) and $M^{(-1,0,1)}_i =P_{i},B_{i},K_{i}$ ($P_i, B_i$ and $K_i$ represent spatial momentum, Galilean boost and spatial SCT). The full GCA is given as
\bea{GCA}
&& [L^{(n)}, L^{(m)}] = (n-m) L^{(n+m)}, \quad [L^{(n)}, M^{(m)}_i] = (n-m) M_i^{(n+m)},\non\\
&& [M^{(n)}_i, M^{(m)}_j]=0, \quad [L^{(n)}, J_{ij}] = 0, \quad [J_{ij},M^{(n)}_k] = M^{(n)}_{[j} \delta_{i]k}.
\eea
The interesting point to be noted here is that the above algebra closes, even if the index $n$ runs over all integers. Hence,  it is concluded that GCA has an infinite extension in all spacetime dimensions, which is very unlike relativistic conformal algebras for $d>2$.

\medskip

\subsection*{Galilean Yang-Mills}

\noindent In \cite{Bagchi:2017yvj}, a detailed analysis of Galilean conformal field theories was performed including free as well as interacting theories. The last example presented in this paper was Galilean $SU(2)$ Yang-Mills theory coupled to fermions. The various sectors of $\mathcal{O}(1500)$ emerged as sub-limits of EEE, EEM, EMM and MMM. However, using some consistency requirements the possible sectors were brought down to $19$ (including both interacting and free sectors). This significant reduction in the number of sectors hinted towards something deeper coming into the scenario, which is unfortunately yet to be explained.  Another interesting observation was that, the equations of motion in all the sub-limits of $SU(2)$ YM with fermions turned out to be invariant under GCA.

The purpose of this appendix is to generalise the existing calculations of $SU(2)$ case \cite{Bagchi:2017yvj} to non-relativistic limit of $SU(N)$ Yang-Mills theory coupled to fermions. We would look into the structure of equations of motion and  try to find the constraints on parameters $a_m, b_m$ appearing in the theory. Ultimately, we would check the invariance of a representative sector under GCA.  

\medskip

We begin by writing down the equations of motion of parent relativistic $SU(N)$ coupled to fermions theory:
\be{}\label{eomsymm}  \p_{\mu}F^{\mu \nu a}+gf^{abc}A^{b}_{\mu}F^{\mu\nu c}+g\bar{\psi}_{m} \gamma^{\nu}T^{a}_{mn}\psi_{n}=0,~~
i\gamma^{\nu}(D_{\nu}\psi)_n =0.  \ee
where $ D_\mu \equiv \partial_\mu-igT^{a}A_{\mu}^{a}$ is the non abelian gauge covariant derivative  and $F_{\mu\nu}^{a}= \partial_{\mu} A_{\nu}^{a} - \partial_\nu A_{\mu}^{a}+gf^{abc}A^{b}_{\mu}A^{c}_{\nu}$ is the non abelian field strength tensor.
The relativistic equations of motion under the decomposition of Dirac fermion (\ref{ddc}) are  
\bes{}
\bea{} 
&& {\hspace{-0.5cm}} \p_{t}F_{tj}^{a}-\p_{i}F_{ij}^{a} + gf^{abc}(A^{b}_{t}F_{tj}^{c} - A^{b}_{i}F_{ij}^{c}) - g(\phi^{\dagger}_{m} \sigma_{j}T^{a}_{mn}\chi_{n} +\chi^{\dagger}_{m} \sigma_{j}T^{a}_{mn}\phi_{n})=0,~~~\\
&& \p_{i}F_{it}^{a}+gf^{abc}A^{b}_{i}F_{it}^{c} -g(\phi^{\dagger}_m T^{a}_{mn}\phi_{n} +\chi^{\dagger}_{m} T^{a}_{mn}\chi_{n})=0,\\
&& i\p_{t}\phi_m +gT^{a}_{mn}A_{t}^{a}\phi_{n} +i\sigma_{i}\p_{i}\chi_{m} +g\sigma_{i}T^{a}_{mn}A_{i}^a \chi_{n} =0,\\
 && i\p_{t}\chi_{m}+gT^{a}_{mn}A_{t}^{a}\chi_{n} +i\sigma_{i}\p_{i}\phi_{m} +g\sigma_{i}T^{a}_{mn}A_{i}^a \phi_{n} =0.
\eea\ees 
The next step is to find the appropriate scaling on gauge fields and fermions. The scaling of $A_{\mu}$ and ($\phi_m, \chi_m$) are explained in details in Sec[\ref{cfdf}] for the Carrollian case. Here we follow the same scaling keeping in mind that the change in the two systems happens as the underlying spacetime scales differently \refb{ged} as opposed to \refb{stc}. Under these scalings, the generalised equations of motion are given by: 
\begin{itemize} 
\item Case 1: $\mtD -p_0 +1 \leq a \leq \mtD$:
\bes{}
\bea{}
&&\p_i \p_i A_t^\a-g[ \e^{a_m+a_n+2}\phi^{\dagger}_m  T^{\a}_{mn}\phi_n +
\e^{b_m+b_n+2}\chi^{\dagger}_m T^{\a}_{mn}\chi_n]=0,\\ \non \\
&& \p_t \p_j A^{\a}_{t}+ \p_i (\p_i A^{\a}_j - \p_j A^{\a}_i+g f^{\a}_{~JK} A^{J}_i A^{K}_j)+ g f^{\a}_{~\b \gamma} A^\b_t \p_j A^{\gamma}_t \non\\&&+gf^{\a}_{~JK}A^{J}_{i} (\p_i A^{K}_j-\p_j A^{K}_i)
+ g[\e^{a_m+b_n+1}\phi^{\dagger}_m \sigma_j T^{\a}_{mn}\chi_n+ \e^{b_m+a_n+1}\chi^{\dagger}_m \sigma_j T^{\a}_{mn}\phi_n ]=0.\non\\
\eea\ees
\item Case 2: $1 \leq a \leq \mtD-p_0$:
\bes{}
\bea{}
&&\hspace{-1cm}\p_i(\p_i A_t^I -\p_t A^{I}_i +gf^{I}_{~J\a}A^{J}_iA^{\a}_{t}) +gf^{I}_{~J\a}A^{J}_{i}\p_i A^{\a}_{t}\non\\&&\hspace{3cm}-g[ \e^{a_m+a_n+1}\phi^{\dagger}_m  T^{I}_{mn}\phi_n +
\e^{b_m+b_n+1}\chi^{\dagger}_m T^{I}_{mn}\chi_n]=0,\\ \non \\
&&\hspace{-1cm} \p_i (\p_i A^{I}_j - \p_j A^{I}_i)+ g[\e^{a_m+b_n+2}\phi^{\dagger}_m \sigma_j T^{I}_{mn}\chi_n+ \e^{b_m+a_n+2}\chi^{\dagger}_m \sigma_j T^{I}_{mn}\phi_n ]=0.
\eea\ees
\end{itemize}
The Dirac equations are same in both the cases,
\bes
\bea{}&&
i\p_t \phi_m +\e^{a_n -a_m} gT^{\a}_{mn}A^{\a}_{t}\phi_n +\e^{a_n -a_m +1} gT^{I}_{mn}A^{I}_{t}\phi_n +\e^{b_m-a_m -1} i\sigma_{i}\p_i \chi_m \non\\&&\hspace{3.5cm}+\e^{b_n -a_m +1}g\sigma_{i}T^{\a}_{mn} A^{\a}_{i}\chi_n +\e^{b_n -a_m}g\sigma_{i}T^{I}_{mn} A^{I}_{i}\chi_n=0,\\&&
i\p_t \chi_m +\e^{b_n -b_m} gT^{\a}_{mn}A^{\a}_{t}\chi_n +\e^{b_n -b_m +1} gT^{I}_{mn}A^{I}_{t}\chi_n +\e^{a_m-b_m -1} i\sigma_{i}\p_i \phi_m \non\\&&\hspace{3.5cm}+\e^{a_n -b_m +1}g\sigma_{i}T^{\a}_{mn} A^{\a}_{i}\phi_n +\e^{a_n -b_m}g\sigma_{i}T^{I}_{mn} A^{I}_{i}\phi_n=0.
\eea
\ees
Comparing with the free equations, the constraints on ($a_m,b_m$) can be written as
\bea{}a_m +a_n +1 \geq 0,~a_m +b_n +1 \geq 0,~ b_m =a_m+1. \eea
We will not go into details of the number of possible sectors and ways to reduce them in this appendix. What we are really after is the emergence of symmetries in these various sub-sectors. 

\medskip

\noindent{\em{Symmetries of EOM}}

\medskip

\noindent We now pick a representative sector for which we will find the invariance under GCA. We choose a sector randomly, say $a_m=0,b_m=1$. The equations of motion in this sector are given as
\bes{}
\bea{}
&& \p_t \p_j A^{\a}_{t}+ \p_i (\p_i A^{\a}_j - \p_j A^{\a}_i+g f^{\a}_{~JK} A^{J}_i A^{K}_j)+ g f^{\a}_{~\b \gamma} A^\b_t \p_j A^{\gamma}_t \non\\&&\hspace{5cm}+gf^{\a}_{~JK}A^{J}_{i} (\p_i A^{K}_j-\p_j A^{K}_i)=0,\label{sun12}\\
&&\p_i(\p_i A_t^I -\p_t A^{I}_i +gf^{I}_{~J\a}A^{J}_iA^{\a}_{t}) +gf^{I}_{~J\a}A^{J}_{i}\p_i A^{\a}_{t}=0,\label{sun13} \\
&& \p_i (\p_i A^{I}_j - \p_j A^{I}_i)=0,~~\p_i \p_i A_t^\a=0,\\
&&
i\p_t \phi_m +gT^{\a}_{mn}A^{\a}_{t}\phi_n + i\sigma_{i}\p_i \chi_m =0,~~
 i\sigma_{i}\p_i \phi_m =0.
\eea
\ees
 We will now find the invariance of the equations under GCA generators $L^n, M^{n}_{i}$. Before that, let us pause to remind the action of these generators on fields of different spins. They are given as
 \bea{inftr} &&  [L^{(n)},\Phi(t,x)]= (t^{n+1}\p_{t}+(n+1)t^{n}x_{l}\p_{l}
+(n+1)t^{n}\Delta)\Phi(t,x) \non\\&&\hspace{4cm}-t^{n-1}n(n+1)x_{k} U[M_{k}^{(0)},\Phi(0,0)]U^{-1},
\non\\&&
~[M_{l}^{(n)},\Phi(t,x)]= -t^{n+1}\p_{l}\Phi(t,x)+(n+1)t^{n}U[M_{l}^{(0)},\Phi(0,0)]U^{-1}.
\eea  
where  $\Phi =\{\varphi, \phi, \chi, A_t, A_i \}$, $U=e^{tL^{-1}-x_{i}M^{-1}_i}$ and 
\bea{bost}
 &&[M_{k}^{(0)},\Phi(0,0)]= a\varphi_k +b\sigma_{k}\chi +\tilde{b} \sigma_{k}\phi + s A_{k}+rA_{t}\delta_{ki} + \ldots,
\eea
For details on representation theory of GCA, we recommend Section 2.2 of \cite{Bagchi:2017yvj}.
The values of the constants ($r,s,b,\tilde{b}$) are given as, 
\bea{changea}
\Delta_1 =1,~\Delta_2=\frac{3}{2} ,~r^{\a}= -1,~  s^\a= 0,~r^{I}=0,~s^{I}=-1,~b=0,~ \tilde{b}=\frac{1}{2}. 
\eea
where $\Delta_1,~\Delta_2$ are scaling weights of gauge fields and fermionic fields respectively. The equations are trivially invariant under $M^{n}_{i}$. Under $L^{n}$, we have
\bes{}
\bea{}
&&\hspace{-.5cm} [L_{n},\eqref{sun12}]=n(n+1)t^{n-1}(\Delta_1 +2 -\delta_{ii})\p_{j}A^{\alpha}_{t}+(\D_1 -1)(n+1)t^n [gf^{\alpha}_{~\beta\gamma}A^{\beta}_{t}\p_{j} A^{\gamma}_{t}\non\\&&\hspace{5cm}+gf^{\alpha}_{~JK}(A^{K}_{j}\p_{i}A^{J}_{i}-A^{J}_{i}\p_{j}A^{K}_{i}+2A^{J}_{i}\p_{i}A^{K}_{j})],
\\&&\hspace{-.5cm}[L_n, \eqref{sun13}]=(\Delta_1 -1)(n+1)[-nt^{n-1}\p_{i}A^{I}_{i}+gf^{I}_{J \alpha}t^{n}(A^{\alpha}_{t}\p_{i}A^{J}_{i}+2A^{J}_{i}\p_{i} A^{\alpha}_{t})], \\
&&\hspace{-.5cm} [L_n, \p_i (\p_i A^{I}_j - \p_j A^{I}_i)]=0,~~[L_n, \p_i \p_i A_t^\a]=0,~~[L_n,i\sigma_{i}\p_i \phi_m] =0,\\
&&\hspace{-.5cm}
[L_n, i\p_t \phi_m +gT^{\a}_{mn}A^{\a}_{t}\phi_n + i\sigma_{i}\p_i \chi_m] =in(n+1)t^{n-1}(\Delta_2 -\frac{3}{2})\phi_{m} \non\\&&\hspace{6.8cm}+(\Delta_1 -1)t^n (n+1)gT^{\a}_{mn}A^{\a}_{t}\phi_n.
\eea
\ees
From the above set of results, it is easier to conclude that the equations of motion are invariant under the generators of GCA. 

\newpage

\section{Comparison between Galilean and  Carrollian CFTs}
\label{Comparison between Galilean and conformal Carrollian  field theories}
The reader familiar with earlier work \cite{Bagchi:2017yvj,Bagchi:2015qcw,Bagchi:2014ysa} will find many similarities in the methods and indeed the final results of Carrollian and Galilean conformal field theories. Here we attempt to make a comparison between these field theories, which can be thought of as two distinct sub-sectors of a parent relativistic theory. 

The underlying algebras, the CCA and the GCA, are obviously different as they appear from different inequivalent contractions of the relativistic conformal algebra. For the finite algebras, arising solely out of contractions, the number of generators are the same. As we have seen, the infinite extensions in both cases have different flavours and the algebraic structure of the infinite dimensional algebras are quite different. The GCA has the same structure for all dimensions, which the CCA differs significantly depending on dimensions (and whether or not one can give the super-rotation part an infinite lift). 

Moving on to the explicit examples of the field theories, we find quite a few important distinctions. We will begin  the discussion with our first example, the scalar field theory. In \cite{Bagchi:2017yvj}, we found that in Galilean limit \eqref{ged} of $\p^{\mu}\p_{\mu}\varphi =0$, we get 
\bea{}\label{sedds} \p_{i}\p_{i}\varphi =0. \eea 
The above expression \eqref{sedds} contains  spatial derivative only. If we take Carrollian limit \eqref{stsc} on scalar equation,  the result comes out to be 
\bea{}\p_t \p_t \varphi =0, \eea       
which only involve time derivatives. This is a generic feature and obviously has to do with the underlying contraction where in the non-relativistic case, space is scaled down and as a result, spatial derivatives scaled up as opposed to the Carrollian case where time and time-derivatives are scaled down and up respectively. 

For our second example, we will look at massless Dirac field theory. In Galilean case, the equations are given as
\bea{}i\p_t \phi + i\sigma_{i}\p_i \chi=0,~~ i\sigma_{i}\p_i \phi=0, \eea
where $\phi,\chi$ are two component spinors in Pauli-Dirac representation. In Carrollian case, the equations are given by \eqref{eomdf}. In both cases, the scaling of the spinors are taken to be $\phi \rightarrow \phi, \chi \rightarrow \e \chi$. Again, as expected, in the NR case, we get an equation which contain only spatial derivatives whereas in the UR case we have the time derivative acting on the spinor.

\medskip

 Let us now move on to Electrodynamics. The relativistic EOM are: 
\bea{}\label{ddw} \p^{\mu} F_{\mu\nu}=0.\eea 
To take the non-relativistic limit on \eqref{ddw}, we used the non-relativistic scaling
of the coordinates and scaled the gauge field components as
\bes \label{limsi}
\bea{}&&
\mbox{Electric sector:} \quad A_t \to A_t, \, A_i \to \e A_i \label{elimconti}\\&&
\mbox{Magnetic sector:} \quad A_t \to \e A_t, \, A_i \to A_i \label{mlimconti}
\eea
\ees
The reason that we get these two sectors is due to that fact that the contravariant and covariant vectors behave like two independent quantities (because the metric becomes degenerate). This can also be seen clearly as follows
\bes
\bea{}&&\bar{A}^{\mu} = \frac{\p \bar{x}^{\mu}}{\p x^{\a}}A^{\a}  \longrightarrow \Big\{ \bar{A}_t =A_t,~~ \bar{A}_i =\epsilon A_i\Big\},\label{ddwww}\\&&
\bar{A}_{\mu} = \frac{\p x^{\a}}{\p \bar{x}^{\mu}}A_{\a}  \longrightarrow \Big\{ \bar{A}_t =\e A_t,~~ \bar{A}_i = A_i\Big\},\label{ddwwww}
\eea
\ees
 where we have used $\bar{x}_i \rightarrow \e x_i, \bar{t} \rightarrow t$ in the intermediate steps. We see that \eqref{ddwww} can be associated with electric sector, whereas \eqref{ddwwww} denotes the magnetic sector. In case of ultra-relativistic limit, the gauge fields are scaled in a similar fashion \eqref{limsi}. But contravariant vectors gives rise to magnetic sector (see \eqref{contravary}) and the covariant ones give the electric sector. This is because of opposite scaling of spacetime  coordinates ($\bar{x}_i \rightarrow x_i, \bar{t} \rightarrow \e t$) in contrast to Galilean case.    
 
The equations of motion in Galilean case and Carrollian case are also different. Non-relativistic  limit taken on the relativistic equation yields
\bes
\bea{}
\hspace{-2cm}\mbox{{\bf{Electric sector}}:}\qquad &&\p_i \partial_i A_t = 0, \quad \partial _j \partial_j A_i - \partial _i \partial_j A_j + \partial_t \partial _i A_t= 0; \label{Eeom} \\
\mbox{{\bf{Magnetic sector}}:} \qquad &&(\partial _j \partial_j)A_{i}-\partial_{i}\partial_{j}A_{j} =0, \quad (\p_i \partial_i)A_{t}-\partial_{i}\partial_{t}A_{i}= 0. \label{Meom}
\eea \ees
For Carrollian limit, we have
\bes \label{edeom}
\bea{} 
\label{urseleom}\hspace{-2cm}\mbox{{\bf{Electric sector}}:}\qquad && \p_{i}\p_{i}A_{t}-\p_{i}\p_{t}A_{i}=0,~~ \p_{t}\p_{i}A_{t}-\p_{t}\p_{t}A_{i}=0; \\
\label{ursmageom} \mbox{{\bf{Magnetic sector}}:} \qquad && \p_{i}\p_{t}A_{i}=0,~~ \p_{t}\p_{t}A_{i}=0.
\eea
\ees

We now want to focus on the differences that occur when we check for the symmetries of the EOM in both cases.  We have worked in  the scale-spin representation. The scale-spin representations of GCA and CCA are determined by the set $\{\D, j, a, b\}$. The constants $a,b$ are boost labels  on the fields, exclusive to each spin. For Galilean Electrodynamics (GED), the values of the constants are given by $(a^e,b^e)=(1,0)$ for electric limit and $(a^m, b^m) =(0,1)$ for magnetic limit \footnote{In \cite{Bagchi:2014ysa}, the values of $(a^e,b^e)$ and $(a^m,b^m)$ were taken to be $a^e=-1,b^e=0$ and $a^m=0,b^m=-1$. These sign differences in the two cases arise from sign conventions of the representation theory. When chosen consistently, the signs of the constants turn out to be the same.}. For Carrollian Electrodynamics case, the value of the constants are given by $(a^e,b^e)=(0,1)$ for electric limit whereas for magnetic limit, the values are $(a^m, b^m)=(1,0)$. We see that the boost labels are interchanged between the sectors when we compare both limits. For example, the values of constants in electric sector of GED are same as that in magnetic sector of Carrollian ED. 

Moving on to gauge fields coupled to matter, we also see some differences.  In GED coupled to fermions, the magnetic limit is devoid of any interactions between the fields
\be{msse} \p_{i}\p_{i}A_{t} - \p_{i}\p_{t} A_{i} =0,~~ \p_{i}\p_{i}A_{j} - \p_{i}\p_{j} A_{i} =0,~~
i\sigma_{i}\p_{i}\chi =0 ,~~ i\sigma_{i}\p_{i}\phi +i\p_{t}\chi=0.\ee
 In Carrollian ED coupled to fermions, only the sectors with $(\a>0)$ of electric limit is devoid of the interaction terms and can be seen explicitly from the equations
\begin{eqnarray}\hspace{-.5cm}
 \p_i (\partial_i A_t-\p_t A_i)=0,~
\p_t (\p_t A^j-\p_j A_t)=0,
i\p_t \phi=0,~
i\p_t \chi+ i\sigma_{i}\p_i \phi=0.
\end{eqnarray}
Another point of difference is that while taking generalised scaling of fermions ($\phi, \chi$) as
\bea{} \phi \rightarrow \e^{\a} \phi,~\chi \rightarrow \e^{\b} \chi, \eea
the minimum value of $\a$  for electric limit turns out to be $\a =-1$ and for magnetic limit  $\a=-\frac{1}{2}$ in Galilean case, whereas in Carrollian case, the values are given as $\a=0$ (for electric limit) and $\a=-\frac{1}{2}$ (for magnetic limit). 
These minimum values of $\a$ show that  any value below this, fails to reproduce the free equations (kinetic terms). 
 
Lastly, we consider the example of $SU(N)$ YM theory.  For the case, $\mathfrak{D}-p_0 +1 \leq a \leq \mathfrak{D}$ in $SU(N)$ Carrollian YM  (section \ref{eer}), we have two separate scalar equations because of the magnetic index $J$. 
For $J=1$, the equation is denoted by $\mathcal{F}$ \eqref{fff} and for $J>1$, we obtain $\mathcal{E}$ \refb{ffd}.
Correspondingly, we get two different constraints on $a_m$ for $SU(N)$ Carrollian YM coupled to fermions. They are given as
\bea{} a_m \geq 0 ~~\text{for $J=1$ and}~~ a_m \geq -\frac{1}{2} ~~\text{for $J>1$}. \eea
The sectors where the $\mathcal{E}$ hold, are not ``nice" sectors, as these equations have only interaction pieces and no kinetic terms. This means that we need to discard all those sectors where $\mathcal{E}$ has a non-trivial realisation. In fact, the only surviving sector with $J>1$ is the purely magnetic one as $\mathcal{E}$ drops off there. 
 In $SU(N)$ Gallilean YM, we do not have any such divisions because we do not get any equation similar to $\mathcal{E}$. We only have single scalar equation for  $\mathfrak{D}-p_0 +1 \leq a \leq \mathfrak{D}$ case and it is given as $\p_i \p_i A_t^{ \alpha}=0$. On the one hand, this is a simplification in terms of the underlying equations. On the other hand, the absence of the equivalent of equation $\mathcal{E}$ means that one has a huge number of possible sectors to deal with in the Galilean YM theories. 

\bigskip
\noindent For the convenience of the reader, we have listed the points we discussed above in an extensive table in the next page. 

\newpage
\begin{center}
\begin{longtable}{ |p{0.5cm}|p{7cm}|p{7cm}|  }
\hline
\multicolumn{3}{|c|}{Comparison between Galilean and Carrollian Conformal theories} \\
\hline\hline
& \quad\textbf{Galilean Conformal Algebra} & \quad\textbf{Carrollian Conformal Algebra} \\
\hline
1. & Scaling of the coordinates: $x_{i} \rightarrow \e x_i,~t \rightarrow t$ along with $c=1$ and $\e \rightarrow 0$. 
Lightcones flatten out. 
& Scaling of the coordinates: $x_{i} \rightarrow x_i,~t \rightarrow \e t$  along with $c=1$ and $\e \rightarrow 0$. Lightocones close up. \\ \hline
2. & Presence of  Infinite extension of the algebra in all dimensions. & Presence of Infinite extension of the algebra in $d=2,3$ case and partial infinite extension in $d=4,5,..$ cases.  \\ \hline
3.&Galilean boost and the SCT are $B_i=t\p_i, K_i =t^2\p_i, K=-t^2\p_t -2tx_i \p_i$. & Carrollian boost and SCT have the form $B_i =x_i\p_t, K_i =-2x_i (t\p_t +x_k \p_k)+x^2\p_i ,K=x^2 \p_t$. \\\hline
4. &We will write some of the commutators to show the difference between two algebras. They are written down as $$[B_i,P_j]=0,~[K_i,P_j]=0.$$ & Similarly, for CCA we have $$[B_i,P_j]=-\delta_{ij}H,~[K_i,P_j]=-2D\delta_{ij}-2J_{ij}.$$ \\\hline
5. &We can have representations labelled by the scale and boost in all dimensions. & Here the scale-boost representation is only possible in $d=2$ case. For $d>2$ case, there are no non-trivial representation that are labelled by weights under scale and boost. \\\hline\hline
\hline
& \textbf{Galilean Conformal Field theories} & \textbf{Carrollian Conformal Field theories} \\
\hline
6. & In case of scalars, the equation of motion is given as $\p_i\p_i\varphi =0$.  Here, we have no time derivatives. & For scalars, the equation is given as $\p_t\p_t\varphi=0$. We see the absence of spatial derivatives.  \\\hline
7. & In fermion case, one of the equations just contains the spatial derivatives and is given by $i\sigma_i \p_i \phi=0$& Whereas, in this case, one of the equations contains only the time derivative and is given as $i\p_t \phi=0$. \\\hline
8.& In GED case, the contravariant vectors are associated with electric sector, whereas the covariant vectors give rise to the magnetic sector. & In Carrollian ED case, the contravariant vectors are associated with the magnetic sector, whereas the covariant vectors denotes the electric sector. \\\hline
9. & The constants that defines the scale-spin representation are given by $a,b$. For GED, the values of the constants are given by $a^e=1,b^e=0$ for electric limit and $a^m =0,b^m =1$ for magnetic limit.& In Carrollian ED case, the values of the constants are given by $a^e=0,b^e=1$ for electric limit whereas for magnetic limit, the values are $a^m =1, b^m=0$. \\\hline
10. & In GED coupled with fermions, the magnetic limit is devoid of any interactions between the fields.& In Carrollian ED coupled with fermions, the electric limit (only the sectors with $\a>0$) is devoid of the interaction terms. \\\hline
11. & In $SU(2)$ GYM, the self interactions among gauge fields are present in EEE, EMM sector.& In $SU(2)$ Carrollian YM, the self interactions among gauge fields are present in EEM, MMM sector. \\\hline
12. & 
In $SU(N)$ GYM, we do not have any such divisions because we do not get any equation similar to $\mathcal{E}$. 
We only have a single scalar equation for the $\mathcal{D}-p_0 +1 \leq a \leq \mathcal{D}$ case.  & For the case $\mathcal{D}-p_0 +1 \leq a \leq \mathcal{D}$ in $SU(N)$ Carrollian YM, we have two separate scalar equation because of the magnetic index $J$. For $J=1$, the equation is denoted by $\mathcal{F}$ and for $J>1$, we have $\mathcal{E}$. Due to such division, we get two different constraints on $a_m$ for $SU(N)$ Carrollian YM coupled with fermions. For $J=1$, we have $a_m \geq 0$ and $a_m \geq -\frac{1}{2}$ for $J>1$.\\\hline

13. & Number of sectors in GYM is large as there are no sectors we discard for non-existence of a kinetic term & The number of sectors in Carrollian YM is more limited due to the existence of sectors where there are EOM like $\mathcal{E}$. This class of EOM only contain interaction terms. So all sectors with non-trivial realisations of this class of EOM are discarded. \\\hline

\end{longtable}
\end{center}
\newpage
\bibliographystyle{JHEP}
\bibliography{ccft}

\end{document}